\documentclass{aa}

\usepackage[varg]{txfonts}
\usepackage{graphicx}
\usepackage{algorithm}
\usepackage{algorithmic}
\usepackage{color}
\usepackage{placeins}

\usepackage{natbib,twoopt}
\usepackage[breaklinks=true]{hyperref} 
\bibpunct{(}{)}{;}{a}{}{,} 
\newcommand{\Treyer}{Treyer et al. (in prep.)}

\begin{document}

\title{Photometric Redshift Estimation with Convolutional Neural Networks and Galaxy Images: A Case Study of Resolving Biases in Data-Driven Methods}

\author{Q. Lin\inst{\ref{inst1}}
\and D. Fouchez\inst{\ref{inst1}}
\and J. Pasquet\inst{\ref{inst2}, \ref{inst3}}
\and M. Treyer\inst{\ref{inst4}}
\and R. Ait Ouahmed\inst{\ref{inst4}}
\and S. Arnouts\inst{\ref{inst4}}
\and O. Ilbert\inst{\ref{inst4}}}

\institute{Aix-Marseille Univ., CNRS/IN2P3, CPPM, Marseille, France\label{inst1}
\and UMR TETIS, Univ. Montpellier, France\label{inst2}
\and AgroParisTech, Cirad, CNRS, Irstea, Montpellier, France\label{inst3}
\and Aix Marseille Univ., CNRS, CNES, LAM, Marseille, France\label{inst4}}

\date{Received; accepted}

\abstract{Deep Learning models have been increasingly exploited in astrophysical studies, yet such data-driven algorithms are prone to producing biased outputs detrimental for subsequent analyses. In this work, we investigate two major forms of biases, i.e., class-dependent residuals and mode collapse, in a case study of estimating photometric redshifts as a classification problem using Convolutional Neural Networks (CNNs) and galaxy images with spectroscopic redshifts. We focus on point estimates and propose a set of consecutive steps for resolving the two biases based on CNN models, involving representation learning with multi-channel outputs, balancing the training data and leveraging soft labels. The residuals can be viewed as a function of spectroscopic redshifts or photometric redshifts, and the biases with respect to these two definitions are incompatible and should be treated in a split way. We suggest that resolving biases in the spectroscopic space is a prerequisite for resolving biases in the photometric space. Experiments show that our methods possess a better capability in controlling biases compared to benchmark methods, and exhibit robustness under varying implementing and training conditions provided with high-quality data. Our methods have promises for future cosmological surveys that require a good constraint of biases, and may be applied to regression problems and other studies that make use of data-driven models. Nonetheless, the bias-variance trade-off and the demand on sufficient statistics suggest the need for developing better methodologies and optimizing data usage strategies.}

\keywords{Galaxies: distance and redshifts -- Surveys -- Methods: data analysis -- Techniques: image processing}

\maketitle

\section{Introduction} \label{sec:intro}

Estimation of galaxy redshifts is crucial for the studies of galaxy evolution and cosmology. While redshifts obtained by spectroscopic measurements (spec-\textit{z}) typically have high accuracy, they are highly time-intensive and not ideal for an extremely large size of data from ongoing or future imaging surveys (e.g., DES, Euclid, HSC, LSST, WFIRST). Therefore, photometric redshifts (photo-\textit{z}) have become an alternative for galaxies without spectroscopically confirmed redshifts.

There are two broad categories of methods to estimate photometric redshifts for individual galaxies --- template-fitting methods and data-driven methods (see \citet{Salvato2019} for a review). Template-fitting methods model the galaxy spectral energy distribution (SED) and infer redshifts by fitting the galaxy photometry based on the SED templates \citep[e.g.,][]{Arnouts1999, Feldmann2006, Ilbert2006, Greisel2015, Leistedt2019}.

Data-driven or empirical methods predict redshifts by establishing a function or a density estimator that maps the input photometry or images to redshift estimates. These include Self-organizing Maps \citep[SOMs; e.g.,][]{Way2012, Carrasco2014, Speagle2017, Buchs2019, Wilson2020}, Artificial Neural Networks \citep[ANNs; e.g.,][]{Collister2004, Bonnett2015, Hoyle2016, DP2018, Pasquet2019, Mu2020photoz, Ansari2021, Schuldt2021} and other Machine Learning algorithms \citep[e.g.,][]{Gerdes2010, CB2013, Rau2015, Beck2016, Jones2017, Hatfield2020}. In particular, studies such as \citet{Pasquet2019} develop models to predict redshift probability densities for individual galaxies, which are trained with stamp images of galaxies as input and categorical labels converted from spectroscopic redshifts\footnote{We note that spectroscopic redshifts are not \textit{true} redshifts, but we use them as training labels and the reference for testing as they are generally measured with a good accuracy.}, i.e., turning the problem of redshift estimation into a classification problem. Images of galaxies may be superior over other forms of input such as galaxy photometry, since the model may further incorporate the spatial flux information of galaxies and avoid biases inherited from photometry extraction. Unlike template-fitting methods, data-driven methods do not depend on any theoretical modeling as all information is learned from data.

Besides individual estimates, calibrating redshift sample distributions is important for applications such as weak lensing analysis \citep[e.g.,][]{Mandelbaum2008, Hemmati2019, Euclid2021}. In addition to assembling individual estimates, one can calibrate the redshift distribution for a photometric sample by conducting spatial cross correlation with a reference sample that has accurate spectroscopic redshift information \citep[e.g.,][]{Newman2008, Davis2017, Morrison2017, Kovetz2017, Gatti2022}.

In general, the aforementioned methods suffer from systematic effects coming from different sources. For example, template-fitting methods would be problematic if SED templates do not match the physical properties of the to-be-fitted galaxy sample, while data-driven methods would fail outside the photometric parameter space covered by the training sample. The hybridization of empirical-template or cross correlation-template methods has the potential to improve the estimation accuracy \citep[e.g.,][]{Cavuoti2016, Sanchez2018, Alarcon2020, Soo2021}, and it is promising to develop composite likelihood inference frameworks for photometric redshift estimation \citep[e.g.,][]{Rau2022}.

Data-driven methods such as Deep Learning neural networks have been increasingly utilized in many astrophysical applications mainly in the manner of supervised learning. Apart from photometric redshift estimation discussed above, there have also been other cases in which data-driven methods outperform benchmark techniques, such as estimating the mass of the Local Group \citep{McLeod2017}, measuring galaxy cluster masses \citep{Ntampaka2015, Armitage2019, Ntampaka2019, Yan2020}, measuring galaxy properties \citep{Dominguez2018, Bonjean2019, Wu2019}, estimating cosmological parameters \citep{Ravanbakhsh2016b, Gupta2018, Ribli2019} and analyzing forming planets \citep{Alibert2019}. Nonetheless, purely data-driven methods may not be able to perfectly capture salient information concerned for a certain task and the output estimates would easily suffer from biases of various kinds. This issue has been noticed in a few studies \citep[e.g.,][]{Voigt2014, Zhang2018imbalance, Hosenie2020, Burhanudin2021, Cranmer2021}, though there have not been extensive investigations on resolving biases in data-driven models in the field of astrophysics.

Using photometric redshift estimation as an example, we tackle the following forms of biases in our work\footnote{We note that the biases discussed in this work would exist for both the training set and the test set no matter whether they follow the same distribution in the parameter space. Other than those biases, there is another commonly-known bias, namely mismatch or covariate shift between the training set and the test set, meaning that the two sets differ in the input domain. This would be a particularly challenging problem if the training set has incomplete \textit{coverage} of the parameter space represented by the test set, typically faint objects without spectroscopic follow-ups. We expect that our bias correction methods (discussed in Section~\ref{sec:methods}) are robust to some degree of mismatch in number density between the two sets, as long as there is no coverage mismatch. A full discussion on coverage mismatch is beyond the scope of this work.}:

\textbf{(a) Class-dependent residuals.} This is known as the \textit{long-tail} recognition problem in the domain of computer science in which training data follows an imbalanced distribution with a low number density in the tail classes. A network trained with such imbalanced data tends to give predictions skewed towards \textit{over-populated} classes, resulting in class-dependent residuals. In this work, each spectroscopic redshift bin $[z_{spec}, z_{spec}+\delta z_{spec}]$ is considered as a class. This bias is presented as non-zero dependence of mean residuals as a function of redshifts. There have been several techniques for tackling this bias outside astrophysics, including data-level re-sampling \citep[e.g.,][]{Chawla2002, Han2005, Garcia2012, Buda2018}, algorithm-level exploitation of class-dependent losses or techniques \citep[e.g.,][]{Huang2016imbalance, Khan2017, Cao2019NEURIPS, Cui2019CVPR, Hayat2019, Khan2019, Huang2020imbalance, Duarte2021, wu2021adversarial} and feature transfer or augmentation \citep[e.g.,][]{Liu2019imbalance, Tong2019imbalance, Yin2019CVPR, Okerinde2021}.
We note that it might be non-trivial to apply a correction with a tenable statistical basis or determine representative features for astronomical data, and it is more difficult to use these techniques for the redshift estimation problem than a classic classification problem, because there are non-uniform correlations between redshift bins. With these considerations, this work adopts and extends the technique presented by \citet{Kang2020Decoupling} at the data level without introducing complex algorithms or loss functions. This will also serve as a framework for resolving mode collapse (discussed below).

Furthermore, although one can under-sample data to suppress \textit{over-populated} classes, there could be \textit{under-populated} classes for which the number density is too low or even zero, precluding the reconstruction of a flat distribution via re-sampling. In particular, there are virtually ``zero-populated classes'' beyond the boundaries set by the finite coverage of data. Depending on the amplitude of estimation errors, there may be significant skewness close to the boundaries of the output. This effect exists even when the training data follows a perfectly uniform distribution inside the coverage of data.

In our work on photometric redshift estimation, mean residuals can be computed with two different views; one is defined in the spectroscopic space while the other is defined in the photometric space:
(i) \textit{Mean residuals as a function of spectroscopic redshifts}, i.e., defined in each spectroscopic redshift bin $[z_{spec}, z_{spec}+\delta z_{spec}]$, quantifying the deviations relative to ground-truth values.
(ii) \textit{Mean residuals as a function of photometric redshifts}, i.e., defined in each photometric redshift bin $[z_{photo}, z_{photo}+\delta z_{photo}]$, quantifying the deviations relative to measured or expected values. The correction of such residuals is essential for applications such as weak lensing analysis when a sample is divided in tomographic redshift bins \citep[e.g.,][]{Mandelbaum2008, Hemmati2019, Euclid2021}.
We note that the biases given by these two definitions are inter-correlated and cannot be eliminated simultaneously in the presence of inhomogeneous estimation errors that have different projections in the spectroscopic space and the photometric space. In other words, correcting biases defined with one view will cause biases with the other if the measurements are imperfect. Therefore, the biases with these two views have to be considered separately according to the specific requirement of the study. While \citet{Pasquet2019} does not find strong $z_{photo}$-dependent residuals in simple cases where training data and test data are randomly sampled from the same parent distribution, we are interested in controlling biases in more general scenarios. In the methods presented in this paper, one must first resolve biases in the spectroscopic space to produce a least-biased reference sample before calibrating the measurements in the photometric space.

\textbf{(b) Mode collapse.} This typically refers to the phenomenon that the output sample distribution has collapsed modes that are not representative of the full diversity of the actual distribution\footnote{A conceptually similar term is boundary distortion \citep{Santurkar2018}, referring to the phenomenon that a classifier trained with GAN-generated data instances has a skewed decision boundary due to a lack of diversity compared with those trained with real data, which limits the performance of the classifier.}.
The problem of mode collapse is widely discussed in the context of Generative Adversarial Networks \citep[GANs; e.g.,][]{Arjovsky2017, Kodali2017, Srivastava2017, Zhang2018mode, Jia2019mode, Bhagyashree2020, Thanh2020, Li2021gan}, but it also exists in other situations \citep[e.g.,][]{Nguyen2018, Ruff2018, Chong2020mode}. In our work, we use this term to indicate the inability of the classifier to reproduce the diversity of the target output domain.

Collapsed modes can be understood as local regions with over-confidence or degeneracies, meaning that estimated values are not smoothly distributed but concentrated at discrete positions, resulting in a series of spikes that make the estimated sample distribution deviate from the actual distribution. Mode collapse may be caused by noise and contamination from nearby sources on images that pollute the salient information necessary for determining correct output estimates. Meanwhile, there are correlations between adjacent redshift bins, which may hinder the establishment of a clear discrimination among similar redshift values and lead to ``fuzzy” decision boundaries between bins. In addition, the network may be under-trained or lack complexity so that it is unable to fully capture salient information in data. Mode collapse implies an imbalance between the locally collapsed output due to an imperfect mapping from the input and the perfect target output imposed by hard (one-hot) labels. Hence, we will deal with mode collapse by leveraging soft labels.

The goal of this work is not to achieve a better global accuracy or a lower catastrophic rate compared to state-or-the-art results; rather, we attempt to develop methodologies to examine and resolve biases, which are critical to photometric redshift estimation for cosmological applications \citep[e.g.,][]{Laureijs2011}. Our methods are also expected to be applicable to a wider variety of cases involving data-driven models. Due to unresolved prior in density estimates produced by neural networks, we only focus on point estimates in our work. The outline of this paper is as follows. In Section~\ref{sec:data}, we describe the data used in this work, including galaxy images and spectroscopic redshift labels. The images are collected from two surveys, i.e., the Sloan Digital Sky Survey (SDSS) and the Canada–France–Hawaii Telescope Legacy Survey (CFHTLS). In Section~\ref{sec:methods}, we introduce our methods for tackling class-dependent residuals and mode collapse with respect to both spectroscopic redshifts and photometric redshifts. Section~\ref{sec:results} presents our main experiments on bias correction for photometric redshift estimation and makes a link to cosmological applications. In Section~\ref{sec:discussions_short}, we focus on the behaviors of the biases, investigating how the biases would evolve with varying implementing and training details and testing the validity of our methods. Detailed discussions are presented in Appendix~\ref{sec:discussions}. In Section~\ref{sec:comparison}, we compare our estimated photometric redshifts with those obtained by other studies. Finally, we conclude in Section~\ref{sec:conclusion} and discuss prospects for future research.

\section{Data} \label{sec:data}

\subsection{Sloan Digital Sky Survey (SDSS)}

The Sloan Digital Sky Survey (SDSS) is a wide-field imaging and spectroscopic survey that covers roughly 1/3 of the celestial sphere. The first dataset in our work consists of stamp images of 496,524 galaxies and associated spectroscopically determined redshifts selected by \citet{Pasquet2019} from SDSS Data Release 12 \citep{Alam2015}. Each stamp image covers one of the five photometric passbands ($ugriz$), and has $64\times64$ pixels in spatial dimensions, corresponding to a pixel scale of 0.396 arcsec. A galaxy is aligned to the center of each image using its angular coordinates. The pair of five images in the five passbands and the associated spectroscopic redshift forms a data instance. The dataset covers $z<0.4$ in spectroscopic redshift and $r<17.8$ in dereddened $r$-band petrosian magnitude. Due to the good data quality and the relatively restricted parameter space (i.e., low redshifts and high brightnesses), the SDSS dataset is the main source of data utilized for developing and evaluating our bias correction methodologies. Similar to \citet{Pasquet2019}, we randomly select 393,219 instances as a training sample and 103,305 instances as a test sample. More detailed descriptions can be found in \citet{Pasquet2019} and \citet{Alam2015}.

\subsection{Canada–France–Hawaii Telescope Legacy Survey (CFHTLS)}

In order to apply our methods in a larger parameter space (i.e., high redshifts and faint magnitudes), we also exploit multi-band ($ugriz$) galaxy images acquired from the two components of the Canada–France–Hawaii Telescope Legacy Survey (CFHTLS; see \citet{CFHTLST07} for the latest data release) --- a Deep Survey covering 4 deg$^2$ in four disconnected fields, reaching a depth of $r = 25.6$ defined as the 80\% completeness limit for stellar sources; and a Wide Survey covering four fields over a total area of 155 deg$^2$ with a depth of $r = 25.0$.

The spectroscopic redshifts for the CFHTLS images are a compilation of data from released wide and/or deep spectroscopic surveys. These include the CLAMATO survey \citep[Data Release 1;][]{KGLee2018},
the DEEP2 survey \citep[Data Release 4;][]{Newman2013},
the GAMA survey \citep[Data Release 3;][]{Baldry2018},
the SDSS-BOSS Data Release 12,
the UDS survey \citep{McLure2013, Bradshaw2013}
the VANDELS survey \citep[Data Release 4;][]{Garilli2021},
the VIPERS survey \citep[Data Release 2;][]{Scodeggio2018},
the VUDS survey \citep{LeFevre2015},
the VVDS Wide and Deep surveys \citep{LeFevre2013},
the WiggleZ survey \citep[Final Release;][]{Drinkwater2018} and the z-COSMOS survey \citep{Lilly2007}. From the redshift surveys above, we only take the most secure spectroscopic redshifts, identified with high signal-to-noise ratios (SNRs) or measured with multiple spectral features (equivalent to the VIPERS or VVDS redshift flags, i.e., $3\le \mathrm{zFlag} \le$4), which we refer to as high-quality (HQ) redshifts.

We also include the most secure low-resolution prism redshift measurements for bright sources ($\mathrm{zFlag}=4$, $I\le22.5$) from the PRIMUS survey \citep[Data Release 1;][]{Coil2011, Cool2013}, and the secure grism redshift measurements from the 3D-HST survey \citep[Data Release v4.1.5;][]{Skelton2014, Momcheva2016}. They are referred to as low-resolution (LR) spectroscopic redshifts.
 
We select 26,249 galaxies from the CFHTLS Deep fields and 108,510 galaxies from the Wide fields as two separate datasets (denoted with ``CFHTLS-DEEP'' and ``CFHTLS-WIDE'' respectively) with spectroscopic redshifts ranging up to $z \sim 4.0$ and dereddened $r$-band magnitudes in a range of $12.3<r<27.2$. Like the SDSS dataset, the extracted stamp images are in the format of $64\times64\times5$ pixels, corresponding to a pixel scale of 0.187 arcsec (described in detail in \Treyer{}). We randomly select 10,000 instances from the Deep fields and 10,000 instances from the Wide fields with high-quality spectroscopic redshifts as separate test samples. They are not mixed since the Deep fields and the Wide fields have potential systematics. The remaining 16,249 instances from the Deep fields and 98,510 instances from the Wide fields are regarded as training samples. They are separately used in training when applying bias correction.

\subsection{Data processing}

We reduce the pixel intensity of the CFHTLS images by a factor of 1,000 so that the majority of pixel values lie below one. The pixel intensity of all images is then rescaled with the following equation
\begin{equation}
I = \begin{cases}
-\sqrt{-I_0 + 1.0} + 1.0, \quad I_0 < 0\\
\sqrt{I_0 + 1.0} - 1.0, \quad I_0 > 0
\end{cases}
\label{eq:rescaling}
\end{equation}
where $I$ and $I_0$ stand for the rescaled intensity and the original intensity, respectively. This rescaling operation reduces the disparity between the galaxy peak flux and the remaining flux so that the network can have comparable responses to more pixels.

\section{Methods} \label{sec:methods}

\subsection{Artificial neural networks}

As \citet{Pasquet2019} and \Treyer{}, we use artificial neural networks (ANNs) to estimate photometric redshifts in the manner of supervised learning. An ANN can be viewed as a multi-variable function or a conditional probability density estimator that maps the input to the target output, which consists of a sequence of layers inspired by biological neural networks. In vanilla ANNs, the layers are fully-connected (tailored for one-dimensional input) and composed of neurons that usually contain free parameters (or weights) and non-linear activation functions (e.g., Sigmoid, ReLU, PReLU, etc.). In particular, for classification tasks, a Softmax function is usually applied to the output layer, producing a vector normalized to unity and thus mimicking a probability density distribution. The network is usually trained iteratively with mini-batches of training data. With a loss function defined for a given task, gradients with respect to the weights of each neuron throughout the network are calculated via backpropagation. The weights are updated via gradient descent in order to minimize the loss. Once the loss converges or does not vary rapidly, a parametrization of the network is established that encodes the relationship between the input and the target output, which is collectively determined by the training data and the model implementation strategies including the network architecture, the loss function and the training procedure. For purely data-driven methods such as ANNs, the training data and the model are major sources of prior information that affect the prediction on the test data.

Convolutional neural networks (CNNs) form a special category of ANNs. Unlike vanilla ANNs, the input of CNNs could be high-dimensional such as multi-band images. CNNs possess convolutional layers with kernels (or filters) composed of only a few neurons that are designed for scanning over the whole input image and extracting local information, outputting image-like feature maps. The extracted features are becoming more abstract from layer to layer, evolving gradually from the pixel level to the object level. \citet{Pasquet2019} utilizes CNNs with galaxy images instead of photometric information in order to incorporate the spatial flux distribution and avoid biases from photometry extraction, producing results with a better accuracy compared to benchmark methods \citep[e.g.,][]{Beck2016}, though using images instead of photometry as input increases the dimensionality of the input and introduces the risk of fitting irrelevant spatial information, thus more training data is required.

\subsection{Bias correction procedure}

\begin{figure}[ht]
\begin{center}
\centerline{\includegraphics[width=\columnwidth]{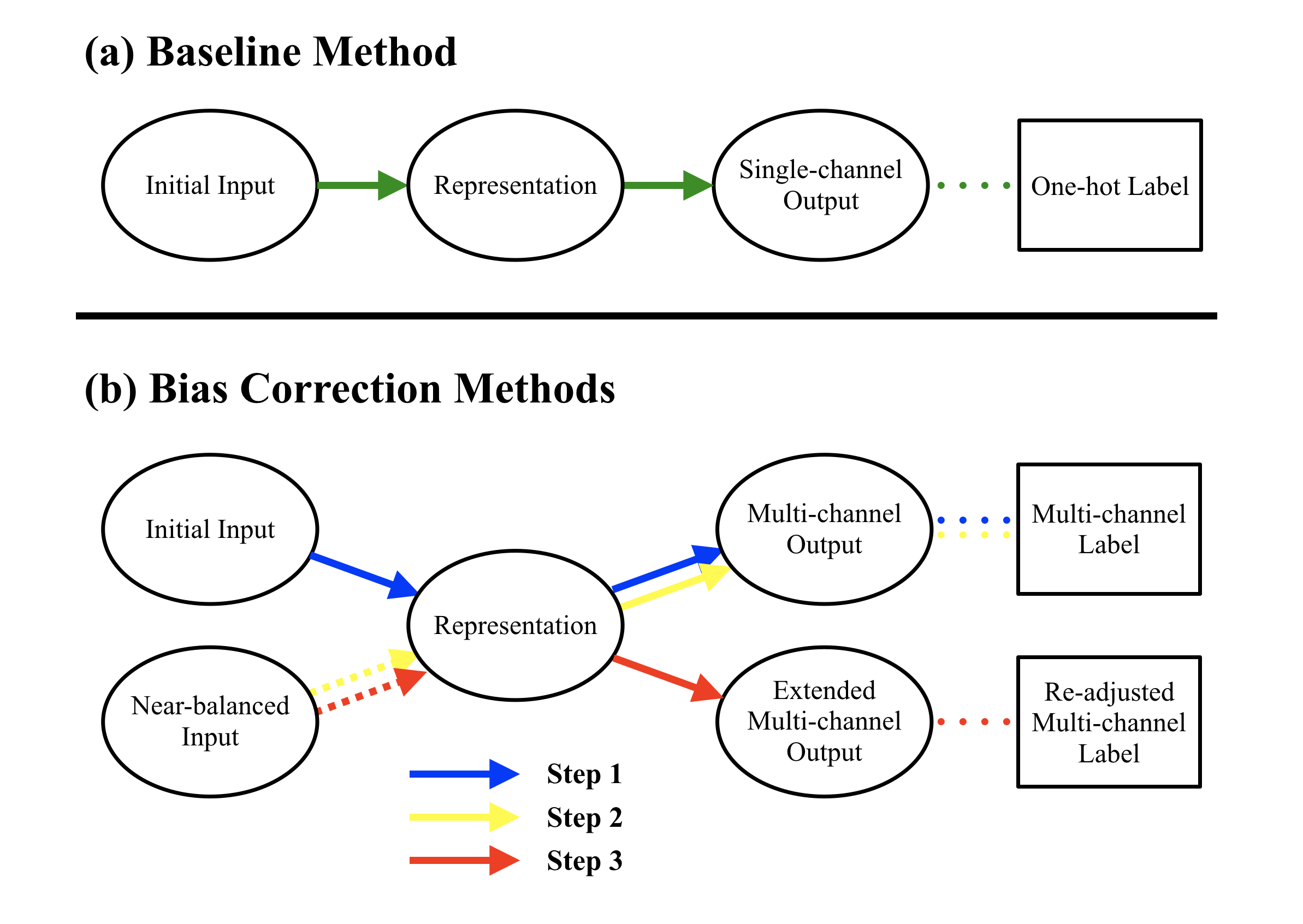}}
\caption{Paradigms of the Baseline method and our methods for correcting $z_{spec}$-dependent residuals and mode collapse. The Baseline method does not explicitly specify a representation and produces single-channel output estimates that only require one-hot labels for training. In contrast, our methods split representation learning and classification into three consecutive steps and produces multi-channel output estimates. The labels for training the model match the dimensions of the output estimates and they are softened and re-adjusted in Step 3 according to the biases. The dashed arrows indicate that the part of the model prior to the representation is fixed in Steps 2 and 3.}
\label{fig:paradigms_pz}
\end{center}
\end{figure}

\begin{figure}[ht]
\begin{center}
\centerline{\includegraphics[width=\columnwidth]{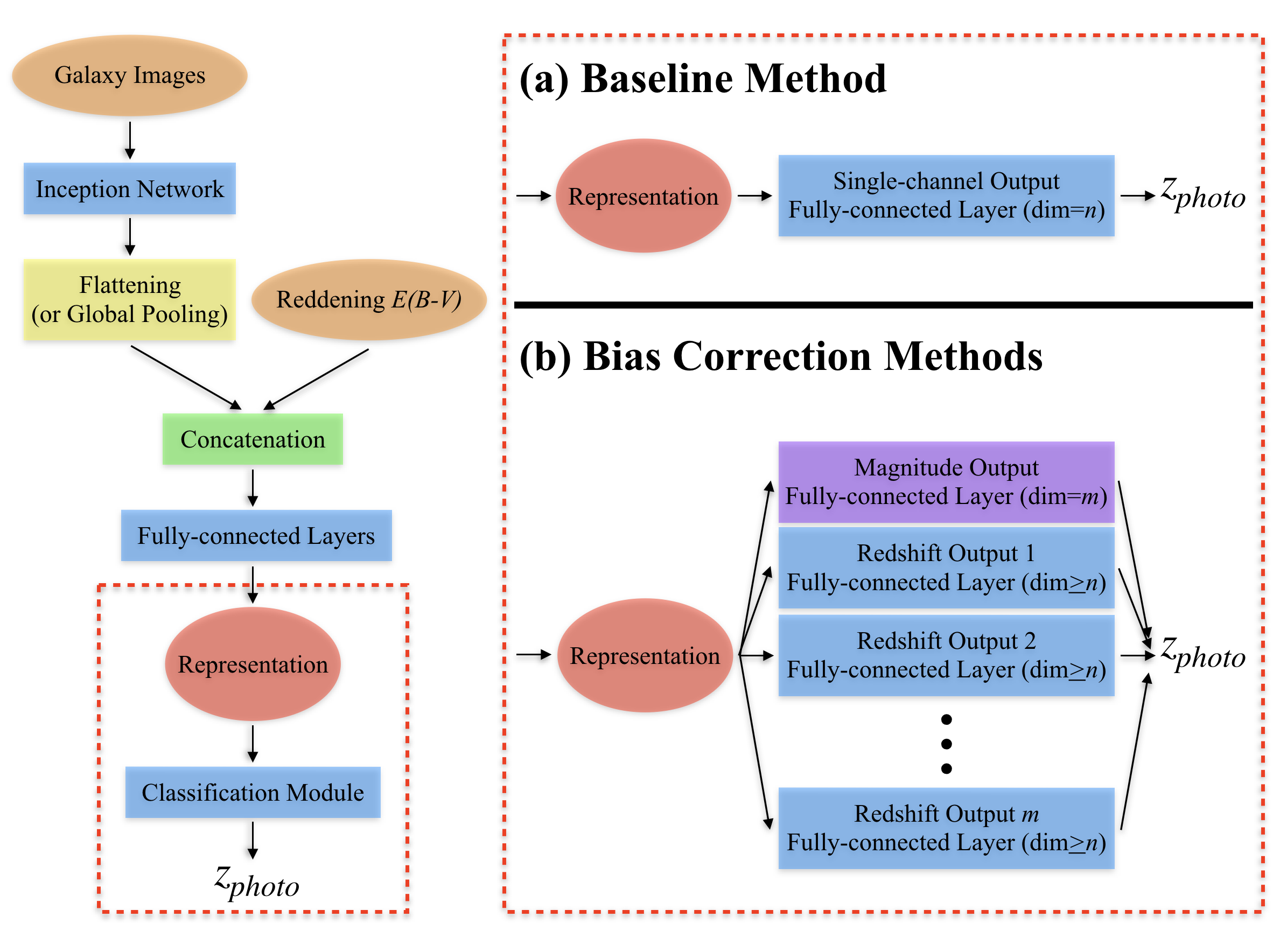}}
\caption{Graphic illustration of the CNN models for the Baseline method and our methods for correcting $z_{spec}$-dependent residuals and mode collapse. We use the network architecture adopted by \citet{Pasquet2019}, in which multi-scale inception modules \citep{Szegedy2015} are utilized and the galactic reddening of galaxies is inputted in addition to images. The representation is obtained at the penultimate fully-connected layer. The classification module is elaborated in the red rectangle: while the Baseline method takes a final fully-connected layer as a single-channel output for redshifts, our methods take parallel fully-connected layers as multiple output channels to predict redshifts which also incorporate the magnitude information (discussed in Section~\ref{sec:step1}). The labels ``$n$'' and ``$m$'' indicate the dimensions of the output layers, whose values are given in Tab.~\ref{tab:coverage} for the SDSS data and the CFHTLS data. ``$\geq n$'' means that additional bins are appended to the redshift outputs in Step 3.}
\label{fig:nets_pz}
\end{center}
\end{figure}

In this subsection, we present our proposed procedure for correcting the two biases (summarized in Algorithm~\ref{alg:steps}). As mentioned in Section~\ref{sec:intro}, residuals can be defined as a function of either spectroscopic redshifts or photometric redshifts, which have to be treated separately. Specifically, in our bias correction methods, Steps 1 -- 3 are applied to correct $z_{spec}$-dependent residuals and mode collapse (Fig.~\ref{fig:paradigms_pz}). These steps are built on CNN models with a statistical basis (Appendix~\ref{sec:formalism}) leveraging pre-estimated redshifts for analyzing and resolving biases, which should be applied in advance of Step 4. Step 4 is a direct calibration applied to correct $z_{photo}$-dependent residuals, which may not be necessary depending on actual requirements on bias correction. These four steps will be compared with a baseline method that is not corrected for biases. Although we will discuss this procedure only in the setting of photometric redshift estimation, we note that it is potentially applicable to any study that involves data-driven models.

As will be discussed in Section~\ref{sec:step2}, correcting $z_{spec}$-dependent residuals requires a selection of data from the original training dataset. It is impractical to train a deep network using only a selected small subset of data. However, as illustrated by \citet{Kang2020Decoupling}, the learning procedure of a model can be split into \textit{representation learning} and \textit{classification}\footnote{We only focus on classification problems in this work. It is simpler to apply our bias correction methods to classification problems than regression problems, as the probability density predicted by classification is easier to modify than a single collapsed point estimate predicted by regression. Extensions to regression problems may be explored in future work.}. In this idea, the whole network is first trained with all training data to obtain a representation at a chosen hidden layer, which is usually a low-dimensional embedding of information from input data (see, e.g., \citet{Bengio2013} for a review on representation learning); then the classification output can be fine-tuned or re-trained with the representation fixed. Having representation learning and classification decoupled makes it possible to apply the bias correction techniques solely in the phase of classification even in the presence of a biased representation, as long as full representativeness is essentially maintained in the representation. Fixing the representation also makes it possible to apply bias correction using a small subset of data without strong overfitting.

For Steps 1 -- 3 in our main experiments, we adopt the network architecture from \citet{Pasquet2019} (denoted with \texttt{Net\_P}; illustrated in Fig.~\ref{fig:nets_pz}) that makes use of the multi-scale inception modules \citep[e.g.,][]{Szegedy2015} to extract features at different resolution scales. The feature maps produced by a set of convolutional and pooling layers, expected to encapsulate the extracted salient information, are flattened, concatenated with an additional input of the galactic reddening $E(B-V)$, and fed into fully-connected layers for the final prediction. The network from \Treyer{} (denoted with \texttt{Net\_T}) has fewer trainable parameters but a deeper architecture. We note that overfitting is likely to occur for both networks as a result of limited training data. For applying our bias correction methods, the penultimate fully-connected layer of the network, which has 1,024 nodes, is used to obtain a representation, and the last fully-connected layer is replaced by a multi-channel output unit (discussed in Section~\ref{sec:step1}).

Furthermore, the output of a classification model is a density estimate subject to the prior imposed by the model implementation and the training data. A full posterior distribution should be a marginalization over all such prior and possible parametrizations, which is not tractable. Therefore, we only focus on point estimates in our analysis. While \citet{Pasquet2019} and \Treyer{} take $z_{mean}$ (i.e., the probability-weighted mean redshift) and $z_{median}$ (i.e., the median redshift drawn from the probability distribution) as point estimates respectively, we consider $z_{mode}$ (i.e., the redshift corresponding to the maximum probability density) as point estimates for both our methods and the Baseline method, as it is least biased by the prior (discussed in Appendix~\ref{sec:zcomparison}).

We elaborate on the details of each step below.

\begin{algorithm}[H]
\begin{itemize}
\item \textbf{Step 1. Representation learning:} Replace the last layer with a multi-channel output unit (i.e., a classification module) and train the whole network from scratch using all training data with multi-channel hard labels.
\item \textbf{Step 2. For correcting over-population-induced $z_{spec}$-dependent residuals:} Fix the learned representation and fine-tune the multi-channel output unit using a near-balanced subset of training data with multi-channel hard labels.
\item \textbf{Step 3. For correcting mode collapse and under-population-induced $z_{spec}$-dependent residuals:} Extend the redshift range of the multi-channel output unit. Fix the learned representation and re-train the extended multi-channel output unit using the near-balanced training subset with re-adjusted multi-channel soft labels.
\item \textbf{Step 4. For correcting $z_{photo}$-dependent residuals:} Re-sample from the training data and construct a sample whose $z_{spec}$-magnitude distribution is the same as the $z_{photo}$-magnitude distribution of the test sample (regarded as the underlying $z_{spec}$-magnitude distribution). In each $z_{photo}$-magnitude cell, estimate the mean deviation between $z_{photo}$ and $z_{spec}$ using the re-sampled training data and subtract it from the $z_{photo}$ estimated with the test sample. \textit{Note: this step may be unnecessary depending on actual applications.}
\end{itemize}
\caption{Bias correction steps}
\label{alg:steps}
\end{algorithm}

\subsubsection{Step 1: replacing the last fully-connected layer by a multi-channel unit and learning a representation} \label{sec:step1}

\begin{figure}[ht]
\begin{center}
\centerline{\includegraphics[width=\columnwidth]{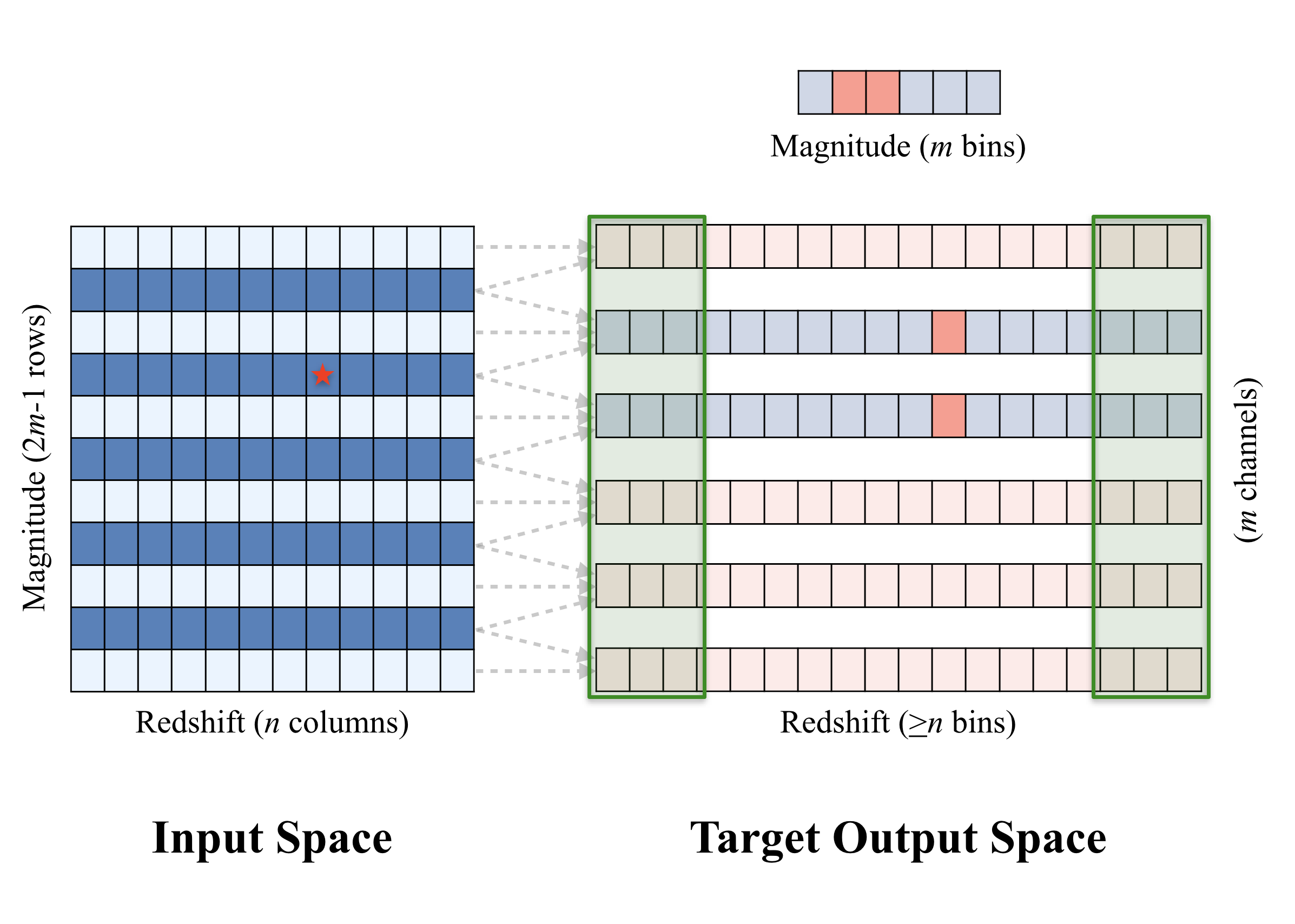}}
\caption{Illustration of the correspondence between the input parameter space and the target output space represented by the multi-channel output unit. The input space is divided into a grid with $2m-1$ rows covering different magnitudes and $n$ columns covering different redshifts. The output unit is comprised of one magnitude output with $m$ bins and $m$ corresponding redshift outputs. Each redshift output has $n$ bins in correspondence with the input space before being extended in Step 3. The values of ``$n$'' and ``$m$'' for the SDSS data and the CFHTLS data are given in Tab.~\ref{tab:coverage}. The shaded green boxes indicate the extended redshift intervals. Each of the rows in the input space is connected to one or two redshift outputs alternately. Or equivalently, each redshift output responds to adjacent three input rows (or two rows on the edges). As an example, in response to a data instance drawn from the input space indicated by the red star, the target magnitude output has two bins activated (shown in dark red) while the other bins remain null (shown in light blue); there is one bin being activated in each of the two corresponding redshift outputs (before using re-adjusted soft labels in Step 3), while the remaining redshift outputs have a flat response over all bins (shown in light red).}
\label{fig:z_io_factorize}
\end{center}
\end{figure}

Since the data for training a model spans a multi-dimensional spectroscopic and photometric parameter space, imbalances in any influencing dimension would bias the redshift prediction. For instance, a model with the standard single-channel output (i.e., the Baseline method) tends to assign larger redshifts to infrequent low-redshift faint galaxies if the galaxy brightness in each redshift bin is not well balanced in the training data, even though the brightness is not explicitly predicted.

Therefore, it is important to explicitly specify and balance such influencing dimensions. For simplicity, we use two dominant variables, \textit{spectroscopic redshift} and \textit{dereddened $r$-band magnitude}, to approximate the multi-dimensional parameter space. We adopt a multi-channel output unit in our methods that consists of one magnitude output with $m$ bins and $m$ associated redshift outputs (Fig.~\ref{fig:z_io_factorize}), where the magnitude information can be explicitly predicted. Before being extended in Step 3 (discussed in Section~\ref{sec:step3}), the redshift outputs preserve the same binning and coverage as the single-channel output (with $n$ bins), each corresponding to a bin in the magnitude output. All the redshift outputs and the magnitude output are implemented as single fully-connected layers with the Softmax activation and connected to a common layer, which is taken as the representation in our analysis (Fig.~\ref{fig:nets_pz}). In other words, the outputs are conditionally independent given the representation. The final redshift output prediction incorporates all the outputs and marginalizes over the magnitude, i.e.,
\begin{equation}
p(z) = \sum_{j=1}^m p(r_{o(j)}) p(z|r_{o(j)})
\label{eq:factorize}
\end{equation}
where $p(r_{o(j)})$ and $p(z|r_{o(j)})$ denote the density distributions predicted by the magnitude output and the redshift outputs, respectively; $r_{o(j)}$ denotes the $j$-th bin in the $r$-band magnitude output; the index $j$ runs over $m$ bins. This multi-channel unit is essentially a classification module that can be fine-tuned or re-trained for bias correction once the representation is obtained. It builds a basis for the successive bias correction steps, and can also be regarded as a way of regularizing the model and preventing overfitting.

As shown in Fig.~\ref{fig:z_io_factorize}, the input space is divided into a grid of $z_{spec}$-magnitude cells, with $2m-1$ rows in magnitude and $n$ bins in redshift. The non-negligible $r$-band magnitude measurement errors might lead data instances to incorrect magnitude rows in the input space. Directly using incorrect magnitudes as labels for the magnitude output would confuse the model. Therefore, we adopt a magnitude output with bins that have \textit{interlacing} coverages over the input space. Specifically, each magnitude row in the input space is accounted for by one or two adjacent bins in the magnitude output in an alternate manner; each magnitude bin is responsible for adjacent three magnitude rows (or two rows on the borders). The rows that contribute to two bins alleviate the potential confusion at the boundary between the bins. In practice, the $n$ redshift bins cover the whole spectroscopic redshift distribution of the data, whereas the $2m-1$ magnitude rows only cover the dominant portion of the $r$-band magnitude distribution. The data instances that lie above or below the grid are re-assigned to the uppermost or lowermost magnitude rows.

While only high-quality data should be used for the subsequent steps, it is possible to utilize data with various systematics and SNRs for representation learning, since the focus at this stage is to build an informative representation rather than control the biases. We train our model with all data from scratch as the Baseline method, except that the last fully-connected layer that gives a single-channel classification output is replaced by the multi-channel output unit. We utilize the Cross Entropy loss function for each output, i.e.,
\begin{equation}
l(p,y) = -\sum_j y_j \log p_j
\label{eq:CE}
\end{equation}
where $p$ stands for the probability density predicted by the Softmax activation for each of the redshift outputs or the magnitude output; $y$ is the corresponding one-channel label, one component of the integrated multi-channel label. The index $j$ runs over all bins in each redshift output or the magnitude output. By convention, both $p$ and $y$ are normalized to unity. The labels for the magnitude output obeys the one-hot or two-hot encoding, depending on whether the magnitude value lies in a row that contributes to one or two magnitude bins in the target output due to the interlacing correspondence. The labels for the redshift output(s) associated with the activated magnitude bin(s) are one-hot (yet they will be softened in Step 3; discussed in Section~\ref{sec:step3}), whereas flat labels are used for the remaining redshift outputs (i.e., $y_j = 1/n$ for all bins where $n$ is the number of bins) in order to impose confusion on the non-activated magnitude bins. The final output combines the density distributions predicted by all the redshift outputs weighted by the corresponding probabilities given by the magnitude output (i.e., Eq.~\ref{eq:factorize}).

\subsubsection{Step 2: for correcting $z_{spec}$-dependent residuals induced by over-populated classes} \label{sec:step2}

\begin{figure}[ht]
\begin{center}
\centerline{\includegraphics[width=\columnwidth]{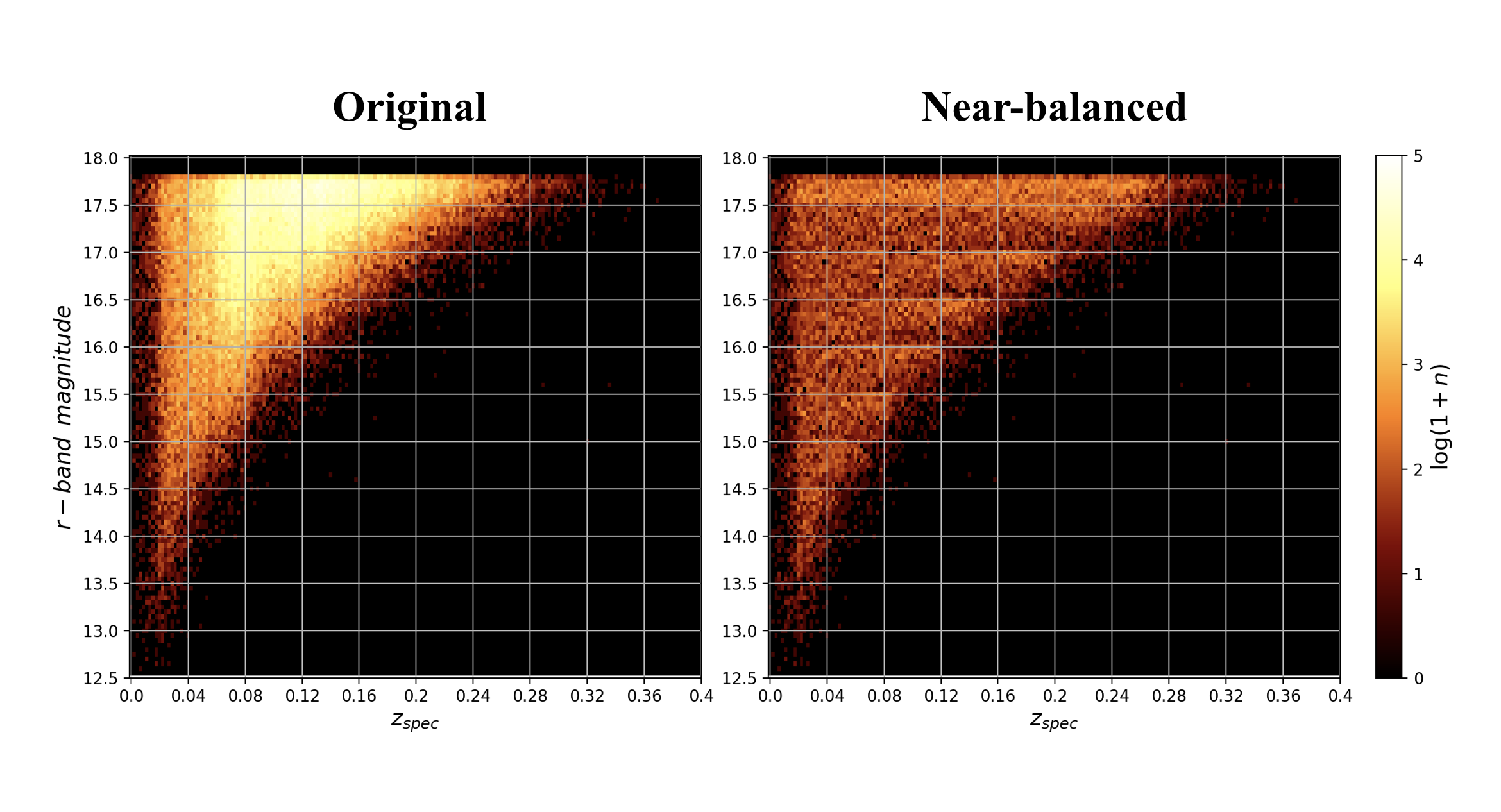}}
\caption{Two-dimensional distributions ($z_{spec}$, $r$-band magnitude) for the SDSS dataset before and after balancing the number density. The color indicates the relative number density in the logarithmic scale. As shown, the data cannot be perfectly balanced due to the existence of under-populated classes.}
\label{fig:step2}
\end{center}
\end{figure}

The $z_{spec}$-dependent residuals are heavily subject to the influence of over-populated classes. We note that simply increasing the rate of occurrences of the instances from other classes in training is not effective for damping the dominance of over-populated classes, presumably because this does not balance the \textit{information content} in different classes. Therefore, we attempt to correct the influence of over-populated classes by constructing a near-balanced training subset via under-sampling. Specifically, in each $z_{spec}$-magnitude cell $(z_{spec}, r_i)$ in the division of the input space, a number of instances from the training dataset are randomly selected no more than a pre-defined balancing threshold $th$; all the cells with number densities above the threshold are thus considered over-populated and trimmed off. In this manner, the balancing is carried out over both the redshift dimension and the magnitude dimension (Fig.~\ref{fig:step2}). Then, the re-sampled subset is exploited for fine-tuning the multi-channel output unit pre-trained in Step 1 with the representation fixed.

In other words, we attempt to create a near-flat prior over all bins so that the trained model will not favor over-populated bins, despite the fact that they may originally hold more information content that is useful for the model prediction. Nonetheless, resampling in a two-dimensional subspace will inevitably overlook high-order information in the intrinsic heterogeneous multi-dimensional parameter space, yet incorporating more dimensions in the output unit will further complicate the model and aggravate the sparsity of data. We therefore only focus on the ``first-order'' correction in the two-dimensional subspace and leave the higher-order correction for future work. It should also be noted that the overall estimation accuracy would be slightly traded off due to a decrease in the size of the training sample.

\subsubsection{Step 3: for correcting mode collapse and $z_{spec}$-dependent residuals induced by under-populated classes} \label{sec:step3}

\begin{figure}[ht]
\begin{center}
\centerline{\includegraphics[width=\columnwidth]{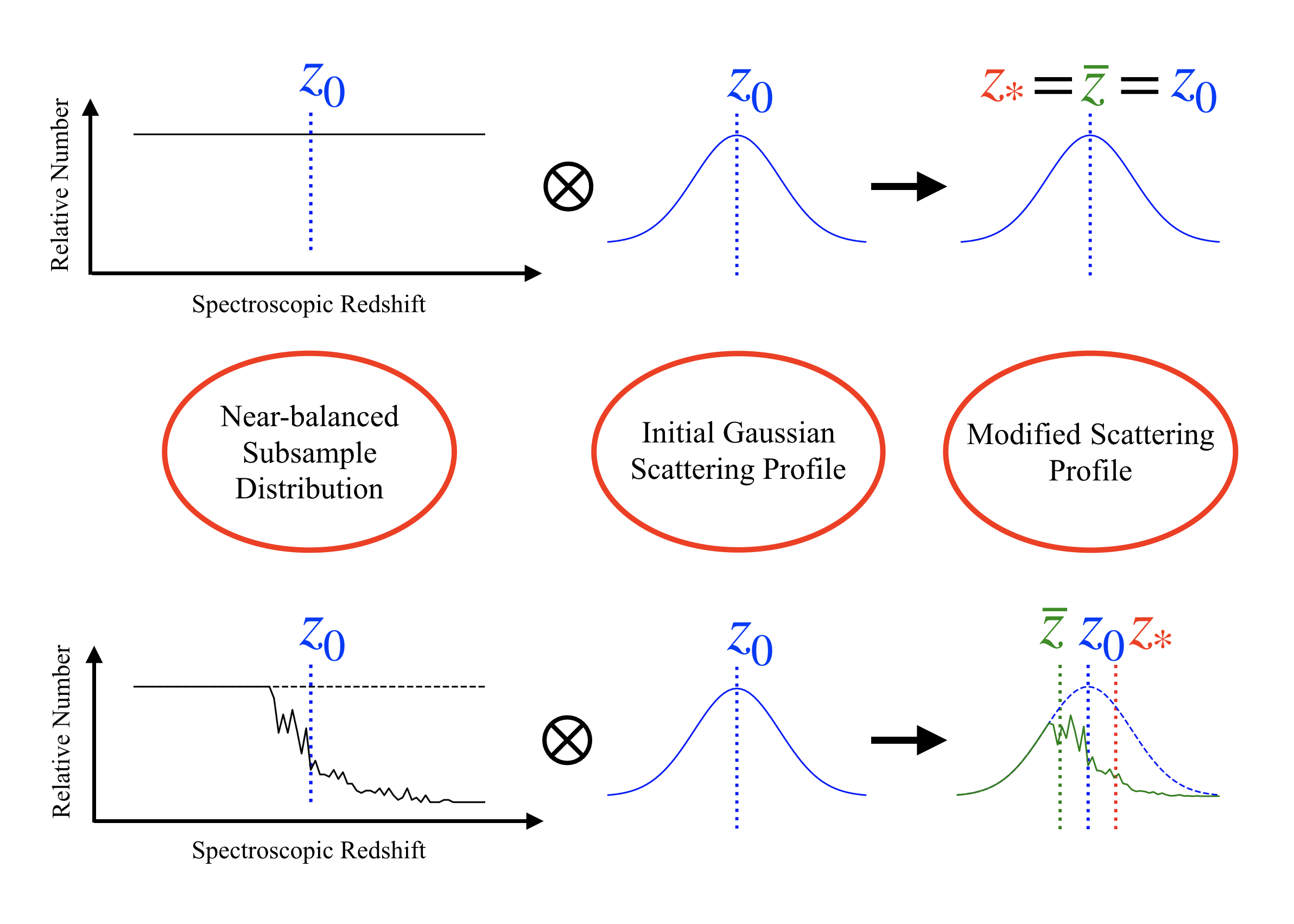}}
\caption{Illustration of determining the means of soft labels. $z_0$, $\overline{z}$ and $z_*$ stand for the spectroscopic redshift of a given galaxy, the mean of its skewed scattering profile and the determined mean of its soft label, respectively, also indicated by the blue, green and red dotted vertical lines. The skewed scattering profile is assumed to be a bin-wise product of the initial Gaussian profile and the shape of the subsample distribution in the given magnitude bin. Ideally, if the subsample distribution is perfectly balanced (top row), the product profile will remain as a Gaussian function such that the mean $\overline{z}$ of the profile and the mean $z_*$ of the soft label align with the spectroscopic redshift value $z_0$. On the other hand, if the subsample distribution has under-populated regions (bottom row), the product profile will be affected by non-flatness and deviate from a Gaussian function (i.e., skewed; shown as the green curve), which produces a shifted mean $\overline{z}$. The mean $z_*$ of the soft label in this case should be determined according to Eq.~\ref{eq:zmean}.}
\label{fig:step3}
\end{center}
\end{figure}

The use of a near-balanced training subset in Step 2 is unlikely to completely eliminate the $z_{spec}$-dependent residuals due to unaddressed under-populated classes, especially the hypothetical zero-populated classes outside the finite parameter space covered by the data\footnote{Over-populated classes and under-populated classes may be relatively defined with the balancing threshold $th$, yet not \textit{all} classes can be reconciled with a low threshold due to limited statistics and/or zero-populated classes.}. Furthermore, collapsed modes exist regardless of the balancing of training data.

Using the same near-balanced training subset and the fixed representation from Step 2, we consider jointly tackling mode collapse and under-population-induced $z_{spec}$-dependent residuals by replacing the one-hot redshift components of the multi-channel labels with soft Gaussian labels for the activated redshift outputs. In particular, the dispersions of the soft Gaussian labels are introduced to suppress mode collapse, and the means are re-adjusted to address under-population-induced $z_{spec}$-dependent residuals. They are determined according to the pre-estimated redshifts from Step 2.

As a prerequisite, we first extend the range of the redshift output with more bins to allow for the softening and re-adjustment of labels especially for data instances close to the boundaries. We note that density or point estimates that exceed the initial boundaries are permitted, in particular ``negative'' estimates if the lower boundary is shifted below zero.

\textbf{(a) Determining the dispersions.} We apply a uniform dispersion to soften all the hard labels for a given redshift output corresponding to a given bin in the magnitude output. The dispersion is denoted with $\sigma_1(r_o)$, which is a function of the given magnitude bin $r_o$. It is fitted by minimizing the Kullback-Leibler divergence between the $z_{spec}$ subsample distribution at the given magnitude bin $r_o$ from Step 2 and the corresponding predicted $z_{photo}$ subsample distribution convolved with a Gaussian function whose dispersion is $\sigma_1(r_o)$, i.e.,
\begin{equation}
\arg\min_{\sigma_1(r_o)} D_{\mathrm{KL}} [p(z_{spec}|r_o) || (p*N_{\sigma_1^2(r_o)}) (z_{photo}|r_o)]
\label{eq:sigma1}
\end{equation}
where $p(z_{spec}|r_o)$ and $p(z_{photo}|r_o)$ denote the $z_{spec}$ and $z_{photo}$ subsample distributions given the magnitude bin $r_o$, respectively; $N_{\sigma_1^2(r_o)}(z_{photo}|r_o)$ denotes the Gaussian function with a dispersion of $\sigma_1(r_o)$. In other words, the fitted dispersions are thought to offer the minimal amount of softening for smoothing out the spikes in the $z_{photo}$ subsample distribution so as to approximate the $z_{spec}$ subsample distribution. The soft Gaussian labels with such dispersions can be leveraged to train the multi-channel output unit and correct mode collapse. 

In order to avoid introducing imbalance along the redshift dimension, we do not use redshift-dependent dispersions. Nonetheless, the fitted dispersions are different for different magnitude bins, which is a sign of varying photometric properties, data quality and statistics. Dividing the input data into the two-dimensional space is thus required for applying separate fits for the dispersions. Additionally, for data instances from a magnitude row in the input space that is linked to two bins in the magnitude output, two different dispersions are fitted and applied to two redshift outputs separately. 

In contrast to hard labels that assign no weight to any non-activated redshift bins, soft\footnote{We use the term ``soft'' in a sense different from the ``smoothing labels'' (i.e., uniform smoothing over all non-activated bins) proposed by \citet{Muller2019}.} labels would naturally account for the correlations between bins since the dispersions implicitly indicate the ``influential scales'' around the central values. However, we are cautious about a trade-off between the use of soft labels and the constraint of estimation errors. The dispersions would be dramatically over-estimated if the pre-estimation of photometric redshifts is at a low accuracy, and applying soft labels with over-estimated dispersions would further boost errors, making the estimation unreliable. In this sense, the proper fit of dispersions (and more generally, the whole procedure of bias correction) relies on a good control of estimation errors.

\textbf{(b) Determining the means.} The use of original spectroscopic redshifts as hard labels in Step 2 yields a ``scattering profile'' for each redshift estimate, which models the probability distribution of being randomly scattered to different values under the influence of the subsample corresponding to a given bin in the magnitude output. For a flat $z_{spec}$ subsample distribution, the scattering profile for each instance is assumed to be a Gaussian function (different from the aforementioned soft Gaussian labels), whose mean or center of mass is aligned with its spectroscopic redshift. In contrast, an imbalanced $z_{spec}$ subsample distribution due to under-populated bins would imprint non-zero skewness on the scattering profile and alter its mean. Therefore, we attempt to compensate such skewness by relocating the soft labels (illustrated in Fig.~\ref{fig:step3}).

We assume that the skewed scattering profile is simply a bin-wise product of the initial Gaussian profile and the shape of the subsample distribution. The validity of this assumption is tested in Appendix~\ref{sec:modeling_underp}.

The dispersion of the initial Gaussian scattering profile is estimated as the local error at a given spectroscopic redshift and a given bin in the magnitude output. Assuming a uniform error distribution along the redshift as $\sigma_1(r_o)$ may be over-simplified. Therefore, we attempt to model the underlying evolving \textit{variance} by determining a redshift-dependent quadratic error curve for each magnitude bin (denoted with $\sigma_2^2(z_{spec}|r_o)$). We first estimate the squared distances $d^2$ between $z_{photo}$ and $z_{spec}$ in each cell $(z_{spec}, r_o)$ and average their mean and median, which gives the \textit{average} squared distances $\overline{d^2}(z_{spec}|r_o)$, i.e.,
\begin{equation}
\begin{aligned}
&\overline{d^2}(z_{spec}|r_o) = \frac{1}{2} [\mathrm{Mean}(d^2(z_{spec}, r_o)) + \mathrm{Median}(d^2(z_{spec}, r_o))] \\
&\mathrm{where} \,\,\, d^2(z_{spec}, r_o) = (z_{photo} - z_{spec})^2 \,\,\, \mathrm{with \, given} \,\,\, (z_{spec}, r_o)
\end{aligned}
\label{eq:avg_sq_dist}
\end{equation}
The reason for averaging the mean and the median is that the former may be driven by outliers while the latter is typically lower than the underlying variance. We also note that if an instance is from a magnitude row in the input space connected to two bins in the magnitude output, it will contribute to two redshift outputs separately for estimating the squared distances.

The \textit{average} squared distances $\overline{d^2}(z_{spec}|r_o)$ can be seen as the raw estimates of variances for the scattering profiles. However, such raw estimates contain random fluctuations and the correlations between redshift bins are not taken into account. Knowing that $\sigma_1(r_o)$ fitted with Eq.~\ref{eq:sigma1} characterizes the ``influential scale'' around each bin, we obtain the quadratic error curve $\sigma_2^2(z_{spec}|r_o)$ by further convolving $\overline{d^2}(z_{spec}|r_o)$ with a Gaussian function with a dispersion of $\sigma_1(r_o)$, i.e.,
\begin{equation}
\sigma_2^2(z_{spec}|r_o) = (\overline{d^2}*N_{\sigma_1^2(r_o)}) (z_{spec}|r_o)
\label{eq:sigma2}
\end{equation}
In addition, since we make use of the labels softened by $\sigma_1(r_o)$ on top of the already existing estimation errors, the final estimates of variances for the scattering profiles should be the quadratic error curves $\sigma_2^2(z_{spec}|r_o)$ expanded by $\sigma_1(r_o)$ in quadrature, i.e.,
\begin{equation}
Var(z_{spec}|r_o) = \sigma_1^2(r_o) + \sigma_2^2(z_{spec}|r_o)
\label{eq:sigma1_plus_sigma2}
\end{equation}

With a scattering profile estimated for each instance, the under-population-induced imbalance (characterized by the shift of the mean of the skewed profile relative to its spectroscopic redshift) can be offset by re-adjusting the mean of the Gaussian label, i.e.,
\begin{equation}
\begin{aligned}
&z_* = z_0 - (\overline{z} - z_0) \\
&\mathrm{where} \,\,\, \overline{z} = \int_{z_{spec}} z_{spec} p(z_{spec}|r_o) N_{z_0,var(z_0|r_o)}(z_{spec}) dz_{spec}
\end{aligned}
\label{eq:zmean}
\end{equation}
$z_*$ stands for the new mean of the Gaussian label; $z_0$ stands for the spectroscopic redshift; $\overline{z}$ stands for the mean of the skewed scattering profile, which is the product of $p(z_{spec}|r_o)$ and $N_{z_0,var(z_0|r_o)}(z_{spec})$ --- the $z_{spec}$ subsample distribution at the given magnitude bin $r_o$ and the initial Gaussian scattering profile with a mean of $z_0$ and a variance of $Var(z_0|r_o)$. In particular, the imbalances present at the boundaries of the finite coverage of data can be corrected in the same manner by assigning zero number density to the hypothetically zero-populated classes exceeding the boundaries. 

To summarize, using the pre-determined redshift point estimates for the near-balanced training subset from Step 2, we estimate the dispersions and the means of the soft Gaussian labels, according to Eq.~\ref{eq:sigma1} and Eq.\ref{eq:zmean} respectively. The new labels are exploited to re-train the multi-channel output unit with an extended redshift range, using the same training subset and keeping the learned representation fixed. With this step, both mode collapse and $z_{spec}$-dependent residuals induced by under-populated classes are expected to be corrected.

\subsubsection{Step 4: for correcting $z_{photo}$-dependent residuals}

This step is unnecessary for studies that require bias correction with respect to $z_{spec}$. However, cosmological applications such as weak lensing analysis require a calibration of the mean photometric redshift in each tomographic bin, which necessitates the correction of $z_{photo}$-dependent residuals. Although the knowledge of $z_{photo}$-dependent residuals is not accessible from the test sample, we may obtain this knowledge from the training sample, as long as the same sampling process applies to each $z_{spec}$-magnitude cell $(z_{spec}, r_i)$ in both samples. We note that $r_i$ in this case stands for a magnitude row in the input space rather than a bin in the magnitude output. In principle, for a test sample with any ``expected'' two-dimensional $z_{spec}$-magnitude distribution, we can construct a sample with the same distribution by resampling from the training data. Then in each $z_{photo}$-magnitude cell $(z_{photo}, r_i)$, the bias (i.e., the deviation between the mean $z_{photo}$ and the mean $z_{spec}$) estimated with the resampled training data can be subtracted from the $z_{photo}$ estimated with the test data. We emphasize that we attempt to match the $z_{spec}$-magnitude distributions rather than the $z_{photo}$-magnitude distributions since the latter cannot guarantee the same $z_{photo}-z_{spec}$ deviations for both samples. Furthermore, this step only requires the reference training data to have the same sampling as the test data, thus it is not limited to CNN models.

Correctly estimating the $z_{spec}$-magnitude distribution of the test sample is important in this procedure, since it enables the resampled training sample and the test sample to have the same bias distribution over all $z_{photo}$-magnitude cells. For simplicity, the estimated $z_{photo}$-magnitude distribution of the test sample corrected for $z_{spec}$-dependent residuals and mode collapse is regarded as its expected $z_{spec}$-magnitude distribution. We note that it is necessary to first apply Steps 1 -- 3 to correct $z_{spec}$-dependent residuals and mode collapse, since these biases (e.g., spikes due to mode collapse) may modify the $z_{photo}$-magnitude distribution. However, the estimated $z_{photo}$-magnitude distribution is still not identical to the $z_{spec}$-magnitude distribution due to the existence of estimation errors. Nonetheless, given the similarity between the $z_{photo}$ and $z_{spec}$ distributions, we do not attempt to correct estimation errors in order to avoid other biases.

\section{Main experiments and results} \label{sec:results}

\subsection{Training protocol}

\begin{table*}\footnotesize
\caption{Data coverage in spectroscopic redshift and $r$-band magnitude.} \label{tab:coverage}
\centering
\begin{tabular}{l | c c c c c}
\hline
   &  $z_{min}$ ($z_{min}'$)\tablefootmark{*}  &  $z_{max}$ ($z_{max}'$)\tablefootmark{*}  &  \# Initial Bins ($n$)  &  Bin Size  &  \# Appended Bins (left/right)\tablefootmark{*}  \\
\hline
SDSS  &  0.0 ($-0.2$)  &  0.4 (0.6)  &  180  &  $2.2\times10^{-3}$  &  90/90 \\
\hline
CFHTLS  &  0.0 ($-2.0$) &  4.0 (8.0) &  1,000  &  $4.0\times10^{-3}$  &  500/1,000  \\
\hline
\hline
   &  $r_{min}$  &  $r_{max}$  &  \# Input Rows ($2m-1$)  &  Row Size  &  \# Output Bins ($m$)  \\
\hline
SDSS  &   12.5  &  18.0  &  11  &  0.5  &  6 \\
\hline
CFHTLS  &  16.0  &  25.5  &  19  &  0.5  &  10 \\
\hline
\end{tabular}

\tablefoot{
\tablefoottext{*}{The redshift range is extended in Step 3 by appending additional bins with the original bin size to both sides of the initial redshift output. $z_{min}'$ and $z_{max}'$ denote the new boundaries after extension.}
}
\end{table*}

\begin{table*}\footnotesize
\caption{Hyperparameters in training.} \label{tab:hyper}
\centering
\begin{tabular}{l | c c c c}
\hline
   &  Learning Rate\tablefootmark{*}  & \# Iterations &  Mini-batch Size &  Balancing Threshold \\
\hline
Baseline \& Step 1 (SDSS)  &  $10^{-4}$   & 120,000  & 128  &  --- \\
\hline
Baseline \& Step 1 (CFHTLS)  &  $10^{-4}$   & 120,000  & 32  &  --- \\
\hline
Step 2 \& Step 3 (SDSS)  &  $2\times10^{-5}$   & 40,000  & 32  &  200 \\
\hline
Step 2 \& Step 3 (CFHTLS-DEEP)  &  $2\times10^{-5}$   & 40,000  & 32  &  5 \\
\hline
Step 2 \& Step 3 (CFHTLS-WIDE)  &  $2\times10^{-5}$   & 40,000  & 32  &  100 \\
\hline
\end{tabular}
\tablefoot{
\tablefoottext{*}{Reduced to $2\times10^{-5}$ after 60,000 iterations for the Baseline and Step 1.}
}
\end{table*}

Throughout this work, we train CNN models and apply our bias correction methods for each dataset (SDSS, CFHTLS-WIDE and CFHTLS-DEEP) using the training data. Unless otherwise noted, we show results only with the test data, assuming that both the training and test samples follow the same sampling process in each redshift-magnitude cell $(z_{spec}, r_i)$ in the input space. Exceptionally, we allow this assumption to be violated in the stage of representation learning where the goal is to expand the information volume learned by the model despite the presence of distinct systematics or sub-structures in the training data. However, the existence of such systematics would jeopardize the balancing of the data distribution, thus the training data must be purified for the later stage of bias correction. For example, the CFHTLS-DEEP data generally have better SNRs than the CFHTLS-WIDE data due to longer exposure time. Given that the model may not be well generalizable over different noise properties, we consider separating these two CFHTLS datasets for bias correction. In our work, the training sample used for bias correction and the test sample are randomly selected from the same parent distribution for each given dataset, though their overall distributions are allowed to be different if they follow the same sampling in each redshift-magnitude cell.

The data coverage and the hyperparameters used in training for each dataset are detailed in Tabs.~\ref{tab:coverage} and \ref{tab:hyper}. To train our models, we adopt the Mini-batch Gradient Descent technique with the default Adam Optimizer \citep{Adam}. In each training iteration, a mini-batch of training instances are randomly selected from the training sample, consisting of image-label pairs. The images are randomly flipped and rotated with 90 deg steps under the assumption of spatial invariance. For the SDSS dataset, the Baseline method and Step 1 are conducted using the whole training sample of 393,219 instances. In Steps 2 and 3, the constructed near-balanced training subset is down-sampled to 60,000 instances for fitting the soft Gaussian labels (Eq.~\ref{eq:sigma1} and Eq.\ref{eq:zmean}). Since we only need to train the output unit of a network with fewer training instances after Step 1, the number of iterations and the mini-batch size are reduced.

For the CFHTLS datasets, the CFHTLS-DEEP training sample of 16,249 instances and the CFHTLS-WIDE training sample of 98,510 instances are mixed for the Baseline method and Step 1, with both high-quality and low-resolution spectroscopic redshifts. Yet for Steps 2 and 3, the CFHTLS-DEEP sample and the CFHTLS-WIDE sample have to be separately considered due to potential systematics, producing two separate models for the two datasets. In addition, only the instances with high-quality spectroscopic redshifts in the training samples are used for bias correction, similar to those in the test samples. There are 11,220 high-quality instances in the near-balanced training subset for the CFHTLS-DEEP sample and down-sampled 60,000 instances for the CFHTLS-WIDE sample.

Finally, we leverage the strategy of averaging an ensemble of models to improve the performance. For the Baseline method, we train the network five times. Each time the training is conducted in the same way except with a different initialization of weights and different selections of mini-batches. The output redshift density estimates from the five individual models are averaged and a point estimate is obtained from each average density estimate. For our methods, similarly, we train five individual models, each with Steps 1 -- 3 consecutively applied. Each time Step 1 starts with a different initialization; then without averaging, the pre-trained model is passed to Steps 2 and 3 with a different but randomly selected near-balanced training subset. For efficiency, the averaging is only applied to the final output estimates in Step 3.

\subsection{Correction of $z_{spec}$-dependent residuals and mode collapse} \label{sec:result1}

\begin{figure*}
\begin{center}
\centerline{\includegraphics[width=0.85\linewidth]{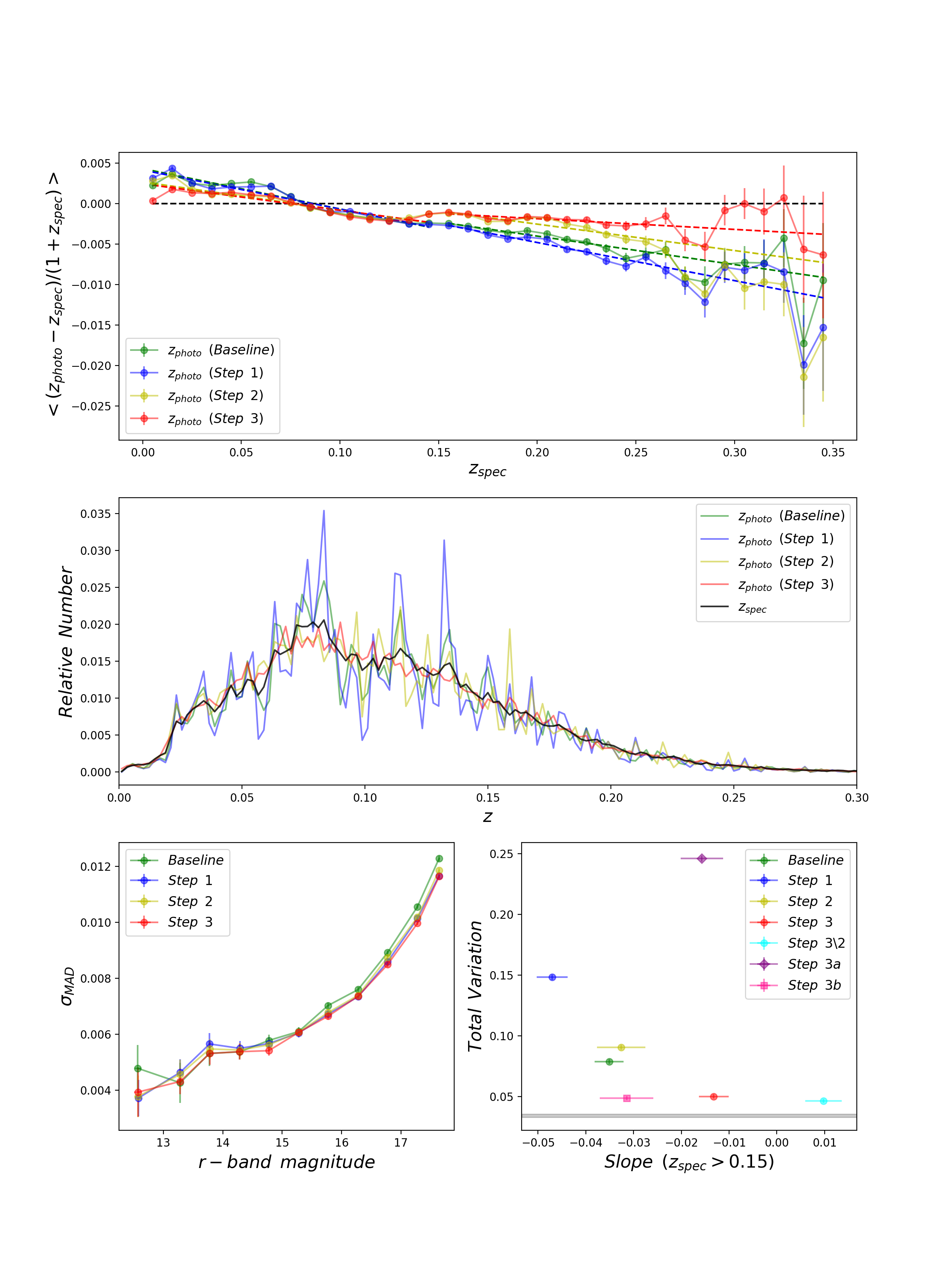}}
\caption{\textbf{Upper Panel:} Mean residuals as a function of spectroscopic redshifts for the Baseline method and Steps 1 -- 3 of our methods with the SDSS data and the network \texttt{Net\_P}. The corresponding $<\Delta z> - \,\, z_{spec}$ piecewise linear fits (Eq.~\ref{eq:resfit}) are shown as the dashed lines with the same colors. Zero residual is indicated by the horizontal dashed black line. \textbf{Middle Panel:} Sample distributions of spectroscopic redshifts and photometric redshifts estimated in different cases. \textbf{Lower Left Panel:} $\sigma_{\mathrm{MAD}}$ (Eq.~\ref{eq:sigma_mad}) as a function of $r$-band magnitudes. \textbf{Lower Right Panel:} Total variation distance between the $z_{photo}$ and $z_{spec}$ sample distributions (Eq.~\ref{eq:d_tv}) as a function of the slope of the $<\Delta z> - \,\, z_{spec}$ piecewise linear fit (Eq.~\ref{eq:resfit}) in the high-redshift interval. The shaded area shows the total variation distance with the $1 \, \sigma$ uncertainty for the simulated $z_{photo}$ sample distribution for the Baseline method that is expected to have no mode collapse. ``Step $3\backslash2$'' denotes a case in which Step 3 is applied without Step 2. ``Step 3a'' denotes a variant case of Step 3 in which hard labels are used instead of soft labels but the means are adjusted. ``Step 3b'' denotes a case in which soft labels are used but the means are not adjusted.}
\label{fig:main_res_SDSS}
\end{center}
\end{figure*}

\begin{figure*}
\begin{center}
\centerline{\includegraphics[width=0.85\linewidth]{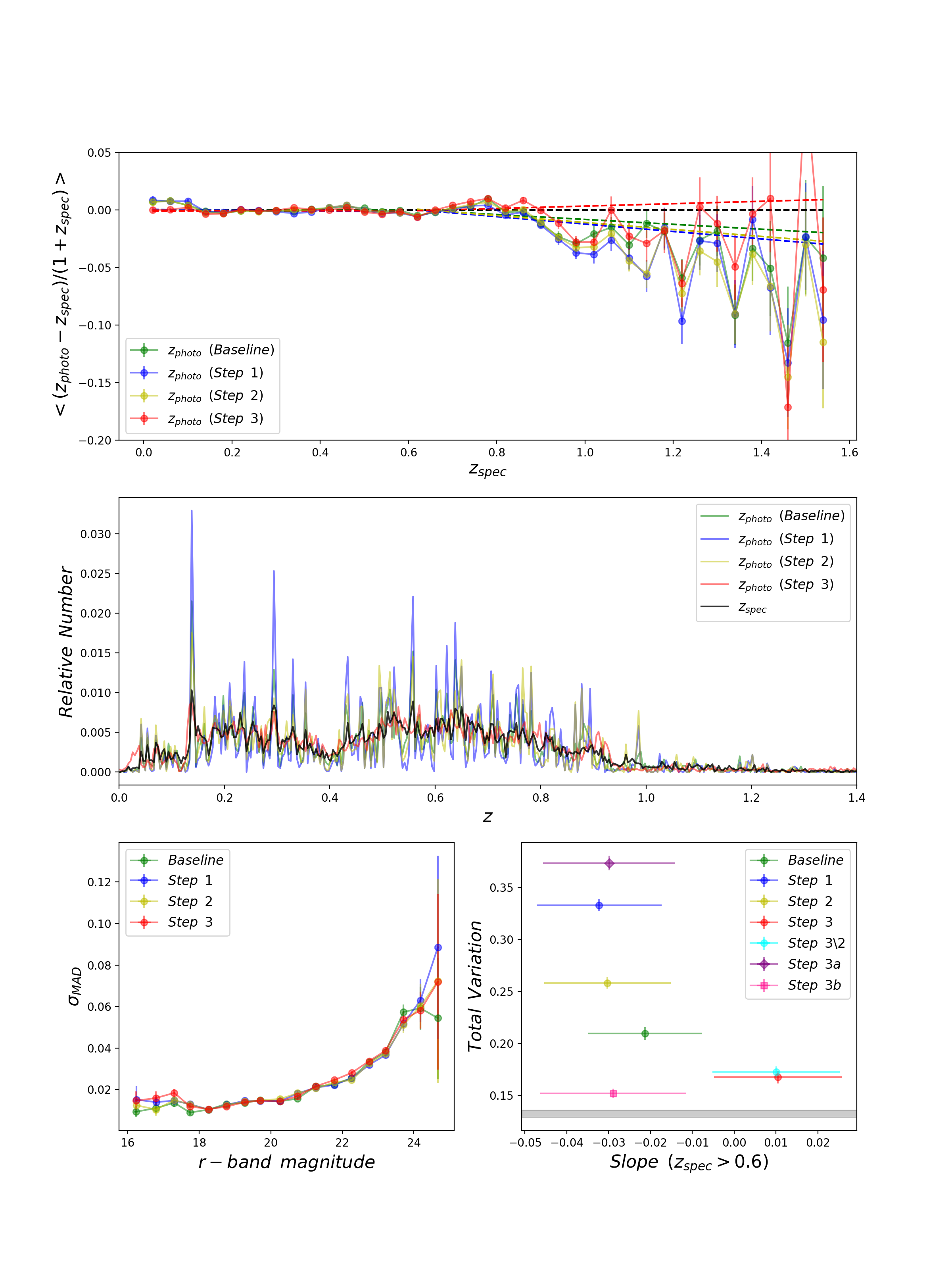}}
\caption{Same as Fig.~\ref{fig:main_res_SDSS}, except with the CFHTLS-WIDE data.}
\label{fig:main_res_CFHTW}
\end{center}
\end{figure*}

\begin{figure*}
\begin{center}
\centerline{\includegraphics[width=1.0\linewidth]{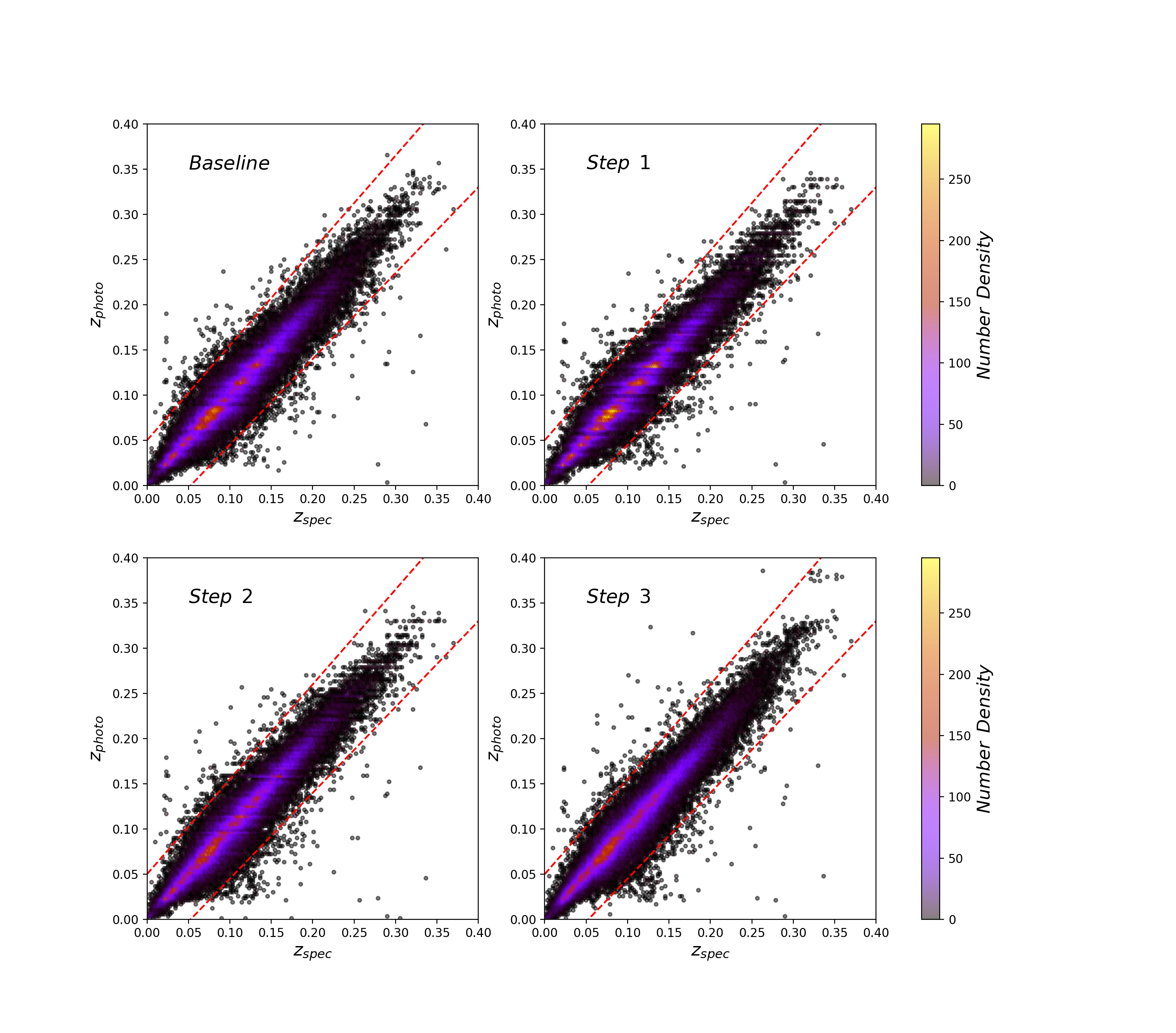}}
\caption{Predicted photometric redshifts as a function of spectroscopic redshifts, color-coded with number density. The Baseline method and Steps 1 -- 3 of our methods are compared using the SDSS data and the network \texttt{Net\_P}. The dashed red lines indicate the boundaries $(z_{photo} - z_{spec}) / (1 + z_{spec}) = \pm 0.05 $ for outliers defined for the SDSS data as in \citet{Pasquet2019} and \Treyer{}.}
\label{fig:photoz_vs_specz}
\end{center}
\end{figure*}

The results of correcting $z_{spec}$-dependent residuals and mode collapse are illustrated in Figs.~\ref{fig:main_res_SDSS} and \ref{fig:main_res_CFHTW} with the SDSS and the CFHTLS-WIDE test data, respectively. For exclusively demonstrating the outcome of bias correction, we do not show the result for the CFHTLS-DEEP dataset due to limited statistics, though we will evaluate the quality of $z_{photo}$ estimation for this dataset in Appendix~\ref{sec:evaluation}. The redshift residual of each data instance indicates the difference between the estimated photometric redshift and the spectroscopic redshift, whose errorbar is estimated via the quadratic error curves $\sigma_2^2(z_{spec}|r_o)$ given by Eq.~\ref{eq:sigma2} (rather than Eq.~\ref{eq:sigma1_plus_sigma2}) with $\sigma_1(r_o)$ fitted on the near-balanced training subset and $\overline{d^2}(z_{spec|r_o})$ estimated on the test data\footnote{If a data instance contributes to two redshift outputs, the two estimated quadratic errors will be averaged.}. The mean residual is defined as the weighted mean in each spectroscopic redshift bin
$[z_{spec}, z_{spec}+\delta z_{spec}]$.

For the Baseline method with either the SDSS data or the CFHTLS-WIDE data, the mean residual generally follows a decreasing trend as a function of the spectroscopic redshift, predominantly resulting from the imbalanced distribution of the training data. To better demonstrate the divergent bias levels in low-redshift and high-redshift regions for each step, we apply the following piecewise linear fit
\begin{equation}
\begin{aligned}
&< \Delta z > = \begin{cases}
A \, z_{spec} + B, \quad z_{spec} < z_b \\
C \, z_{spec} + D, \quad z_{spec} > z_b \\
\end{cases} \\
&\mathrm{where} \,\,\, \Delta z = \frac{z_{photo} - z_{spec}}{1 + z_{spec}}
\end{aligned}
\label{eq:resfit}
\end{equation}
$<>$ denotes the weighted mean. $A$, $B$, $C$ and $D$ are the parameters to be fitted. The boundary $z_b$ defines two redshift intervals ($z_b=0.15, 0.6$ for the SDSS data and the CFHTLS-WIDE data, respectively). As the second interval roughly covers the high-redshift tail of a sample for which the prediction would be easily affected by the low-redshift region, it is more sensitive to the correction of redshift dependence. The slopes of the fitted lines are approaching zero with our correction methods, implying that the dependence on spectroscopic redshifts can be essentially lifted. The change between Step 1 and Step 2 (Step 2 and Step 3) indicates the removal of the influence from the over-populated (under-populated) bins. Additionally, there is no big gap between the Baseline and Step 1, meaning that the biases cannot be resolved in Step 1, yet this step is a prerequisite for the successive bias correction steps.

The consequence of mode collapse can be seen as strong spikes in the sample distribution of estimated photometric redshifts for the Baseline method in contrast to the sample distribution of spectroscopic redshifts. We note that this is not due to the effect of discretized binning, since the typical separation between two spikes is wider than the bin size. Similarly, as presented in Fig.~\ref{fig:photoz_vs_specz}, the data points are distributed along parallel lines as local concentrations, centered at discrete photometric redshift values. In particular, the last mode is located at a redshift lower than the maximum redshift, which may be the reason for the plateau at $z \sim 0.3$ observed by \citet{Pasquet2019} (see Fig. 7 therein) where the instances with the highest spectroscopic redshifts are assigned with relatively lower photometric redshifts. However, with Step 3, the spikes and the concentrated lines can be substantially wiped out.

The intensity of mode collapse can be quantified by the total variation distance
\begin{equation}
D_{\mathrm{TV}} (p_{photo}, p_{spec}) = \frac{1}{2}\int {| p_{photo} (z) - p_{spec} (z) | dz}
\label{eq:d_tv}
\end{equation}
where $p_{photo}(z)$ and $p_{spec}(z)$ are the normalized $z_{photo}$ and $z_{spec}$ sample distributions, respectively. This metric is sensitive to the ``vertical'' separation between the two distributions, especially conspicuous features such as spikes. We also construct $z_{photo}$ sample distributions for which collapsed modes are smoothed out by randomly drawing the same number of instances as the $z_{spec}$ sample based on the $<\Delta z> - \,\, z_{spec}$ curves. These simulated distributions set the lower limits of the total variation distances, which are shown as the shaded areas in Figs.~\ref{fig:main_res_SDSS} and \ref{fig:main_res_CFHTW}.

By plotting the total variation distance against the slope of the linear fit in the second redshift interval for different cases, we demonstrate that our methods up to Step 3 achieve considerable reduction of both $z_{spec}$-dependent residuals and mode collapse. As Step 3 tackles mode collapse and under-population-induced $z_{spec}$-dependent residuals simultaneously, we further illustrate the correction of each of the two biases with two variant cases of Step 3 --- ``Step 3a'' (which leverages hard labels with adjusted means) and ``Step 3b'' (which leverages soft labels with unadjusted means). For the SDSS data, Step 3a flattens the slope yet fails to reduce the total variation distance, whereas Step 3b produces the opposite result, suggesting that softening the labels and re-adjusting the means are both necessary and they account for the correction of the two biases respectively. For the CFHTLS-WIDE data, Step 3a fails to compensate the negative slope probably due to the influence of amplified mode collapse.

In addition, we examine another variant case ``Step $3\backslash2$'' in which Step 3 is applied directly after Step 1 without trimming off over-populated bins by Step 2. For the CFHTLS-WIDE data, the maximum number density (i.e., 268) is comparable with the chosen balancing threshold (i.e., 100), thus the results from Step 3 and Step $3\backslash2$ are close. However, for the SDSS data, the maximum number density (i.e., 2,626) is much higher than the chosen threshold (i.e., 200). It seems that the untrimmed high peak number density is detrimental to the validity of labeling re-adjustment and thus excessively lifts up the high-redshift tail as indicated by the unrealistic positive slope, which suggests the necessity of Step 2 (more is discussed in Appendix~\ref{sec:impact_th}).

We quantify the accuracy of photometric redshift estimation using the estimate of standard deviation based on the median absolute deviation (MAD),
\begin{equation}
\begin{aligned}
&\sigma_{\mathrm{MAD}} = 1.4826 \times \mathrm{Median}|\Delta z - \mathrm{Median}(\Delta z) | \\
&\mathrm{where} \,\,\, \Delta z = \frac{z_{photo} - z_{spec}}{1 + z_{spec}}
\end{aligned}
\label{eq:sigma_mad}
\end{equation}
In all the cases, $\sigma_{\mathrm{MAD}}$ gradually degrades as the magnitude increases, indicating large estimation errors for faint galaxies. Furthermore, we observe that the biases are tightly connected to estimation errors. Stronger biases are usually associated with larger errors, and the correction of stronger biases would further introduce more errors. We note that the multi-channel output unit, the restricted subset of training data and the soft labels would all contribute extra errors compared to the Baseline method. While not strong for the SDSS data, this compromise is more significant for the CFHTLS data due to a larger redshift range and less training data.

\subsection{Correction of $z_{photo}$-dependent residuals} \label{sec:result2}

\begin{figure*}
\begin{center}
\centerline{\includegraphics[width=1.0\linewidth]{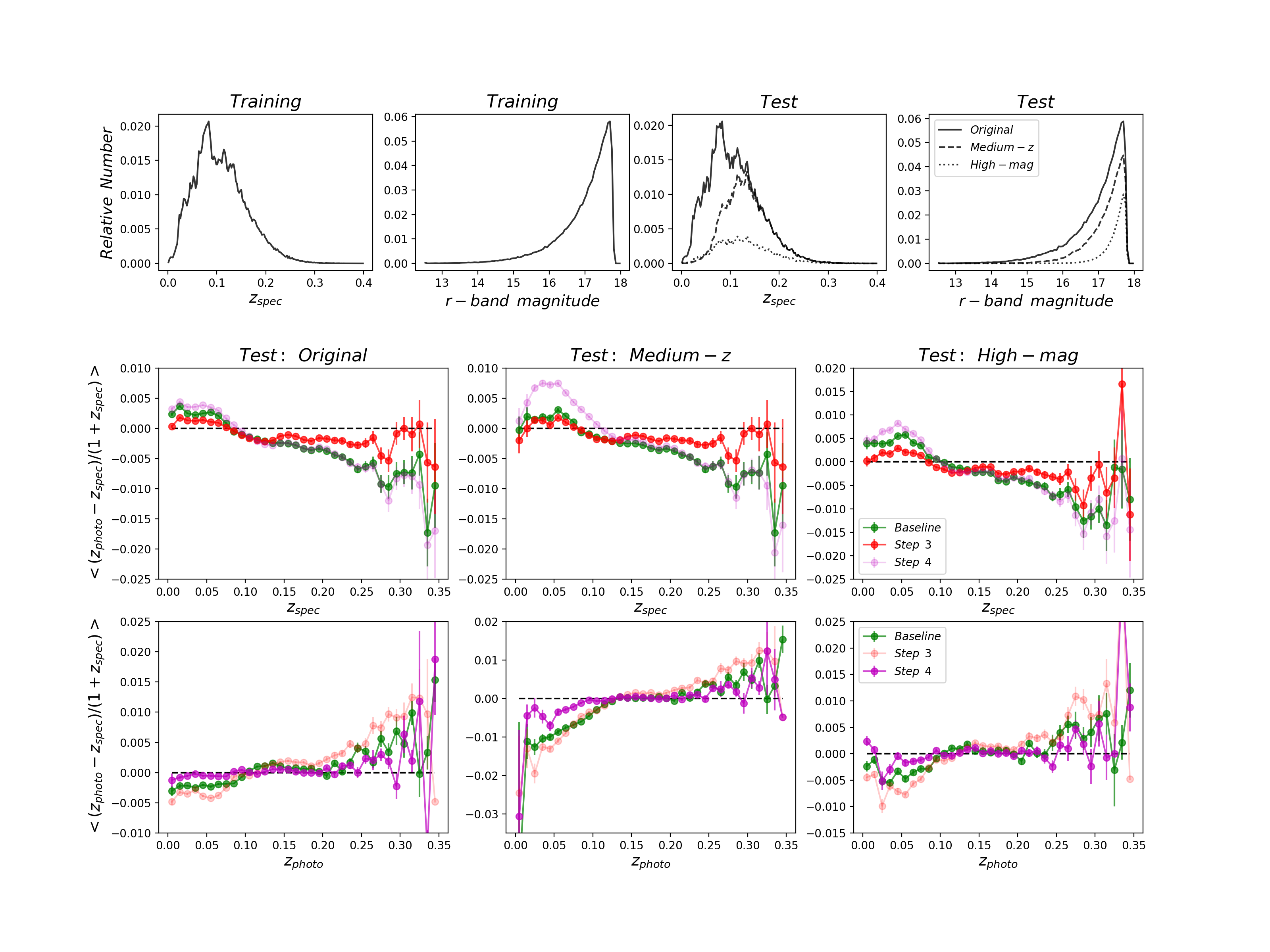}}
\caption{\textbf{Upper Panels:} Spectroscopic redshift and $r$-band magnitude distributions for the SDSS training data and the test data. Besides the original test sample, two other test samples are constructed that are biased towards medium redshifts ($z_{spec} \sim 0.12$) or high magnitudes ($r \sim 17.5$). \textbf{Middle Panels:} Mean residuals as a function of spectroscopic redshifts for the Baseline method and Step 3 with the three test samples and the network \texttt{Net\_P}. The results from Step 4 are also plotted in light purple for comparison. Steps 1 -- 3 are applied in advance of the calibration by Step 4. \textbf{Lower Panels:} Mean residuals as a function of photometric redshifts for the Baseline method and Step 4 with the three test samples and the network \texttt{Net\_P}. The results from Step 3 are also plotted in light red for comparison.}
\label{fig:photoz_corr_SDSSmain}
\end{center}
\end{figure*}

\begin{figure*}
\begin{center}
\centerline{\includegraphics[width=1.0\linewidth]{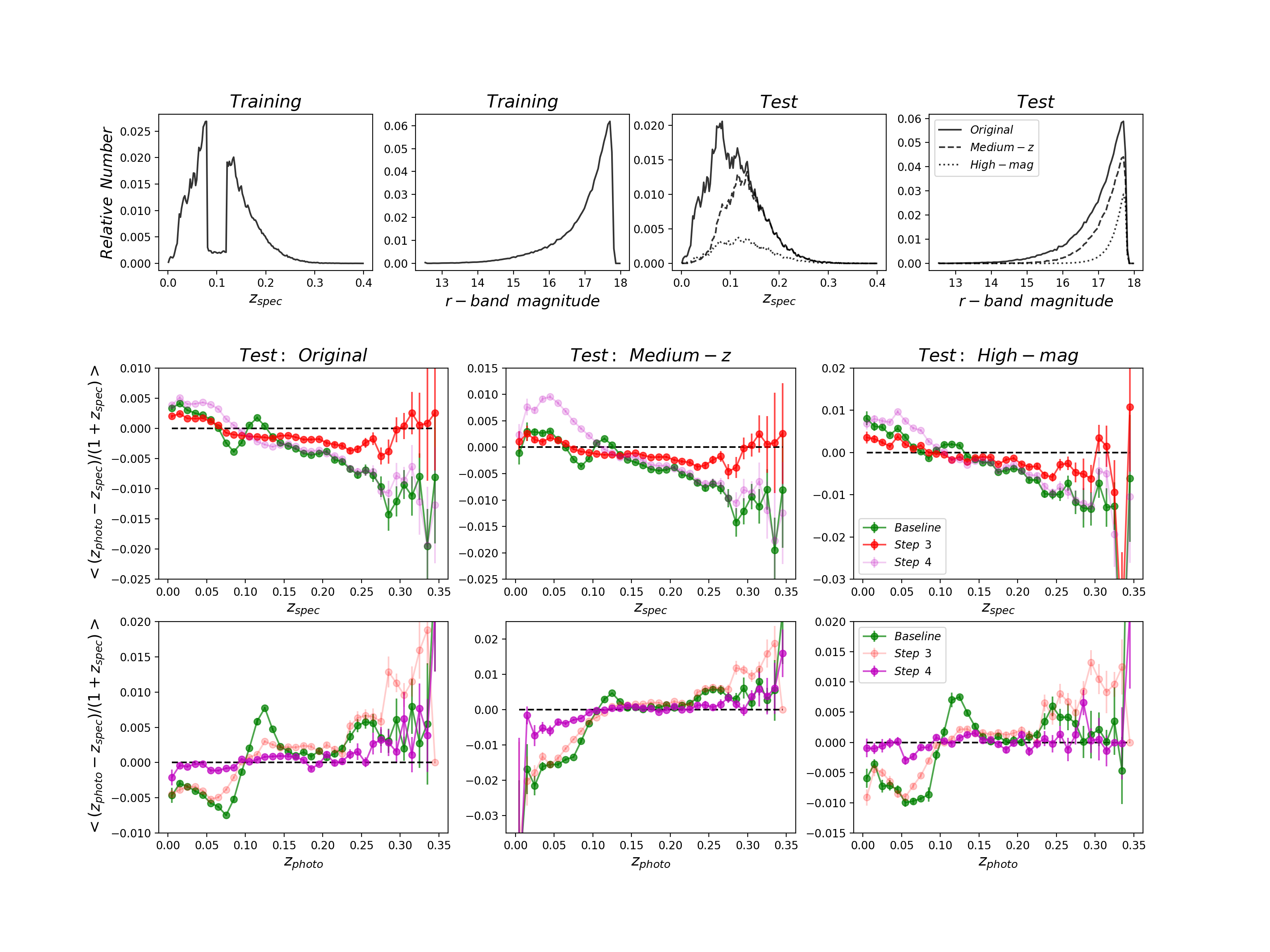}}
\caption{Same as Fig.~\ref{fig:photoz_corr_SDSSmain}, except that the number density within $0.08 < z_{spec} < 0.12$ for the training sample is reduced by 90\%.}
\label{fig:photoz_corr_SDSSmidreduce}
\end{center}
\end{figure*}

Figs.~\ref{fig:photoz_corr_SDSSmain} and \ref{fig:photoz_corr_SDSSmidreduce} show the outcomes of correcting $z_{photo}$-dependent residuals and the comparison with those of correcting $z_{spec}$-dependent residuals using the SDSS test data. While \citet{Pasquet2019} exploits training data and test data whose distributions are identical, we apply our methods in scenarios where there are mismatches between their distributions. We construct three test samples with different distributions in order to mimic mismatches due to selection effects, which are unmodified or biased towards medium redshifts ($z_{spec} \sim 0.12$) or high magnitudes ($r \sim 17.5$). The model is trained with the original training sample or a sample in which 90\% of the medium-redshift instances ($0.08 < z_{spec} < 0.12$) are artificially discarded, mimicking an artifact due to, e.g., combining samples from different sources. In each of the six cases, we perform Step 4 five times; each time the residuals are corrected with a different set of resampled training data that has the same sample size as the test sample. The final output estimates for Step 4 are given by averaging the five folds of corrected $z_{photo}$ estimates. Furthermore, we note that the errorbar of the mean residual as a function of $z_{photo}$ should not be estimated using the quadratic error curves $\sigma_2^2(z_{spec}|r_o)$ given by Eq.~\ref{eq:sigma2}, since the independent variable is $z_{photo}$ rather than $z_{spec}$; thus we take the root-mean-square (RMS) of residuals in each photometric redshift bin $[z_{photo}, z_{photo}+\delta z_{photo}]$ as the errorbar.

As already illustrated in Fig.~\ref{fig:main_res_SDSS}, for the Baseline method in all the cases, a negative slope is detected for the mean residual as a function of $z_{spec}$. This negative slope can be corrected by Step 3, while the correction leads to a positive slope for the mean residual as a function of $z_{photo}$. The positive slope can then be corrected by Step 4, yet this in turn reproduces a negative slope as a function of $z_{spec}$. This is due to the fact that the dependences on $z_{spec}$ and $z_{photo}$ cannot be reconciled (as mentioned in Section~\ref{sec:intro}). Whether applying Step 4 or not relies on the definition of the mean residual required for actual analysis.

Regarding the $z_{photo}$-dependent residuals, although \citet{Pasquet2019} claims that zero dependence on $z_{photo}$ can be achieved without the need for bias correction (in a restricted range $z_{photo}<0.25$; see Fig. 8 therein), we are cautious that this is not always guaranteed. Non-zero dependence would most probably exist if the distributions for the training data and the test data are not identical. For instance, the test samples with the modified distributions produce slopes steeper than those with the original distribution. The artifact in the training sample creates a tilde on top of the mean residual evolving trend, which may only be handled via balancing the training data. Additionally, as discussed in Appendix~\ref{sec:impact_ite}, the dependence of mean residuals on $z_{photo}$ may be evolving depending on how well the model fits data, even when there is no mismatch between the training data and the test data. On the contrary, Steps 3 and 4 of our methods give robust outcomes of removing dependence on $z_{spec}$ or $z_{photo}$ in all the cases.

\subsection{Mean redshift measurement for tomographic analysis}

The mean photometric redshift in each tomographic bin is an important quantity for analyzing cosmic shear tomography, since the cosmic shear signal is primarily dependent on the mean distance of sources. For the Euclid mission, the accuracy of mean redshift measurement is required to be $|\Delta z| < 0.002$ \citep{Laureijs2011}, which sets a stringent demand on the performance of photometric redshift estimation approaches. Therefore, we evaluate our methods in this regard.

The residual of the mean redshift in a tomographic bin is defined as
\begin{equation}
\Delta_{<z>} = \frac{<z_{photo}> - <z_{spec}>}{1 + <z_{spec}>}
\label{eq:res_meanz}
\end{equation}
This is different from the definition of the ``mean residual'' in Eq.~\ref{eq:resfit}. We check the evolution of $\Delta_{<z>}$ with regard to both photometric redshifts and $r$-band magnitudes. Despite that $\Delta_{<z>}$ is already a characterization of the deviation between the $z_{photo}$ and $z_{spec}$ sample distributions, we also use the 1-Wasserstein distance to quantify their difference. The 1-Wasserstein distance between the normalized $z_{photo}$ and $z_{spec}$ probability density distributions $p_{photo}$ and $p_{spec}$ can be formulated as
\begin{equation}
D_{\mathrm{1-Wasserstein}} (p_{photo}, p_{spec}) = \int {| F_{photo} (z) - F_{spec} (z) | dz}
\label{eq:d_w}
\end{equation}
where $F_{photo}(z)$ and $F_{spec}(z)$ are the corresponding cumulative density distributions. This metric is a measure of the ``horizontal'' distance and can be interpreted as an effective redshift separation between the two distributions. Unlike the total variation distance (Eq.~\ref{eq:d_tv}), this metric mainly captures the broad-scale discrepancy and it is insensitive to small fluctuations induced by mode collapse.

\begin{figure*}
\begin{center}
\centerline{\includegraphics[width=1.0\linewidth]{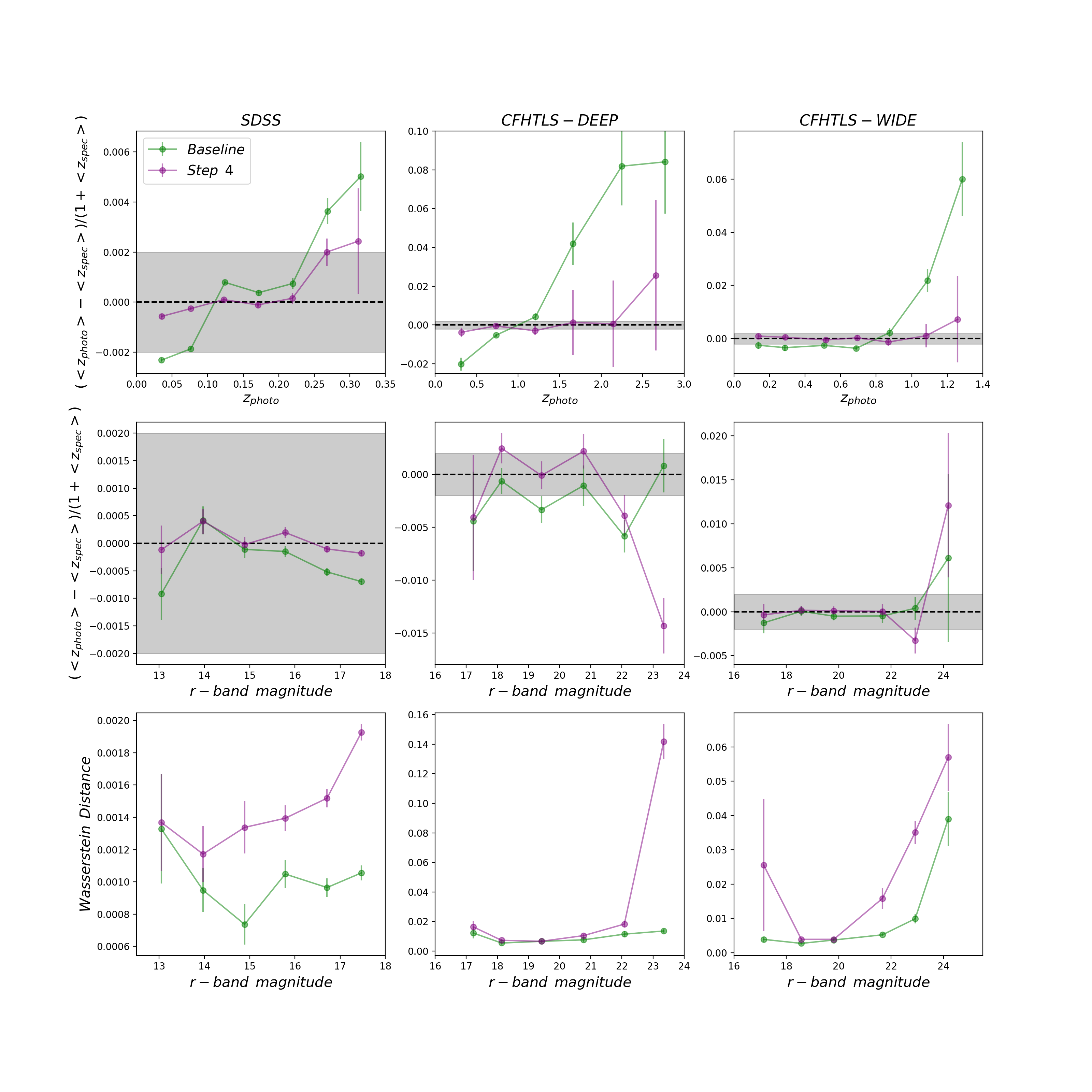}}
\caption{\textbf{Upper Panels:} Residuals of mean redshifts (different from ``mean residuals'') in each $z_{photo}$ tomographic bin for the Baseline method and Step 4 of our methods using the network \texttt{Net\_P} with the SDSS data, the CFHTLS-DEEP data and the CFHTLS-WIDE data, respectively. A magnitude cut $r<24$ is applied to the CFHTLS-DEEP data to make a restricted sample. The shaded areas indicate the requirement for the accuracy of photometric redshift estimation (i.e., $|\Delta z| < 0.002$). Steps 1 -- 3 are applied in advance of the calibration by Step 4. \textbf{Middle Panels:} Same as the Upper Panels, except that residuals of mean redshifts are estimated in $r$-band magnitude bins. \textbf{Lower Panels:} 1-Wasserstein distance between the $z_{photo}$ and $z_{spec}$ sample distributions (Eq.~\ref{eq:d_w}) as a function of $r$-band magnitudes.}
\label{fig:photoz_mean_cal}
\end{center}
\end{figure*}

Fig.~\ref{fig:photoz_mean_cal} shows the mean redshift calibrations for the SDSS data, the CFHTLS-DEEP data and the CFHTLS-WIDE data. We use a restricted CFHTLS-DEEP sample with a magnitude cut $r<24$, which alleviates the impact of faint galaxies. The mean redshift measurement for the SDSS data has a high accuracy. The effective redshift separations indicated by the Wasserstein distances remain small over all magnitudes. The correction conducted by Step 4 lowers the residuals as a function of $z_{photo}$ and meets the $|\Delta z| < 0.002$ requirement. For the CFHTLS datasets, however, this requirement is only fulfilled in the low-magnitude and low-redshift intervals. There are large errors associated with high redshifts or high magnitudes, which is implied by the growth of the Wasserstein distances. Noticeably, the bias correction contributes to a boost of the Wasserstein distances as a result of the compromise between biases and errors. These observations indicate the difficulty of estimating reliable photometric redshifts for the high-redshift regime with data-driven methods mainly due to limited and non-representative spectroscopic data. This difficulty places a demand on optimizing the exploitation of data, updating the existing algorithms and exploring composite methodologies.

\section{Discussions on bias behaviors} \label{sec:discussions_short}

Following our discussions on correcting $z_{spec}$-dependent biases, we investigate the behaviors of biases and the performance of our methods by controlling the CNN models with varying implementing and training conditions. We summarize the main results in this section, while more details can be found in Appendix~\ref{sec:discussions}.

\begin{itemize}
    \item Appendix~\ref{sec:impact_th} discusses the impact of the balancing threshold, which determines the size and the flatness of the near-balanced training subset in Steps 2 and 3. As the threshold varies, we find a trade-off between $z_{spec}$-dependent residuals and the estimation accuracy for Step 2 due to overfitting, while the correction by Step 3 remains robust.
    \item Appendix~\ref{sec:impact_binsize} discusses the impact of the bin size in redshift outputs. Although a large bin size tends to introduce a non-negligible quantization error, the sparsity of data would be enhanced as the bin size shrinks, leading to a compromised accuracy and a tendency of mode collapse.
    \item Appendix~\ref{sec:impact_ite} discusses the impact of the number of training iterations. It indicates how well a model fits data, which is an influencing factor for underfitting/overfitting. In general, there is a transition from underfitting to overfitting as the training proceeds. For the Baseline method, both underfitting and overfitting would increase estimation errors and $z_{spec}$-dependent residuals for test data, and change the behavior of $z_{photo}$-dependent residuals. Underfitting would yield strong mode collapse, whereas overfitting would smooth out collapsed modes by random errors. Our methods are still valid given an underfitted or overfitted representation.
    \item Appendix~\ref{sec:impact_samplesize} discusses the impact of the training sample size. It characterizes the total volume of information provided by the training data, which is another influencing factor for overfitting. The consequence of reducing the sample size is similar to that of having more training iterations. If the original training sample is provided for constructing the near-balanced subset, our methods would still be robust over different degrees of overfitting.
    \item Appendix~\ref{sec:impact_label} discusses the impact of random Gaussian errors added to labels, which determine the fuzziness of the input-output mapping. The labeling errors would boost estimation errors but suppress mode collapse. Our methods would remain robust if the errorless high-quality data is used for bias correction.
    \item Appendix~\ref{sec:impact_model} discusses the impact of the model complexity using a set of networks with different architectures. It characterizes the capacity of a model to acquire information from data. We find that a model with a higher complexity would produce a better estimation accuracy and smaller biases, and sufficient model complexity is necessary for obtaining an informative representation for bias correction. This also implies that the existence of class-dependent residuals may be inherently governed by the data distribution rather than the model, though their amplitudes are strongly affected by the model complexity.
\end{itemize}

\section{Comparison with state-of-the-art results} \label{sec:comparison}

Finally, we check our results with reference to the state-of-the-art results from \citet{Pasquet2019} and \Treyer{} using CNN models, as well as those from \citet{Beck2016} using $k$-Nearest Neighbors (KNNs) applied on galaxy photometry and those obtained by SED fitting using the \textsc{Le Phare} code \citep{Arnouts1999, Ilbert2006}. These results are compared in terms of $z_{spec}$ and $z_{photo}$-dependent residuals. In order to be consistent with our test samples, 103,305 instances in the SDSS sample within $0<z_{spec}<0.4$ from \citet{Beck2016} or \citet{Pasquet2019} and 10,000 instances in the CFHTLS-WIDE sample within $0<z_{spec}<4.0$ from \Treyer{} or the SED fitting method are selected. Furthermore, it should be noted that \citet{Pasquet2019} and \Treyer{} use $z_{mean}$ and $z_{median}$ respectively as the point estimates of photometric redshifts, whereas we use $z_{mode}$ in this work. We directly compare results from these different studies despite the disagreement in the choice of point estimates.

\begin{figure*}
\begin{center}
\centerline{\includegraphics[width=1.0\linewidth]{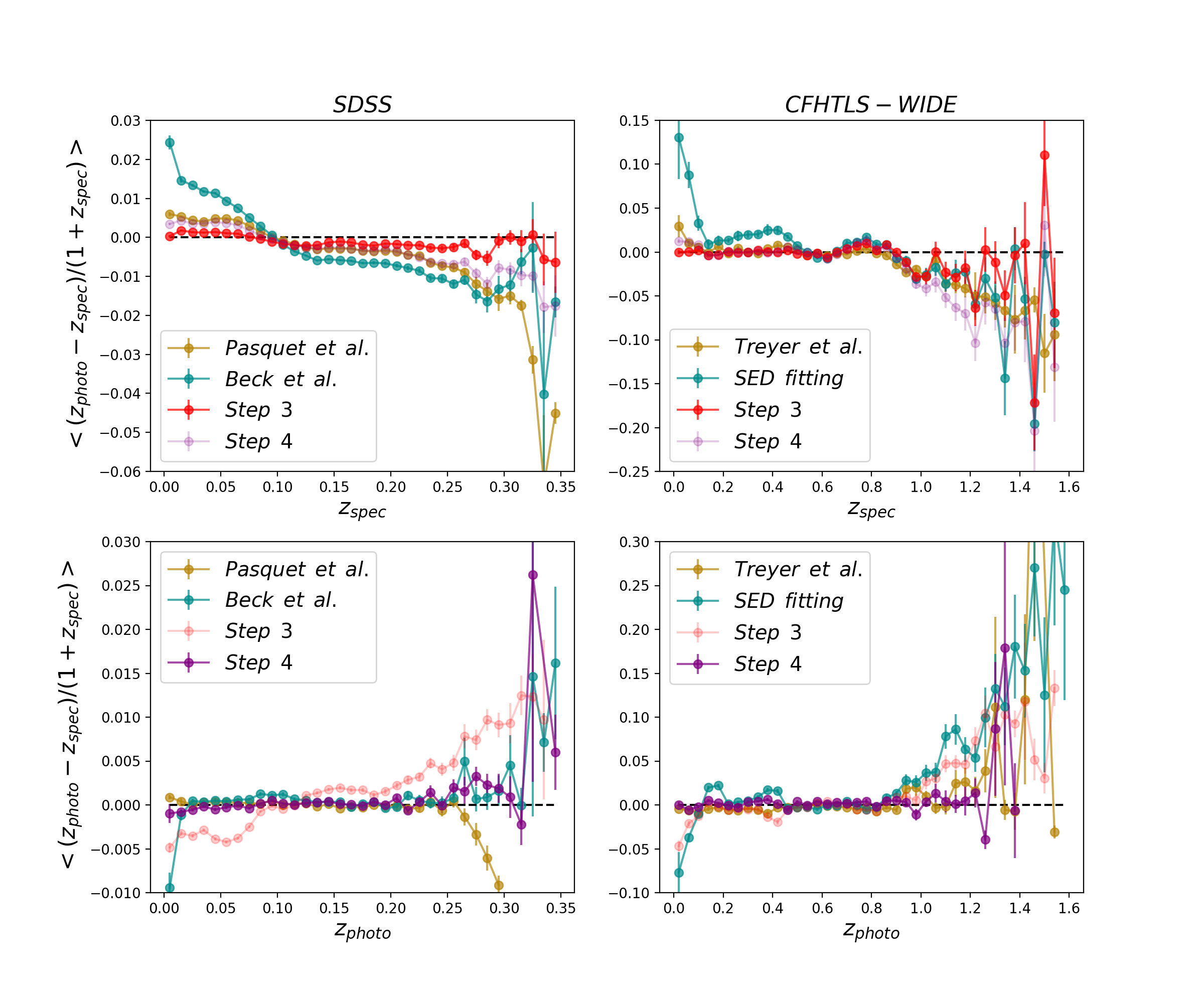}}
\caption{Mean residuals as a function of spectroscopic redshifts or photometric redshifts with the SDSS data and the CFHTLS-WIDE data. The results from Steps 3 and 4 of our methods with the network \texttt{Net\_P} are compared with those produced by data-driven methods from \citet{Pasquet2019}, \citet{Beck2016}, \Treyer{} and SED fitting using the \textsc{Le Phare} code \citep{Arnouts1999, Ilbert2006}, respectively. The results from Step 3 (Step 4) are plotted in light red (purple) for comparison in the lower (upper) panels. Steps 1 -- 3 are applied in advance of the calibration by Step 4. For consistency with our test samples, 103,305 SDSS galaxies and 10,000 CFHTLS-WIDE galaxies in the given redshift ranges are selected from these studies for plotting. We note that we take $z_{mode}$ (i.e., the redshift at the peak probability) as the point estimates of photometric redshifts, whereas \citet{Pasquet2019} takes $z_{mean}$ (i.e., the probability-weighted mean redshift) and \Treyer{} takes $z_{median}$ (i.e., the median redshift drawn from the probability distribution).}
\label{fig:res_compare}
\end{center}
\end{figure*}

As can be seen in Fig.~\ref{fig:res_compare}, the mean residuals as a function of $z_{spec}$ for the reference results from \citet{Beck2016}, \citet{Pasquet2019}, \Treyer{} and the SED fitting method all show significant deviations from zero close to the borders of the redshift range, yet Step 3 of our methods is able to reduce such deviations. In particular, the KNN approach applied by \citet{Beck2016} produces a similar yet worse trend for the SDSS data than the CNN approach by \citet{Pasquet2019}, probably due to larger estimation errors. The SED fitting method may suffer from degeneracies in galaxy photometry or a lack of representative templates in addition to limited accuracy, which may be the reason for the bump around $z_{spec} \sim 0.4$ on top of the non-flat tails for the CFHTLS-WIDE data. The methods we propose in this work for data-driven models may complement SED fitting approaches to deal with such effects.

For the $z_{photo}$-dependent residuals, our results from Step 4 and the reference results from the data-driven approaches are similar, except that the drop at $z_{photo} \sim 0.25$ for the SDSS data from \citet{Pasquet2019} is eliminated by our methods. Meanwhile, the plateau at $z \sim 0.3$ observed by \citet{Pasquet2019} is removed as a result of correcting mode collapse. Although our methods do not result in as strong biases as produced by the SED fitting method for the CFHTLS-WIDE data, there are non-negligible errors at high redshifts that confine the robustness of our methods, suggesting that a good control of estimation errors is a precondition for bias correction.

\section{Conclusion} \label{sec:conclusion}

In this work, we analyze two biases that are generally present in data-driven methods, namely \textit{class-dependent residuals} and \textit{mode collapse}, which are two effects imposed by the prior of training data and the model implementation. Using photometric redshift estimation as a case study, we discuss the statistical basis of bias correction (Appendix~\ref{sec:formalism}) and propose a multi-step procedure for correcting the biases based on Convolutional Neural Networks (CNNs) trained with spectroscopic samples of galaxy images. The first three steps are CNN-based while the last step is a direct calibration. In particular, we suggest that the residuals can be defined as a function of spectroscopic redshifts ($z_{spec}$) or photometric redshifts ($z_{photo}$), which have to be separately taken into consideration according to actual analysis. The correction of biases in the spectroscopic space by the first three steps has to be performed as a prerequisite before applying the calibration in the photometric space by the last step. The highlights of our methods are summarized below:
\begin{itemize}
    \item A multi-channel unit is used as the output in the place of the conventional single-channel fully-connected layer with the Softmax activation, so that $r$-band magnitudes of galaxies can be explicitly predicted.
    \item The learning procedure of a model is split into representation learning and classification. Bias correction is performed solely in the phase of classification with a fixed representation.
    \item A near-balanced subset of training data is constructed via resampling in order to resolve over-population-induced $z_{spec}$-dependent residuals.
    \item Under-population-induced $z_{spec}$-dependent residuals and mode collapse are resolved by exploiting soft Gaussian labels, whose dispersions and means are determined using the pre-estimated redshifts from the preceding step.
    \item $z_{photo}$-dependent residuals are resolved by reconstructing the distribution of the test sample with the training data and subtracting the mean $z_{photo}-z_{spec}$ deviation in each cell of the $z_{photo}$-magnitude grid, which is a direct calibration independent of CNN models.
\end{itemize}

We apply our methods to the SDSS data that covers $0<z_{spec}<0.4$, as well as the CFHTLS data in a larger range of $0<z_{spec}<4.0$ from both the Deep fields and the Wide fields. In terms of bias control, our methods achieve better performance compared to the Baseline method, the KNN method from \citet{Beck2016}, the SED fitting method using the \textsc{Le Phare} code \citep{Arnouts1999, Ilbert2006} and the state-of-the-art CNN models from \citet{Pasquet2019} and \Treyer{}. We also examine a few implementing details and assumptions behind our methods, and analyze their effects on the behaviors of the biases. The results indicate that our methods are generally robust under varying conditions as long as high-quality data is provided for bias correction. We reiterate that representation learning may be less concerned with systematics or low SNRs, whereas bias correction has a higher demand on data quality. Therefore, it is important to identify and understand the internal structures and systematics within data, so that a good control of biases can be ensured.

Despite the merits, we are cautious that the main difficulty of our methods is the trade-off between resolving biases and reducing estimation errors. The use of $z_{mode}$ as point estimates yields the least biases compared to $z_{mean}$ and $z_{median}$ (Appendix~\ref{sec:zcomparison}), yet the accuracy and the catastrophic rate may be compromised (Appendix~\ref{sec:evaluation}). Extra errors may be introduced through the use of the multi-channel output unit and the training with a restricted subset of data and soft labels, yet these are all necessary for bias correction. In addition, biases scale with estimation errors in such a way that correcting large biases caused by large errors would yield even larger errors. It is thus important to have a good constraint of estimation errors before performing bias correction. In spite of increasing errors, we prioritize bias correction in this work, since it is critical to cosmological analysis.

The calibration of residuals with the CFHTLS data in high redshift intervals is unable to meet the accuracy requirement of $|\Delta z| < 0.002$ for a Euclid-like survey. This partial failure for the CFHTLS data points to the difficulty of applying data-driven methods to the high-redshift regime provided with a shortage of spectroscopic training data with good SNRs. Given limited spectroscopic follow-ups for the upcoming imaging surveys, there is an increasing need for optimizing the exploitation of training data, either at the data level (e.g., data augmentation, resampling, etc.) or at the algorithm level (e.g., feature extraction, domain adaptation, etc.). In the stage of representation learning, particularly, the model may incorporate various supervised and unsupervised-learning techniques to maximize the usage of data in divergent forms. Meanwhile, it is worth investigating hybrid methods that take advantage of complementary information from individual approaches. For instance, multiple methods may be explored and integrated for composite likelihood analysis \citep[e.g.,][]{Shuntov2020, Rau2022}. Domain knowledge may also be injected to fill in the gap for under-represented or missing data, which would be essential for alleviating the difficulty in the high-redshift regime.

Although we only discuss our bias correction methods in the context of a supervised-learning classification problem, we note that our work puts forward a framework that can be generalized to regression problems combined with unsupervised-learning components. Indeed, we have shown that the existence of class-dependent residuals may be inherently determined by the unbalanced data distribution rather than the model (Section~\ref{sec:result2} and Appendix~\ref{sec:impact_model}). Our methods are also not limited to the photometric redshift estimation problem. It is promising to apply our methods to similar studies that involve data-driven models. Furthermore, in this work, we deal with point estimates rather than density estimates produced by neural networks due to unresolved prior. The justification for the use of density estimates and the correction of associated biases may be explored in future work.

\begin{acknowledgements}

This project has received funding from the European Union's Horizon 2020 research and innovation programme under the Marie Skłodowska-Curie grant agreement No713750. Also, it has been carried out with the financial support of the Regional Council of Provence-Alpes-C\^{o}te d'Azur and with the financial support of the A*MIDEX (n° ANR-11-IDEX-0001-02), funded by the Investissements d'Avenir project funded by the French Government, managed by the French National Research Agency (ANR).

This work has been carried out thanks to the support of the DEEPDIP ANR project (ANR-19-CE31-0023) and the Programme National Cosmologie et Galaxies (PNCG) of CNRS/INSU with INP and IN2P3, co-funded by CEA and CNES.
 
This work makes use of Sloan Digital Sky Survey (SDSS) data. Funding for SDSS-III has been provided by the Alfred P. Sloan Foundation, the Participating Institutions, the National Science Foundation, and the U.S. Department of Energy Office of Science. The SDSS-III web site is http://www.sdss3.org/. SDSS-III is managed by the Astrophysical Research Consortium for the Participating Institutions of the SDSS-III Collaboration including the University of Arizona, the Brazilian Participation Group, Brookhaven National Laboratory, Carnegie Mellon University, University of Florida, the French Participation Group, the German Participation Group, Harvard University, the Instituto de Astrofisica de Canarias, the Michigan State/Notre Dame/JINA Participation Group, Johns Hopkins University, Lawrence Berkeley National Laboratory, Max Planck Institute for Astrophysics, Max Planck Institute for Extraterrestrial Physics, New Mexico State University, New York University, Ohio State University, Pennsylvania State University, University of Portsmouth, Princeton University, the Spanish Participation Group, University of Tokyo, University of Utah, Vanderbilt University, University of Virginia, University of Washington, and Yale University.

This work is based on observations obtained with MegaPrime/MegaCam, a joint project of CFHT and CEA/DAPNIA, at the Canada-France-Hawaii Telescope (CFHT) which is operated by the National Research Council (NRC) of Canada, the Institut National des Sciences de l'Univers of the Centre National de la Recherche Scientifique (CNRS) of France, and the University of Hawaii. This work is based in part on data products produced at Terapix and the Canadian Astronomy Data Centre as part of the Canada-France-Hawaii Telescope Legacy Survey, a collaborative project of NRC and CNRS.

\end{acknowledgements}

\bibliography{pz_bias}

\begin{thebibliography}{110}
\expandafter\ifx\csname natexlab\endcsname\relax\def\natexlab#1{#1}\fi

\bibitem[{{Alam} {et~al.}(2015){Alam}, {Albareti}, {Allende Prieto}, {Anders},
  {Anderson}, {Anderton}, {Andrews}, {Armengaud}, {Aubourg}, {Bailey}, {Basu},
  {Bautista}, {Beaton}, {Beers}, {Bender}, {Berlind}, {Beutler}, {Bhardwaj},
  {Bird}, {Bizyaev}, {Blake}, {Blanton}, {Blomqvist}, {Bochanski}, {Bolton},
  {Bovy}, {Shelden Bradley}, {Brandt}, {Brauer}, {Brinkmann}, {Brown},
  {Brownstein}, {Burden}, {Burtin}, {Busca}, {Cai}, {Capozzi}, {Carnero
  Rosell}, {Carr}, {Carrera}, {Chambers}, {Chaplin}, {Chen}, {Chiappini},
  {Chojnowski}, {Chuang}, {Clerc}, {Comparat}, {Covey}, {Croft}, {Cuesta},
  {Cunha}, {da Costa}, {Da Rio}, {Davenport}, {Dawson}, {De Lee}, {Delubac},
  {Deshpande}, {Dhital}, {Dutra-Ferreira}, {Dwelly}, {Ealet}, {Ebelke},
  {Edmondson}, {Eisenstein}, {Ellsworth}, {Elsworth}, {Epstein}, {Eracleous},
  {Escoffier}, {Esposito}, {Evans}, {Fan}, {Fern{\'a}ndez-Alvar}, {Feuillet},
  {Filiz Ak}, {Finley}, {Finoguenov}, {Flaherty}, {Fleming}, {Font-Ribera},
  {Foster}, {Frinchaboy}, {Galbraith-Frew}, {Garc{\'\i}a},
  {Garc{\'\i}a-Hern{\'a}ndez}, {Garc{\'\i}a P{\'e}rez}, {Gaulme}, {Ge},
  {G{\'e}nova-Santos}, {Georgakakis}, {Ghezzi}, {Gillespie}, {Girardi},
  {Goddard}, {Gontcho}, {Gonz{\'a}lez Hern{\'a}ndez}, {Grebel}, {Green},
  {Grieb}, {Grieves}, {Gunn}, {Guo}, {Harding}, {Hasselquist}, {Hawley},
  {Hayden}, {Hearty}, {Hekker}, {Ho}, {Hogg}, {Holley-Bockelmann}, {Holtzman},
  {Honscheid}, {Huber}, {Huehnerhoff}, {Ivans}, {Jiang}, {Johnson},
  {Kinemuchi}, {Kirkby}, {Kitaura}, {Klaene}, {Knapp}, {Kneib}, {Koenig},
  {Lam}, {Lan}, {Lang}, {Laurent}, {Le Goff}, {Leauthaud}, {Lee}, {Lee},
  {Licquia}, {Liu}, {Long}, {L{\'o}pez-Corredoira}, {Lorenzo-Oliveira},
  {Lucatello}, {Lundgren}, {Lupton}, {Mack}, {Mahadevan}, {Maia}, {Majewski},
  {Malanushenko}, {Malanushenko}, {Manchado}, {Manera}, {Mao}, {Maraston},
  {Marchwinski}, {Margala}, {Martell}, {Martig}, {Masters}, {Mathur},
  {McBride}, {McGehee}, {McGreer}, {McMahon}, {M{\'e}nard}, {Menzel},
  {Merloni}, {M{\'e}sz{\'a}ros}, {Miller}, {Miralda-Escud{\'e}}, {Miyatake},
  {Montero-Dorta}, {More}, {Morganson}, {Morice-Atkinson}, {Morrison},
  {Mosser}, {Muna}, {Myers}, {Nandra}, {Newman}, {Neyrinck}, {Nguyen},
  {Nichol}, {Nidever}, {Noterdaeme}, {Nuza}, {O'Connell}, {O'Connell},
  {O'Connell}, {Ogando}, {Olmstead}, {Oravetz}, {Oravetz}, {Osumi}, {Owen},
  {Padgett}, {Padmanabhan}, {Paegert}, {Palanque-Delabrouille}, {Pan},
  {Parejko}, {P{\^a}ris}, {Park}, {Pattarakijwanich}, {Pellejero-Ibanez},
  {Pepper}, {Percival}, {P{\'e}rez-Fournon}, {P{\'e}rez-R{\`a}fols},
  {Petitjean}, {Pieri}, {Pinsonneault}, {Porto de Mello}, {Prada}, {Prakash},
  {Price-Whelan}, {Protopapas}, {Raddick}, {Rahman}, {Reid}, {Rich}, {Rix},
  {Robin}, {Rockosi}, {Rodrigues}, {Rodr{\'\i}guez-Torres}, {Roe}, {Ross},
  {Ross}, {Rossi}, {Ruan}, {Rubi{\~n}o-Mart{\'\i}n}, {Rykoff},
  {Salazar-Albornoz}, {Salvato}, {Samushia}, {S{\'a}nchez}, {Santiago},
  {Sayres}, {Schiavon}, {Schlegel}, {Schmidt}, {Schneider}, {Schultheis},
  {Schwope}, {Sc{\'o}ccola}, {Scott}, {Sellgren}, {Seo}, {Serenelli}, {Shane},
  {Shen}, {Shetrone}, {Shu}, {Silva Aguirre}, {Sivarani}, {Skrutskie},
  {Slosar}, {Smith}, {Sobreira}, {Souto}, {Stassun}, {Steinmetz}, {Stello},
  {Strauss}, {Streblyanska}, {Suzuki}, {Swanson}, {Tan}, {Tayar}, {Terrien},
  {Thakar}, {Thomas}, {Thomas}, {Thompson}, {Tinker}, {Tojeiro}, {Troup},
  {Vargas-Maga{\~n}a}, {Vazquez}, {Verde}, {Viel}, {Vogt}, {Wake}, {Wang},
  {Weaver}, {Weinberg}, {Weiner}, {White}, {Wilson}, {Wisniewski},
  {Wood-Vasey}, {Ye`che}, {York}, {Zakamska}, {Zamora}, {Zasowski}, {Zehavi},
  {Zhao}, {Zheng}, {Zhou}, {Zhou}, {Zou}, \& {Zhu}}]{Alam2015}
{Alam}, S., {Albareti}, F.~D., {Allende Prieto}, C., {et~al.} 2015, \apjs, 219,
  12

\bibitem[{Alarcon {et~al.}(2020)Alarcon, Sánchez, Bernstein, \&
  Gaztañaga}]{Alarcon2020}
Alarcon, A., Sánchez, C., Bernstein, G.~M., \& Gaztañaga, E. 2020, \mnras,
  498, 2614

\bibitem[{{Alibert} \& {Venturini}(2019)}]{Alibert2019}
{Alibert}, Y. \& {Venturini}, J. 2019, \aap, 626, A21

\bibitem[{{Ansari} {et~al.}(2021){Ansari}, {Agnello}, \& {Gall}}]{Ansari2021}
{Ansari}, Z., {Agnello}, A., \& {Gall}, C. 2021, \aap, 650, A90

\bibitem[{{Arjovsky} \& {Bottou}(2017)}]{Arjovsky2017}
{Arjovsky}, M. \& {Bottou}, L. 2017, arXiv e-prints, arXiv:1701.04862

\bibitem[{{Armitage} {et~al.}(2019){Armitage}, {Kay}, \&
  {Barnes}}]{Armitage2019}
{Armitage}, T.~J., {Kay}, S.~T., \& {Barnes}, D.~J. 2019, \mnras, 484, 1526

\bibitem[{{Arnouts} {et~al.}(1999){Arnouts}, {Cristiani}, {Moscardini},
  {Matarrese}, {Lucchin}, {Fontana}, \& {Giallongo}}]{Arnouts1999}
{Arnouts}, S., {Cristiani}, S., {Moscardini}, L., {et~al.} 1999, \mnras, 310,
  540

\bibitem[{{Baldry} {et~al.}(2018){Baldry}, {Liske}, {Brown}, {Robotham},
  {Driver}, {Dunne}, {Alpaslan}, {Brough}, {Cluver}, {Eardley}, {Farrow},
  {Heymans}, {Hildebrandt}, {Hopkins}, {Kelvin}, {Loveday}, {Moffett},
  {Norberg}, {Owers}, {Taylor}, {Wright}, {Bamford}, {Bland-Hawthorn},
  {Bourne}, {Bremer}, {Colless}, {Conselice}, {Croom}, {Davies}, {Foster},
  {Grootes}, {Holwerda}, {Jones}, {Kafle}, {Kuijken}, {Lara-Lopez},
  {L{\'o}pez-S{\'a}nchez}, {Meyer}, {Phillipps}, {Sutherland}, {van Kampen}, \&
  {Wilkins}}]{Baldry2018}
{Baldry}, I.~K., {Liske}, J., {Brown}, M.~J.~I., {et~al.} 2018, \mnras, 474,
  3875

\bibitem[{Beck {et~al.}(2016)Beck, Dobos, Budavári, Szalay, \&
  Csabai}]{Beck2016}
Beck, R., Dobos, L., Budavári, T., Szalay, A.~S., \& Csabai, I. 2016, \mnras,
  460, 1371

\bibitem[{Bengio {et~al.}(2013)Bengio, Courville, \& Vincent}]{Bengio2013}
Bengio, Y., Courville, A.~C., \& Vincent, P. 2013, IEEE Transactions on Pattern
  Analysis and Machine Intelligence, 35, 1798

\bibitem[{Bhagyashree {et~al.}(2020)Bhagyashree, Kushwaha, \&
  Nandi}]{Bhagyashree2020}
Bhagyashree, Kushwaha, V., \& Nandi, G.~C. 2020, in 2020 IEEE 4th Conference on
  Information Communication Technology (CICT), 1--6

\bibitem[{{Bonjean} {et~al.}(2019){Bonjean}, {Aghanim}, {Salom{\'e}}, {Beelen},
  {Douspis}, \& {Soubri{\'e}}}]{Bonjean2019}
{Bonjean}, V., {Aghanim}, N., {Salom{\'e}}, P., {et~al.} 2019, \aap, 622, A137

\bibitem[{Bonnett(2015)}]{Bonnett2015}
Bonnett, C. 2015, \mnras, 449, 1043–1056

\bibitem[{{Bradshaw} {et~al.}(2013){Bradshaw}, {Almaini}, {Hartley}, {Smith},
  {Conselice}, {Dunlop}, {Simpson}, {Chuter}, {Cirasuolo}, {Foucaud}, {McLure},
  {Mortlock}, \& {Pearce}}]{Bradshaw2013}
{Bradshaw}, E.~J., {Almaini}, O., {Hartley}, W.~G., {et~al.} 2013, \mnras, 433,
  194

\bibitem[{Buchs {et~al.}(2019)Buchs, Davis, Gruen, DeRose, Alarcon, Bernstein,
  Sánchez, Myles, Roodman, Allen, Amon, Choi, Masters, Miquel, Troxel,
  Wechsler, Abbott, Annis, Avila, Bechtol, Bridle, Brooks, Buckley-Geer, Burke,
  Carnero Rosell, Carrasco Kind, Carretero, Castander, Cawthon, D’Andrea,
  da Costa, De Vicente, Desai, Diehl, Doel, Drlica-Wagner, Eifler, Evrard,
  Flaugher, Fosalba, Frieman, García-Bellido, Gaztanaga, Gruendl, Gschwend,
  Gutierrez, Hartley, Hollowood, Honscheid, James, Kuehn, Kuropatkin, Lima,
  Lin, Maia, March, Marshall, Melchior, Menanteau, Ogando, Plazas, Rykoff,
  Sanchez, Scarpine, Serrano, Sevilla-Noarbe, Smith, Soares-Santos, Sobreira,
  Suchyta, Swanson, Tarle, Thomas, Vikram, \& Collaboration)}]{Buchs2019}
Buchs, R., Davis, C., Gruen, D., {et~al.} 2019, \mnras, 489, 820

\bibitem[{Buda {et~al.}(2018)Buda, Maki, \& Mazurowski}]{Buda2018}
Buda, M., Maki, A., \& Mazurowski, M.~A. 2018, Neural Networks, 106, 249

\bibitem[{Burhanudin {et~al.}(2021)Burhanudin, Maund, Killestein, Ackley, Dyer,
  Lyman, Ulaczyk, Cutter, Mong, Steeghs, Galloway, Dhillon, O’Brien, Ramsay,
  Noysena, Kotak, Breton, Nuttall, Pallé, Pollacco, Thrane, Awiphan, Chote,
  Chrimes, Daw, Duffy, Eyles-Ferris, Gompertz, Heikkilä, Irawati, Kennedy,
  Levan, Littlefair, Makrygianni, Mata-Sánchez, Mattila, McCormac, Mkrtichian,
  Mullaney, Sawangwit, Stanway, Starling, Strøm, Tooke, \&
  Wiersema}]{Burhanudin2021}
Burhanudin, U.~F., Maund, J.~R., Killestein, T., {et~al.} 2021, \mnras, 505,
  4345

\bibitem[{Cao {et~al.}(2019)Cao, Wei, Gaidon, Arechiga, \& Ma}]{Cao2019NEURIPS}
Cao, K., Wei, C., Gaidon, A., Arechiga, N., \& Ma, T. 2019, in Advances in
  Neural Information Processing Systems, ed. H.~Wallach, H.~Larochelle,
  A.~Beygelzimer, F.~d\textquotesingle Alch\'{e}-Buc, E.~Fox, \& R.~Garnett,
  Vol.~32 (Curran Associates, Inc.)

\bibitem[{{Carrasco Kind} \& {Brunner}(2013)}]{CB2013}
{Carrasco Kind}, M. \& {Brunner}, R.~J. 2013, \mnras, 432, 1483

\bibitem[{Carrasco~Kind \& Brunner(2014)}]{Carrasco2014}
Carrasco~Kind, M. \& Brunner, R.~J. 2014, \mnras, 438, 3409

\bibitem[{Cavuoti {et~al.}(2016)Cavuoti, Tortora, Brescia, Longo, Radovich,
  Napolitano, Amaro, Vellucci, La~Barbera, Getman, \& et~al.}]{Cavuoti2016}
Cavuoti, S., Tortora, C., Brescia, M., {et~al.} 2016, \mnras, 466, 2039–2053

\bibitem[{Chawla {et~al.}(2002)Chawla, Bowyer, Hall, \&
  Kegelmeyer}]{Chawla2002}
Chawla, N.~V., Bowyer, K.~W., Hall, L.~O., \& Kegelmeyer, W.~P. 2002, Journal
  of Artificial Intelligence Research, 16, 321–357

\bibitem[{Chong {et~al.}(2020)Chong, Ruff, Kloft, \& Binder}]{Chong2020mode}
Chong, P., Ruff, L., Kloft, M., \& Binder, A. 2020, in 2020 International Joint
  Conference on Neural Networks (IJCNN), IEEE, 1--9

\bibitem[{{Coil} {et~al.}(2011){Coil}, {Blanton}, {Burles}, {Cool},
  {Eisenstein}, {Moustakas}, {Wong}, {Zhu}, {Aird}, {Bernstein}, {Bolton}, \&
  {Hogg}}]{Coil2011}
{Coil}, A.~L., {Blanton}, M.~R., {Burles}, S.~M., {et~al.} 2011, \apj, 741, 8

\bibitem[{{Collister} \& {Lahav}(2004)}]{Collister2004}
{Collister}, A.~A. \& {Lahav}, O. 2004, \pasp, 116, 345

\bibitem[{{Cool} {et~al.}(2013){Cool}, {Moustakas}, {Blanton}, {Burles},
  {Coil}, {Eisenstein}, {Wong}, {Zhu}, {Aird}, {Bernstein}, {Bolton}, {Hogg},
  \& {Mendez}}]{Cool2013}
{Cool}, R.~J., {Moustakas}, J., {Blanton}, M.~R., {et~al.} 2013, \apj, 767, 118

\bibitem[{{Cranmer} {et~al.}(2021){Cranmer}, {Tamayo}, {Rein}, {Battaglia},
  {Hadden}, {Armitage}, {Ho}, \& {Spergel}}]{Cranmer2021}
{Cranmer}, M., {Tamayo}, D., {Rein}, H., {et~al.} 2021, Proceedings of the
  National Academy of Science, 118, 2026053118

\bibitem[{Cui {et~al.}(2019)Cui, Jia, Lin, Song, \& Belongie}]{Cui2019CVPR}
Cui, Y., Jia, M., Lin, T.-Y., Song, Y., \& Belongie, S. 2019, in Proceedings of
  the IEEE/CVF Conference on Computer Vision and Pattern Recognition (CVPR)

\bibitem[{{Davis} {et~al.}(2017){Davis}, {Gatti}, {Vielzeuf}, {Cawthon},
  {Rozo}, {Alarcon}, {Bernstein}, {Bonnett}, {Carnero Rosell}, {Castander},
  {Chang}, {da Costa}, {Davis}, {De Vicente}, {DeRose}, {Drlica-Wagner},
  {Elvin-Poole}, {Gaztanaga}, {Gruen}, {Gschwend}, {Hartley}, {Hoyle}, {Lin},
  {Maia}, {Miquel}, {Ogando}, {Rau}, {Roodman}, {Rykoff}, {Sevilla-Noarbe},
  {Troxel}, {Wechsler}, {Abbott}, {Abdalla}, {Allam}, {Annis}, {Bechtol},
  {Benoit-L{\'e}vy}, {Brooks}, {Buckley-Geer}, {Burke}, {Carrasco Kind},
  {Carretero}, {Crocce}, {Cunha}, {Desai}, {Diehl}, {Doel}, {Eifler},
  {Flaugher}, {Frieman}, {Garc{\'\i}a-Bellido}, {Gerdes}, {Gruendl},
  {Gutierrez}, {Honscheid}, {James}, {Jeltema}, {Krause}, {Kron}, {Kuehn},
  {Kuropatkin}, {Lahav}, {Lima}, {March}, {Marshall}, {Menanteau}, {Nichol},
  {Nord}, {Plazas}, {Sanchez}, {Scarpine}, {Schindler}, {Smith},
  {Soares-Santos}, {Sobreira}, {Suchyta}, {Swanson}, {Tarle}, {Thomas},
  {Tucker}, {Vikram}, {Walker}, \& {Zuntz}}]{Davis2017}
{Davis}, C., {Gatti}, M., {Vielzeuf}, P., {et~al.} 2017, arXiv e-prints,
  arXiv:1710.02517

\bibitem[{{D'Isanto} \& {Polsterer}(2018)}]{DP2018}
{D'Isanto}, A. \& {Polsterer}, K.~L. 2018, \aap, 609, A111

\bibitem[{{Dom{\'\i}nguez S{\'a}nchez} {et~al.}(2018){Dom{\'\i}nguez
  S{\'a}nchez}, {Huertas-Company}, {Bernardi}, {Tuccillo}, \&
  {Fischer}}]{Dominguez2018}
{Dom{\'\i}nguez S{\'a}nchez}, H., {Huertas-Company}, M., {Bernardi}, M.,
  {Tuccillo}, D., \& {Fischer}, J.~L. 2018, \mnras, 476, 3661

\bibitem[{{Drinkwater} {et~al.}(2018){Drinkwater}, {Byrne}, {Blake},
  {Glazebrook}, {Brough}, {Colless}, {Couch}, {Croton}, {Croom}, {Davis},
  {Forster}, {Gilbank}, {Hinton}, {Jelliffe}, {Jurek}, {Li}, {Martin},
  {Pimbblet}, {Poole}, {Pracy}, {Sharp}, {Smillie}, {Spolaor}, {Wisnioski},
  {Woods}, {Wyder}, \& {Yee}}]{Drinkwater2018}
{Drinkwater}, M.~J., {Byrne}, Z.~J., {Blake}, C., {et~al.} 2018, \mnras, 474,
  4151

\bibitem[{Duarte {et~al.}(2021)Duarte, Rawat, \& Shah}]{Duarte2021}
Duarte, K., Rawat, Y., \& Shah, M. 2021, in 2021 IEEE/CVF Conference on
  Computer Vision and Pattern Recognition Workshops (CVPRW), 2733--2742

\bibitem[{{Euclid Collaboration} {et~al.}(2021){Euclid Collaboration},
  {Ilbert}, {de la Torre}, {Martinet}, {Wright}, {Paltani}, {Laigle},
  {Davidzon}, {Jullo}, {Hildebrandt}, {Masters}, {Amara}, {Conselice},
  {Andreon}, {Auricchio}, {Azzollini}, {Baccigalupi},
  {Balaguera-Antol{\'\i}nez}, {Baldi}, {Balestra}, {Bardelli}, {Bender},
  {Biviano}, {Bodendorf}, {Bonino}, {Borgani}, {Boucaud}, {Bozzo}, {Branchini},
  {Brescia}, {Burigana}, {Cabanac}, {Camera}, {Capobianco}, {Cappi}, {Carbone},
  {Carretero}, {Carvalho}, {Casas}, {Castander}, {Castellano}, {Castignani},
  {Cavuoti}, {Cimatti}, {Cledassou}, {Colodro-Conde}, {Congedo}, {Conversi},
  {Copin}, {Corcione}, {Costille}, {Coupon}, {Courtois}, {Cropper}, {Cuby}, {Da
  Silva}, {Degaudenzi}, {Di Ferdinando}, {Dubath}, {Duncan}, {Dupac}, {Dusini},
  {Ealet}, {Fabricius}, {Farrens}, {Ferreira}, {Finelli}, {Fosalba},
  {Fotopoulou}, {Franceschi}, {Franzetti}, {Galeotta}, {Garilli}, {Gillard},
  {Gillis}, {Giocoli}, {Gozaliasl}, {Graci{\'a}-Carpio}, {Grupp}, {Guzzo},
  {Haugan}, {Holmes}, {Hormuth}, {Jahnke}, {Keihanen}, {Kermiche}, {Kiessling},
  {Kirkpatrick}, {Kunz}, {Kurki-Suonio}, {Ligori}, {Lilje}, {Lloro}, {Maino},
  {Maiorano}, {Marggraf}, {Markovic}, {Marulli}, {Massey}, {Maturi}, {Mauri},
  {Maurogordato}, {McCracken}, {Medinaceli}, {Mei}, {Metcalf}, {Moresco},
  {Morin}, {Moscardini}, {Munari}, {Nakajima}, {Neissner}, {Niemi},
  {Nightingale}, {Padilla}, {Pasian}, {Patrizii}, {Pedersen}, {Pello},
  {Pettorino}, {Pires}, {Polenta}, {Poncet}, {Popa}, {Potter}, {Pozzetti},
  {Raison}, {Renzi}, {Rhodes}, {Riccio}, {Romelli}, {Roncarelli}, {Rossetti},
  {Saglia}, {S{\'a}nchez}, {Sapone}, {Schneider}, {Schrabback}, {Scottez},
  {Secroun}, {Seidel}, {Serrano}, {Sirignano}, {Sirri}, {Stanco}, {Sureau},
  {Tallada Cresp{\'a}}, {Tenti}, {Teplitz}, {Tereno}, {Toledo-Moreo},
  {Torradeflot}, {Tramacere}, {Valentijn}, {Valenziano}, {Valiviita},
  {Vassallo}, {Wang}, {Welikala}, {Weller}, {Whittaker}, {Zacchei}, {Zamorani},
  {Zoubian}, \& {Zucca}}]{Euclid2021}
{Euclid Collaboration}, {Ilbert}, O., {de la Torre}, S., {et~al.} 2021, \aap,
  647, A117

\bibitem[{{Feldmann} {et~al.}(2006){Feldmann}, {Carollo}, {Porciani}, {Lilly},
  {Capak}, {Taniguchi}, {Le F{\`e}vre}, {Renzini}, {Scoville}, {Ajiki},
  {Aussel}, {Contini}, {McCracken}, {Mobasher}, {Murayama}, {Sanders},
  {Sasaki}, {Scarlata}, {Scodeggio}, {Shioya}, {Silverman}, {Takahashi},
  {Thompson}, \& {Zamorani}}]{Feldmann2006}
{Feldmann}, R., {Carollo}, C.~M., {Porciani}, C., {et~al.} 2006, \mnras, 372,
  565

\bibitem[{García {et~al.}(2012)García, Sánchez, \& Mollineda}]{Garcia2012}
García, V., Sánchez, J., \& Mollineda, R. 2012, Knowledge-Based Systems, 25,
  13, special Issue on New Trends in Data Mining

\bibitem[{{Garilli} {et~al.}(2021){Garilli}, {McLure}, {Pentericci},
  {Franzetti}, {Gargiulo}, {Carnall}, {Cucciati}, {Iovino}, {Amorin},
  {Bolzonella}, {Bongiorno}, {Castellano}, {Cimatti}, {Cirasuolo}, {Cullen},
  {Dunlop}, {Elbaz}, {Finkelstein}, {Fontana}, {Fontanot}, {Fumana}, {Guaita},
  {Hartley}, {Jarvis}, {Juneau}, {Maccagni}, {McLeod}, {Nandra}, {Pompei},
  {Pozzetti}, {Scodeggio}, {Talia}, {Calabr{\`o}}, {Cresci}, {Fynbo}, {Hathi},
  {Hibon}, {Koekemoer}, {Magliocchetti}, {Salvato}, {Vietri}, {Zamorani},
  {Almaini}, {Balestra}, {Bardelli}, {Begley}, {Brammer}, {Bell}, {Bowler},
  {Brusa}, {Buitrago}, {Caputi}, {Cassata}, {Charlot}, {Citro}, {Cristiani},
  {Curtis-Lake}, {Dickinson}, {Fazio}, {Ferguson}, {Fiore}, {Franco},
  {Georgakakis}, {Giavalisco}, {Grazian}, {Hamadouche}, {Jung}, {Kim},
  {Khusanova}, {Le F{\`e}vre}, {Longhetti}, {Lotz}, {Mannucci}, {Maltby},
  {Matsuoka}, {Mendez-Hernandez}, {Mendez-Abreu}, {Mignoli}, {Moresco},
  {Nonino}, {Pannella}, {Papovich}, {Popesso}, {Roberts-Borsani}, {Rosario},
  {Saldana-Lopez}, {Santini}, {Saxena}, {Schaerer}, {Schreiber}, {Stark},
  {Tasca}, {Thomas}, {Vanzella}, {Wild}, {Williams}, \& {Zucca}}]{Garilli2021}
{Garilli}, B., {McLure}, R., {Pentericci}, L., {et~al.} 2021, \aap, 647, A150

\bibitem[{{Gatti} {et~al.}(2022){Gatti}, {Giannini}, {Bernstein}, {Alarcon},
  {Myles}, {Amon}, {Cawthon}, {Troxel}, {DeRose}, {Everett}, {Ross}, {Rykoff},
  {Elvin-Poole}, {Cordero}, {Harrison}, {Sanchez}, {Prat}, {Gruen}, {Lin},
  {Crocce}, {Rozo}, {Abbott}, {Aguena}, {Allam}, {Annis}, {Avila}, {Bacon},
  {Bertin}, {Brooks}, {Burke}, {Rosell}, {Kind}, {Carretero}, {Castander},
  {Choi}, {Conselice}, {Costanzi}, {Crocce}, {da Costa}, {Pereira}, {Dawson},
  {Desai}, {Diehl}, {Eckert}, {Eifler}, {Evrard}, {Ferrero}, {Flaugher},
  {Fosalba}, {Frieman}, {Garc{\'\i}a-Bellido}, {Gaztanaga}, {Giannantonio},
  {Gruendl}, {Gschwend}, {Hinton}, {Hollowood}, {Honscheid}, {Hoyle},
  {Huterer}, {James}, {Kuehn}, {Kuropatkin}, {Lahav}, {Lima}, {MacCrann},
  {Maia}, {March}, {Marshall}, {Melchior}, {Menanteau}, {Miquel}, {Mohr},
  {Morgan}, {Ogando}, {Palmese}, {Paz-Chinch{\'o}n}, {Percival}, {Plazas},
  {Rodriguez-Monroy}, {Roodman}, {Rossi}, {Samuroff}, {Sanchez}, {Scarpine},
  {Secco}, {Serrano}, {Sevilla-Noarbe}, {Smith}, {Soares-Santos}, {Suchyta},
  {Swanson}, {Tarle}, {Thomas}, {To}, {Varga}, {Weller}, {Wilkinson},
  {Wilkinson}, \& {DES Collaboration}}]{Gatti2022}
{Gatti}, M., {Giannini}, G., {Bernstein}, G.~M., {et~al.} 2022, \mnras, 510,
  1223

\bibitem[{{Gerdes} {et~al.}(2010){Gerdes}, {Sypniewski}, {McKay}, {Hao},
  {Weis}, {Wechsler}, \& {Busha}}]{Gerdes2010}
{Gerdes}, D.~W., {Sypniewski}, A.~J., {McKay}, T.~A., {et~al.} 2010, \apj, 715,
  823

\bibitem[{{Greisel} {et~al.}(2015){Greisel}, {Seitz}, {Drory}, {Bender},
  {Saglia}, \& {Snigula}}]{Greisel2015}
{Greisel}, N., {Seitz}, S., {Drory}, N., {et~al.} 2015, \mnras, 451, 1848

\bibitem[{{Gupta} {et~al.}(2018){Gupta}, {Zorrilla Matilla}, {Hsu}, \&
  {Haiman}}]{Gupta2018}
{Gupta}, A., {Zorrilla Matilla}, J.~M., {Hsu}, D., \& {Haiman}, Z. 2018, \prd,
  97, 103515

\bibitem[{Han {et~al.}(2005)Han, Wang, \& Mao}]{Han2005}
Han, H., Wang, W.-Y., \& Mao, B.-H. 2005, in Advances in Intelligent Computing,
  ed. D.-S. Huang, X.-P. Zhang, \& G.-B. Huang (Berlin, Heidelberg: Springer
  Berlin Heidelberg), 878--887

\bibitem[{Hatfield {et~al.}(2020)Hatfield, Almosallam, Jarvis, Adams, Bowler,
  Gomes, Roberts, \& Schreiber}]{Hatfield2020}
Hatfield, P.~W., Almosallam, I.~A., Jarvis, M.~J., {et~al.} 2020, \mnras, 498,
  5498–5510

\bibitem[{Hayat {et~al.}(2019)Hayat, Khan, Zamir, Shen, \& Shao}]{Hayat2019}
Hayat, M., Khan, S., Zamir, W., Shen, J., \& Shao, L. 2019, Max-margin Class
  Imbalanced Learning with Gaussian Affinity

\bibitem[{{Hemmati} {et~al.}(2019){Hemmati}, {Capak}, {Masters}, {Davidzon},
  {Dor{\`e}}, {Kruk}, {Mobasher}, {Rhodes}, {Scolnic}, \&
  {Stern}}]{Hemmati2019}
{Hemmati}, S., {Capak}, P., {Masters}, D., {et~al.} 2019, \apj, 877, 117

\bibitem[{Hosenie {et~al.}(2020)Hosenie, Lyon, Stappers, Mootoovaloo, \&
  McBride}]{Hosenie2020}
Hosenie, Z., Lyon, R., Stappers, B., Mootoovaloo, A., \& McBride, V. 2020,
  \mnras, 493, 6050–6059

\bibitem[{{Hoyle}(2016)}]{Hoyle2016}
{Hoyle}, B. 2016, Astronomy and Computing, 16, 34

\bibitem[{Huang {et~al.}(2016)Huang, Li, Loy, \& Tang}]{Huang2016imbalance}
Huang, C., Li, Y., Loy, C.~C., \& Tang, X. 2016, in 2016 IEEE Conference on
  Computer Vision and Pattern Recognition (CVPR), 5375--5384

\bibitem[{Huang {et~al.}(2020)Huang, Li, Loy, \& Tang}]{Huang2020imbalance}
Huang, C., Li, Y., Loy, C.~C., \& Tang, X. 2020, IEEE Transactions on Pattern
  Analysis and Machine Intelligence, 42, 2781

\bibitem[{{Hudelot} {et~al.}(2012){Hudelot}, {Cuillandre}, {Withington},
  {Goranova}, {McCracken}, {Magnard}, {Mellier}, {Regnault}, {Betoule},
  {Aussel}, {Kavelaars}, {Fernique}, {Bonnarel}, {Ochsenbein}, \&
  {Ilbert}}]{CFHTLST07}
{Hudelot}, P., {Cuillandre}, J.~C., {Withington}, K., {et~al.} 2012, VizieR
  Online Data Catalog, II/317

\bibitem[{{Ilbert} {et~al.}(2006){Ilbert}, {Arnouts}, {McCracken},
  {Bolzonella}, {Bertin}, {Le F{\`e}vre}, {Mellier}, {Zamorani}, {Pell{\`o}},
  {Iovino}, {Tresse}, {Le Brun}, {Bottini}, {Garilli}, {Maccagni}, {Picat},
  {Scaramella}, {Scodeggio}, {Vettolani}, {Zanichelli}, {Adami}, {Bardelli},
  {Cappi}, {Charlot}, {Ciliegi}, {Contini}, {Cucciati}, {Foucaud}, {Franzetti},
  {Gavignaud}, {Guzzo}, {Marano}, {Marinoni}, {Mazure}, {Meneux}, {Merighi},
  {Paltani}, {Pollo}, {Pozzetti}, {Radovich}, {Zucca}, {Bondi}, {Bongiorno},
  {Busarello}, {de La Torre}, {Gregorini}, {Lamareille}, {Mathez}, {Merluzzi},
  {Ripepi}, {Rizzo}, \& {Vergani}}]{Ilbert2006}
{Ilbert}, O., {Arnouts}, S., {McCracken}, H.~J., {et~al.} 2006, \aap, 457, 841

\bibitem[{Jia \& Zhao(2019)}]{Jia2019mode}
Jia, J. \& Zhao, Q. 2019, in 2019 12th International Congress on Image and
  Signal Processing, BioMedical Engineering and Informatics (CISP-BMEI), 1--6

\bibitem[{Jones \& Singal(2017)}]{Jones2017}
Jones, E. \& Singal, J. 2017, A\&A, 600, A113

\bibitem[{Kang {et~al.}(2020)Kang, Xie, Rohrbach, Yan, Gordo, Feng, \&
  Kalantidis}]{Kang2020Decoupling}
Kang, B., Xie, S., Rohrbach, M., {et~al.} 2020, in International Conference on
  Learning Representations

\bibitem[{Khan {et~al.}(2019)Khan, Hayat, Zamir, Shen, \& Shao}]{Khan2019}
Khan, S., Hayat, M., Zamir, S.~W., Shen, J., \& Shao, L. 2019, in 2019 IEEE/CVF
  Conference on Computer Vision and Pattern Recognition (CVPR), 103--112

\bibitem[{Khan {et~al.}(2018)Khan, Hayat, Bennamoun, Sohel, \&
  Togneri}]{Khan2017}
Khan, S.~H., Hayat, M., Bennamoun, M., Sohel, F.~A., \& Togneri, R. 2018, IEEE
  Transactions on Neural Networks and Learning Systems, 29, 3573

\bibitem[{Kingma \& Ba(2015)}]{Adam}
Kingma, D.~P. \& Ba, J. 2015, in 3rd International Conference on Learning
  Representations, {ICLR} 2015, San Diego, CA, USA, May 7-9, 2015, Conference
  Track Proceedings, ed. Y.~Bengio \& Y.~LeCun

\bibitem[{{Kodali} {et~al.}(2017){Kodali}, {Abernethy}, {Hays}, \&
  {Kira}}]{Kodali2017}
{Kodali}, N., {Abernethy}, J., {Hays}, J., \& {Kira}, Z. 2017, arXiv e-prints,
  arXiv:1705.07215

\bibitem[{Kovetz {et~al.}(2017)Kovetz, Raccanelli, \& Rahman}]{Kovetz2017}
Kovetz, E.~D., Raccanelli, A., \& Rahman, M. 2017, \mnras, 468, 3650–3656

\bibitem[{{Laureijs} {et~al.}(2011){Laureijs}, {Amiaux}, {Arduini},
  {Augu{\`e}res}, {Brinchmann}, {Cole}, {Cropper}, {Dabin}, {Duvet}, {Ealet},
  {Garilli}, {Gondoin}, {Guzzo}, {Hoar}, {Hoekstra}, {Holmes}, {Kitching},
  {Maciaszek}, {Mellier}, {Pasian}, {Percival}, {Rhodes}, {Saavedra Criado},
  {Sauvage}, {Scaramella}, {Valenziano}, {Warren}, {Bender}, {Castander},
  {Cimatti}, {Le F{\`e}vre}, {Kurki-Suonio}, {Levi}, {Lilje}, {Meylan},
  {Nichol}, {Pedersen}, {Popa}, {Rebolo Lopez}, {Rix}, {Rottgering},
  {Zeilinger}, {Grupp}, {Hudelot}, {Massey}, {Meneghetti}, {Miller}, {Paltani},
  {Paulin-Henriksson}, {Pires}, {Saxton}, {Schrabback}, {Seidel}, {Walsh},
  {Aghanim}, {Amendola}, {Bartlett}, {Baccigalupi}, {Beaulieu}, {Benabed},
  {Cuby}, {Elbaz}, {Fosalba}, {Gavazzi}, {Helmi}, {Hook}, {Irwin}, {Kneib},
  {Kunz}, {Mannucci}, {Moscardini}, {Tao}, {Teyssier}, {Weller}, {Zamorani},
  {Zapatero Osorio}, {Boulade}, {Foumond}, {Di Giorgio}, {Guttridge}, {James},
  {Kemp}, {Martignac}, {Spencer}, {Walton}, {Bl{\"u}mchen}, {Bonoli},
  {Bortoletto}, {Cerna}, {Corcione}, {Fabron}, {Jahnke}, {Ligori}, {Madrid},
  {Martin}, {Morgante}, {Pamplona}, {Prieto}, {Riva}, {Toledo}, {Trifoglio},
  {Zerbi}, {Abdalla}, {Douspis}, {Grenet}, {Borgani}, {Bouwens}, {Courbin},
  {Delouis}, {Dubath}, {Fontana}, {Frailis}, {Grazian}, {Koppenh{\"o}fer},
  {Mansutti}, {Melchior}, {Mignoli}, {Mohr}, {Neissner}, {Noddle}, {Poncet},
  {Scodeggio}, {Serrano}, {Shane}, {Starck}, {Surace}, {Taylor},
  {Verdoes-Kleijn}, {Vuerli}, {Williams}, {Zacchei}, {Altieri}, {Escudero
  Sanz}, {Kohley}, {Oosterbroek}, {Astier}, {Bacon}, {Bardelli}, {Baugh},
  {Bellagamba}, {Benoist}, {Bianchi}, {Biviano}, {Branchini}, {Carbone},
  {Cardone}, {Clements}, {Colombi}, {Conselice}, {Cresci}, {Deacon}, {Dunlop},
  {Fedeli}, {Fontanot}, {Franzetti}, {Giocoli}, {Garcia-Bellido}, {Gow},
  {Heavens}, {Hewett}, {Heymans}, {Holland}, {Huang}, {Ilbert}, {Joachimi},
  {Jennins}, {Kerins}, {Kiessling}, {Kirk}, {Kotak}, {Krause}, {Lahav}, {van
  Leeuwen}, {Lesgourgues}, {Lombardi}, {Magliocchetti}, {Maguire}, {Majerotto},
  {Maoli}, {Marulli}, {Maurogordato}, {McCracken}, {McLure}, {Melchiorri},
  {Merson}, {Moresco}, {Nonino}, {Norberg}, {Peacock}, {Pello}, {Penny},
  {Pettorino}, {Di Porto}, {Pozzetti}, {Quercellini}, {Radovich}, {Rassat},
  {Roche}, {Ronayette}, {Rossetti}, {Sartoris}, {Schneider}, {Semboloni},
  {Serjeant}, {Simpson}, {Skordis}, {Smadja}, {Smartt}, {Spano}, {Spiro},
  {Sullivan}, {Tilquin}, {Trotta}, {Verde}, {Wang}, {Williger}, {Zhao},
  {Zoubian}, \& {Zucca}}]{Laureijs2011}
{Laureijs}, R., {Amiaux}, J., {Arduini}, S., {et~al.} 2011, arXiv e-prints,
  arXiv:1110.3193

\bibitem[{{Le F{\`e}vre} {et~al.}(2013){Le F{\`e}vre}, {Cassata}, {Cucciati},
  {Garilli}, {Ilbert}, {Le Brun}, {Maccagni}, {Moreau}, {Scodeggio}, {Tresse},
  {Zamorani}, {Adami}, {Arnouts}, {Bardelli}, {Bolzonella}, {Bondi},
  {Bongiorno}, {Bottini}, {Cappi}, {Charlot}, {Ciliegi}, {Contini}, {de la
  Torre}, {Foucaud}, {Franzetti}, {Gavignaud}, {Guzzo}, {Iovino}, {Lemaux},
  {L{\'o}pez-Sanjuan}, {McCracken}, {Marano}, {Marinoni}, {Mazure}, {Mellier},
  {Merighi}, {Merluzzi}, {Paltani}, {Pell{\`o}}, {Pollo}, {Pozzetti},
  {Scaramella}, {Tasca}, {Vergani}, {Vettolani}, {Zanichelli}, \&
  {Zucca}}]{LeFevre2013}
{Le F{\`e}vre}, O., {Cassata}, P., {Cucciati}, O., {et~al.} 2013, \aap, 559,
  A14

\bibitem[{{Le F{\`e}vre} {et~al.}(2015){Le F{\`e}vre}, {Tasca}, {Cassata},
  {Garilli}, {Le Brun}, {Maccagni}, {Pentericci}, {Thomas}, {Vanzella},
  {Zamorani}, {Zucca}, {Amorin}, {Bardelli}, {Capak}, {Cassar{\`a}},
  {Castellano}, {Cimatti}, {Cuby}, {Cucciati}, {de la Torre}, {Durkalec},
  {Fontana}, {Giavalisco}, {Grazian}, {Hathi}, {Ilbert}, {Lemaux}, {Moreau},
  {Paltani}, {Ribeiro}, {Salvato}, {Schaerer}, {Scodeggio}, {Sommariva},
  {Talia}, {Taniguchi}, {Tresse}, {Vergani}, {Wang}, {Charlot}, {Contini},
  {Fotopoulou}, {L{\'o}pez-Sanjuan}, {Mellier}, \& {Scoville}}]{LeFevre2015}
{Le F{\`e}vre}, O., {Tasca}, L.~A.~M., {Cassata}, P., {et~al.} 2015, \aap, 576,
  A79

\bibitem[{{Lee} {et~al.}(2018){Lee}, {Krolewski}, {White}, {Schlegel},
  {Nugent}, {Hennawi}, {M{\"u}ller}, {Pan}, {Prochaska}, {Font-Ribera},
  {Suzuki}, {Glazebrook}, {Kacprzak}, {Kartaltepe}, {Koekemoer}, {Le
  F{\`e}vre}, {Lemaux}, {Maier}, {Nanayakkara}, {Rich}, {Sanders}, {Salvato},
  {Tasca}, \& {Tran}}]{KGLee2018}
{Lee}, K.-G., {Krolewski}, A., {White}, M., {et~al.} 2018, \apjs, 237, 31

\bibitem[{Leistedt {et~al.}(2019)Leistedt, Hogg, Wechsler, \&
  DeRose}]{Leistedt2019}
Leistedt, B., Hogg, D.~W., Wechsler, R.~H., \& DeRose, J. 2019, \apj, 881, 80

\bibitem[{Li {et~al.}(2021)Li, Wang, Zhang, Hu, \& Ouyang}]{Li2021gan}
Li, Y., Wang, Q., Zhang, J., Hu, L., \& Ouyang, W. 2021, Neurocomputing, 435,
  26

\bibitem[{{Lilly} {et~al.}(2007){Lilly}, {Le F{\`e}vre}, {Renzini}, {Zamorani},
  {Scodeggio}, {Contini}, {Carollo}, {Hasinger}, {Kneib}, {Iovino}, {Le Brun},
  {Maier}, {Mainieri}, {Mignoli}, {Silverman}, {Tasca}, {Bolzonella},
  {Bongiorno}, {Bottini}, {Capak}, {Caputi}, {Cimatti}, {Cucciati}, {Daddi},
  {Feldmann}, {Franzetti}, {Garilli}, {Guzzo}, {Ilbert}, {Kampczyk}, {Kovac},
  {Lamareille}, {Leauthaud}, {Le Borgne}, {McCracken}, {Marinoni}, {Pello},
  {Ricciardelli}, {Scarlata}, {Vergani}, {Sanders}, {Schinnerer}, {Scoville},
  {Taniguchi}, {Arnouts}, {Aussel}, {Bardelli}, {Brusa}, {Cappi}, {Ciliegi},
  {Finoguenov}, {Foucaud}, {Franceschini}, {Halliday}, {Impey}, {Knobel},
  {Koekemoer}, {Kurk}, {Maccagni}, {Maddox}, {Marano}, {Marconi}, {Meneux},
  {Mobasher}, {Moreau}, {Peacock}, {Porciani}, {Pozzetti}, {Scaramella},
  {Schiminovich}, {Shopbell}, {Smail}, {Thompson}, {Tresse}, {Vettolani},
  {Zanichelli}, \& {Zucca}}]{Lilly2007}
{Lilly}, S.~J., {Le F{\`e}vre}, O., {Renzini}, A., {et~al.} 2007, \apjs, 172,
  70

\bibitem[{Liu {et~al.}(2019)Liu, Miao, Zhan, Wang, Gong, \&
  Yu}]{Liu2019imbalance}
Liu, Z., Miao, Z., Zhan, X., {et~al.} 2019, in Proceedings of the IEEE/CVF
  Conference on Computer Vision and Pattern Recognition, 2537--2546

\bibitem[{{Malz}(2021)}]{Malz_Alex2021}
{Malz}, A.~I. 2021, \prd, 103, 083502

\bibitem[{{Malz} \& {Hogg}(2020)}]{Malz2020}
{Malz}, A.~I. \& {Hogg}, D.~W. 2020, arXiv e-prints, arXiv:2007.12178

\bibitem[{Mandelbaum {et~al.}(2008)Mandelbaum, Seljak, Hirata, Bardelli,
  Bolzonella, Bongiorno, Carollo, Contini, Cunha, Garilli, \&
  et~al.}]{Mandelbaum2008}
Mandelbaum, R., Seljak, U., Hirata, C.~M., {et~al.} 2008, \mnras, 386,
  781–806

\bibitem[{McLeod {et~al.}(2017)McLeod, Libeskind, Lahav, \&
  Hoffman}]{McLeod2017}
McLeod, M., Libeskind, N., Lahav, O., \& Hoffman, Y. 2017, Journal of Cosmology
  and Astroparticle Physics, 2017, 034–034

\bibitem[{{McLure} {et~al.}(2013){McLure}, {Pearce}, {Dunlop}, {Cirasuolo},
  {Curtis-Lake}, {Bruce}, {Caputi}, {Almaini}, {Bonfield}, {Bradshaw},
  {Buitrago}, {Chuter}, {Foucaud}, {Hartley}, \& {Jarvis}}]{McLure2013}
{McLure}, R.~J., {Pearce}, H.~J., {Dunlop}, J.~S., {et~al.} 2013, \mnras, 428,
  1088

\bibitem[{{Momcheva} {et~al.}(2016){Momcheva}, {Brammer}, {van Dokkum},
  {Skelton}, {Whitaker}, {Nelson}, {Fumagalli}, {Maseda}, {Leja}, {Franx},
  {Rix}, {Bezanson}, {Da Cunha}, {Dickey}, {F{\"o}rster Schreiber},
  {Illingworth}, {Kriek}, {Labb{\'e}}, {Ulf Lange}, {Lundgren}, {Magee},
  {Marchesini}, {Oesch}, {Pacifici}, {Patel}, {Price}, {Tal}, {Wake}, {van der
  Wel}, \& {Wuyts}}]{Momcheva2016}
{Momcheva}, I.~G., {Brammer}, G.~B., {van Dokkum}, P.~G., {et~al.} 2016, \apjs,
  225, 27

\bibitem[{Morrison {et~al.}(2017)Morrison, Hildebrandt, Schmidt, Baldry,
  Bilicki, Choi, Erben, \& Schneider}]{Morrison2017}
Morrison, C.~B., Hildebrandt, H., Schmidt, S.~J., {et~al.} 2017, \mnras, 467,
  3576

\bibitem[{{Mu} {et~al.}(2020){Mu}, {Qiu}, {Zhang}, {Ma}, \&
  {Fan}}]{Mu2020photoz}
{Mu}, Y.-H., {Qiu}, B., {Zhang}, J.-N., {Ma}, J.-C., \& {Fan}, X.-D. 2020,
  Research in Astronomy and Astrophysics, 20, 089

\bibitem[{M\"{u}ller {et~al.}(2019)M\"{u}ller, Kornblith, \&
  Hinton}]{Muller2019}
M\"{u}ller, R., Kornblith, S., \& Hinton, G.~E. 2019, in Advances in Neural
  Information Processing Systems, ed. H.~Wallach, H.~Larochelle,
  A.~Beygelzimer, F.~d\textquotesingle Alch\'{e}-Buc, E.~Fox, \& R.~Garnett,
  Vol.~32 (Curran Associates, Inc.)

\bibitem[{{Newman}(2008)}]{Newman2008}
{Newman}, J.~A. 2008, \apj, 684, 88

\bibitem[{{Newman} {et~al.}(2013){Newman}, {Cooper}, {Davis}, {Faber}, {Coil},
  {Guhathakurta}, {Koo}, {Phillips}, {Conroy}, {Dutton}, {Finkbeiner}, {Gerke},
  {Rosario}, {Weiner}, {Willmer}, {Yan}, {Harker}, {Kassin}, {Konidaris},
  {Lai}, {Madgwick}, {Noeske}, {Wirth}, {Connolly}, {Kaiser}, {Kirby},
  {Lemaux}, {Lin}, {Lotz}, {Luppino}, {Marinoni}, {Matthews}, {Metevier}, \&
  {Schiavon}}]{Newman2013}
{Newman}, J.~A., {Cooper}, M.~C., {Davis}, M., {et~al.} 2013, \apjs, 208, 5

\bibitem[{Nguyen {et~al.}(2018)Nguyen, Mukkamala, \& Hein}]{Nguyen2018}
Nguyen, Q., Mukkamala, M.~C., \& Hein, M. 2018, in Proceedings of Machine
  Learning Research, Vol.~80, Proceedings of the 35th International Conference
  on Machine Learning, ed. J.~Dy \& A.~Krause (PMLR), 3740--3749

\bibitem[{{Ntampaka} {et~al.}(2015){Ntampaka}, {Trac}, {Sutherland},
  {Battaglia}, {P{\'o}czos}, \& {Schneider}}]{Ntampaka2015}
{Ntampaka}, M., {Trac}, H., {Sutherland}, D.~J., {et~al.} 2015, \apj, 803, 50

\bibitem[{Ntampaka {et~al.}(2019)Ntampaka, ZuHone, Eisenstein, Nagai,
  Vikhlinin, Hernquist, Marinacci, Nelson, Pakmor, Pillepich, Torrey, \&
  Vogelsberger}]{Ntampaka2019}
Ntampaka, M., ZuHone, J., Eisenstein, D., {et~al.} 2019, The Astrophysical
  Journal, 876, 82

\bibitem[{Okerinde {et~al.}(2021)Okerinde, Hsu, Theis, Nafi, \&
  Shamir}]{Okerinde2021}
Okerinde, A., Hsu, W., Theis, T., Nafi, N., \& Shamir, L. 2021, in Computer
  Analysis of Images and Patterns, ed. N.~Tsapatsoulis, A.~Panayides,
  T.~Theocharides, A.~Lanitis, C.~Pattichis, \& M.~Vento (Cham: Springer
  International Publishing), 322--331

\bibitem[{{Pasquet} {et~al.}(2019){Pasquet}, {Bertin}, {Treyer}, {Arnouts}, \&
  {Fouchez}}]{Pasquet2019}
{Pasquet}, J., {Bertin}, E., {Treyer}, M., {Arnouts}, S., \& {Fouchez}, D.
  2019, \aap, 621, A26

\bibitem[{{Rau} {et~al.}(2022){Rau}, {Morrison}, {Schmidt}, {Wilson},
  {Mandelbaum}, {Mao}, {Mao}, \& {LSST Dark Energy Science
  Collaboration}}]{Rau2022}
{Rau}, M.~M., {Morrison}, C.~B., {Schmidt}, S.~J., {et~al.} 2022, \mnras, 509,
  4886

\bibitem[{Rau {et~al.}(2015)Rau, Seitz, Brimioulle, Frank, Friedrich, Gruen, \&
  Hoyle}]{Rau2015}
Rau, M.~M., Seitz, S., Brimioulle, F., {et~al.} 2015, \mnras, 452, 3710

\bibitem[{Ravanbakhsh {et~al.}(2016)Ravanbakhsh, Oliva, Fromenteau, Price, Ho,
  Schneider, \& P\'{o}czos}]{Ravanbakhsh2016b}
Ravanbakhsh, S., Oliva, J., Fromenteau, S., {et~al.} 2016, in Proceedings of
  the 33rd International Conference on International Conference on Machine
  Learning - Volume 48, ICML'16 (JMLR.org), 2407–2416

\bibitem[{{Ribli} {et~al.}(2019){Ribli}, {Pataki}, \& {Csabai}}]{Ribli2019}
{Ribli}, D., {Pataki}, B.~{\'A}., \& {Csabai}, I. 2019, Nature Astronomy, 3, 93

\bibitem[{Ruff {et~al.}(2018)Ruff, Vandermeulen, Goernitz, Deecke, Siddiqui,
  Binder, M{\"u}ller, \& Kloft}]{Ruff2018}
Ruff, L., Vandermeulen, R., Goernitz, N., {et~al.} 2018, in Proceedings of
  Machine Learning Research, Vol.~80, Proceedings of the 35th International
  Conference on Machine Learning, ed. J.~Dy \& A.~Krause (PMLR), 4393--4402

\bibitem[{{Salvato} {et~al.}(2019){Salvato}, {Ilbert}, \&
  {Hoyle}}]{Salvato2019}
{Salvato}, M., {Ilbert}, O., \& {Hoyle}, B. 2019, Nature Astronomy, 3, 212

\bibitem[{S\'{a}nchez \& Bernstein(2018)}]{Sanchez2018}
S\'{a}nchez, C. \& Bernstein, G.~M. 2018, \mnras, 483, 2801

\bibitem[{Santurkar {et~al.}(2018)Santurkar, Schmidt, \& Madry}]{Santurkar2018}
Santurkar, S., Schmidt, L., \& Madry, A. 2018, in International Conference on
  Machine Learning, PMLR, 4480--4489

\bibitem[{{Schuldt} {et~al.}(2021){Schuldt}, {Suyu}, {Ca{\~n}ameras},
  {Taubenberger}, {Meinhardt}, {Leal-Taix{\'e}}, \& {Hsieh}}]{Schuldt2021}
{Schuldt}, S., {Suyu}, S.~H., {Ca{\~n}ameras}, R., {et~al.} 2021, \aap, 651,
  A55

\bibitem[{{Scodeggio} {et~al.}(2018){Scodeggio}, {Guzzo}, {Garilli}, {Granett},
  {Bolzonella}, {de la Torre}, {Abbas}, {Adami}, {Arnouts}, {Bottini}, {Cappi},
  {Coupon}, {Cucciati}, {Davidzon}, {Franzetti}, {Fritz}, {Iovino}, {Krywult},
  {Le Brun}, {Le F{\`e}vre}, {Maccagni}, {Ma{\l}ek}, {Marchetti}, {Marulli},
  {Polletta}, {Pollo}, {Tasca}, {Tojeiro}, {Vergani}, {Zanichelli}, {Bel},
  {Branchini}, {De Lucia}, {Ilbert}, {McCracken}, {Moutard}, {Peacock},
  {Zamorani}, {Burden}, {Fumana}, {Jullo}, {Marinoni}, {Mellier}, {Moscardini},
  \& {Percival}}]{Scodeggio2018}
{Scodeggio}, M., {Guzzo}, L., {Garilli}, B., {et~al.} 2018, \aap, 609, A84

\bibitem[{{Shuntov} {et~al.}(2020){Shuntov}, {Pasquet}, {Arnouts}, {Ilbert},
  {Treyer}, {Bertin}, {de la Torre}, {Dubois}, {Fouchez}, {Kraljic}, {Laigle},
  {Pichon}, \& {Vibert}}]{Shuntov2020}
{Shuntov}, M., {Pasquet}, J., {Arnouts}, S., {et~al.} 2020, \aap, 636, A90

\bibitem[{{Skelton} {et~al.}(2014){Skelton}, {Whitaker}, {Momcheva}, {Brammer},
  {van Dokkum}, {Labb{\'e}}, {Franx}, {van der Wel}, {Bezanson}, {Da Cunha},
  {Fumagalli}, {F{\"o}rster Schreiber}, {Kriek}, {Leja}, {Lundgren}, {Magee},
  {Marchesini}, {Maseda}, {Nelson}, {Oesch}, {Pacifici}, {Patel}, {Price},
  {Rix}, {Tal}, {Wake}, \& {Wuyts}}]{Skelton2014}
{Skelton}, R.~E., {Whitaker}, K.~E., {Momcheva}, I.~G., {et~al.} 2014, \apjs,
  214, 24

\bibitem[{Soo {et~al.}(2021)Soo, Joachimi, Eriksen, Siudek, Alarcon, Cabayol,
  Carretero, Casas, Castander, Fernández, García-Bellido, Gaztanaga,
  Hildebrandt, Hoekstra, Miquel, Padilla, Sánchez, Serrano, \&
  Tallada-Crespí}]{Soo2021}
Soo, J. Y.~H., Joachimi, B., Eriksen, M., {et~al.} 2021, \mnras, 503, 4118

\bibitem[{Speagle \& Eisenstein(2017)}]{Speagle2017}
Speagle, J.~S. \& Eisenstein, D.~J. 2017, \mnras, 469, 1205

\bibitem[{Srivastava {et~al.}(2017)Srivastava, Valkov, Russell, Gutmann, \&
  Sutton}]{Srivastava2017}
Srivastava, A., Valkov, L., Russell, C., Gutmann, M.~U., \& Sutton, C. 2017, in
  Proceedings of the 31st International Conference on Neural Information
  Processing Systems, NIPS'17 (Red Hook, NY, USA: Curran Associates Inc.),
  3310–3320

\bibitem[{Szegedy {et~al.}(2015)Szegedy, Liu, Jia, Sermanet, Reed, Anguelov,
  Erhan, Vanhoucke, \& Rabinovich}]{Szegedy2015}
Szegedy, C., Liu, W., Jia, Y., {et~al.} 2015, in Proceedings of the IEEE
  Conference on Computer Vision and Pattern Recognition (CVPR)

\bibitem[{Thanh-Tung \& Tran(2020)}]{Thanh2020}
Thanh-Tung, H. \& Tran, T. 2020, in 2020 International Joint Conference on
  Neural Networks (IJCNN), IEEE, 1--10

\bibitem[{{Tong} {et~al.}(2019){Tong}, {Liu}, {Wang}, \&
  {Li}}]{Tong2019imbalance}
{Tong}, H., {Liu}, B., {Wang}, S., \& {Li}, Q. 2019, arXiv e-prints,
  arXiv:1901.08429

\bibitem[{Voigt {et~al.}(2014)Voigt, Fried, Backes, \& Rhode}]{Voigt2014}
Voigt, T., Fried, R., Backes, M., \& Rhode, W. 2014, Advances in Data Analysis
  and Classification, 8, 195–216

\bibitem[{Way \& Klose(2012)}]{Way2012}
Way, M.~J. \& Klose, C.~D. 2012, Publications of the Astronomical Society of
  the Pacific, 124, 274

\bibitem[{Wilson {et~al.}(2020)Wilson, Nayyeri, Cooray, \&
  Häu{\ss}ler}]{Wilson2020}
Wilson, D., Nayyeri, H., Cooray, A., \& Häu{\ss}ler, B. 2020, \apj, 888, 83

\bibitem[{{Wu} \& {Boada}(2019)}]{Wu2019}
{Wu}, J.~F. \& {Boada}, S. 2019, \mnras, 484, 4683

\bibitem[{Wu {et~al.}(2021)Wu, Liu, Huang, Wang, \& Lin}]{wu2021adversarial}
Wu, T., Liu, Z., Huang, Q., Wang, Y., \& Lin, D. 2021, in Proceedings of the
  IEEE/CVF Conference on Computer Vision and Pattern Recognition, 8659--8668

\bibitem[{Yan {et~al.}(2020)Yan, Mead, Van~Waerbeke, Hinshaw, \&
  McCarthy}]{Yan2020}
Yan, Z., Mead, A.~J., Van~Waerbeke, L., Hinshaw, G., \& McCarthy, I.~G. 2020,
  \mnras, 499, 3445–3458

\bibitem[{Yin {et~al.}(2019)Yin, Yu, Sohn, Liu, \& Chandraker}]{Yin2019CVPR}
Yin, X., Yu, X., Sohn, K., Liu, X., \& Chandraker, M. 2019, in Proceedings of
  the IEEE/CVF Conference on Computer Vision and Pattern Recognition (CVPR)

\bibitem[{{Zhang} {et~al.}(2018){Zhang}, {Zhang}, \&
  {Zhao}}]{Zhang2018imbalance}
{Zhang}, J., {Zhang}, Y., \& {Zhao}, Y. 2018, \aj, 155, 108

\bibitem[{Zhang {et~al.}(2018)Zhang, Li, \& Yu}]{Zhang2018mode}
Zhang, Z., Li, M., \& Yu, J. 2018, in SIGGRAPH Asia 2018 Technical Briefs, SA
  '18 (New York, NY, USA: Association for Computing Machinery)

\end{thebibliography}

\begin{appendix}

\section{Formalism for photometric redshift estimation} \label{sec:formalism}

This appendix presents a statistical basis for the bias correction methods discussed in our work. Photometric redshift estimation with neural networks can be viewed as a probability density estimation problem. In general, the density of $z_{spec}$ given the data $D$ is mapped to the density of $z_{photo}$ via the following formula
\begin{equation}
p' (z_{photo} | D, \Phi) = \int {q (z_{photo} | z_{spec}, D, \Phi) \, p (z_{spec} | D) \, dz_{spec}}
\label{eq:p2p}
\end{equation}
where $q (z_{photo} | z_{spec}, D, \Phi)$ encodes the potential prior information imposed by the data $D$ and the model implementation $\Phi$ \citep[see also][]{Malz2020, Malz_Alex2021}. Ideally, we expect $p' (z_{photo} | D, \Phi)$ to be close to $p(z_{spec} | D)$. However, Eq.~\ref{eq:p2p} is not an identity mapping between the two densities since $q(z_{photo} | z_{spec}, D, \Phi)$ is different from $\delta(z_{photo} - z_{spec})$; $p' (z_{photo} | D, \Phi)$ will suffer from biases induced by $q (z_{photo} | z_{spec}, D, \Phi)$.

As it is non-trivial to marginalize over the prior from the data $D$ and the model implementation $\Phi$, we introduce an auxiliary term $\tilde{q} (z'_{photo} | z_{photo}, D, \Phi)$ to ``re-estimate'' the density of $z_{photo}$ that is supposed to neutralize the influence from $q(z_{photo} | z_{spec}, D, \Phi)$.
\begin{equation}
p'' (z'_{photo} | D, \Phi) = \int {\tilde{q} (z'_{photo} | z_{photo}, D, \Phi) \, p' (z_{photo} | D, \Phi) \, dz_{photo}}
\label{eq:p2p_r}
\end{equation}
The re-estimated density $p'' (z'_{photo} | D, \Phi)$ is expected to resemble the actual density $p(z_{spec} | D)$. In other words, the exploitation of bias correction techniques can be represented by the introduction of the term $\tilde{q} (z'_{photo} | z_{photo}, D, \Phi)$. Following this idea, we exploit a multi-step procedure in our work in which bias correction methods are applied based on a pre-estimation of the $z_{photo}$ density, i.e., $p' (z_{photo} | D, \Phi)$ (or multiple pre-estimations if necessary).

\FloatBarrier

\section{Comparison of point estimates} \label{sec:zcomparison}

In our work, probability density distributions of photometric redshifts are predicted by the classification model. Such density estimates are not full posterior distributions since the prior imposed by the training data and the model implementation has not been marginalized over. To alleviate the influence from the prior, we use the point estimates derived from the density estimates in our analysis. 

\begin{figure*}
\begin{center}
\centerline{\includegraphics[width=1.0\linewidth]{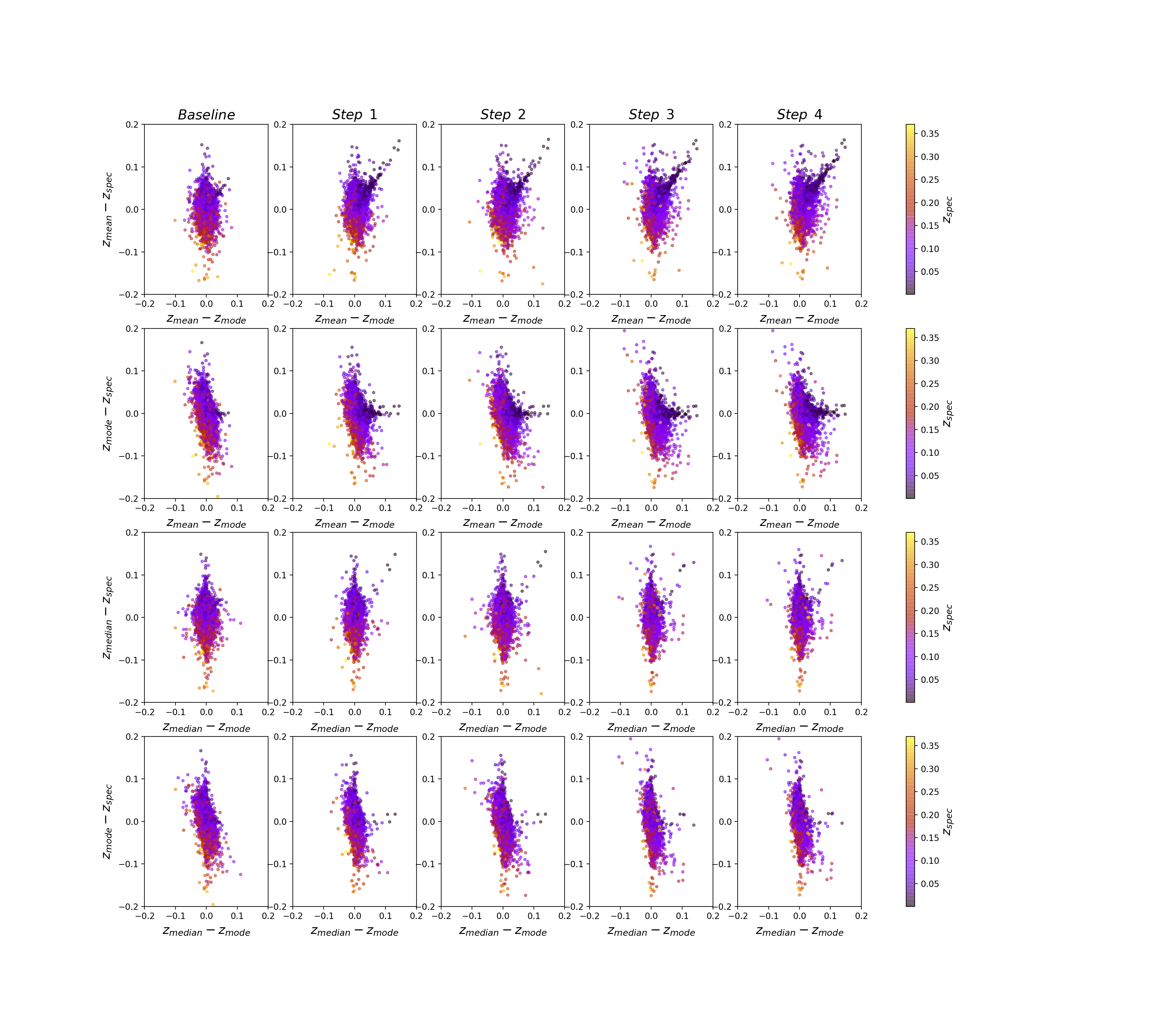}}
\caption{Comparison among three choices of point estimates --- $z_{mode}$, $z_{mean}$ and $z_{median}$ --- for the Baseline method and Steps 1 -- 4 of our methods using the SDSS data and the network \texttt{Net\_P}. All data points are color-coded with $z_{spec}$. As the three point estimates for each data instance are shifted by the same amount in Step 4, the differences between any two point estimates remain unchanged as in Step 3.}
\label{fig:z_comparison}
\end{center}
\end{figure*}

We consider three choices of point estimates: \textbf{(a) $z_{mode}$} --- the redshift at the bin with the maximum probability. \textbf{(b) $z_{mean}$} --- the probability-weighted mean redshift over all bins. \textbf{(c) $z_{median}$} --- the median redshift drawn from the probability distribution over all bins. Different choices of point estimates would result in dissimilar biases. In Fig.~\ref{fig:z_comparison}, we find that $z_{mean} - z_{spec}$ (or $z_{median} - z_{spec}$) evolves with $z_{spec}$. The measures of $z_{mean}$ (or $z_{median}$) are on average over-estimated for the low-$z$ instances but under-estimated for the high-$z$ instances. In particular, $z_{mean} - z_{spec}$ (or $z_{median} - z_{spec}$) is correlated with $z_{mean} - z_{mode}$ (or $z_{median} - z_{mode}$) for the instances with the lowest redshifts, which is due to the ``window function'' imposed by the finite redshift range of the density estimates set by the model, since it makes $z_{mean}$ and $z_{median}$ more concentrated on the central value of the redshift range. This correlation is more pronounced for our methods as the predicted density estimates are widened compared to the Baseline method\footnote{We note that Step 3 (and subsequently, Step 4) is performed based on pre-estimated $z_{mode}$, but the overall behaviors would be similar for $z_{mean}$-based or $z_{median}$-based bias correction.}. Similar trends can also be found with the CFHTLS datasets. Therefore, we suggest that $z_{mean}$ and $z_{median}$ are not statistically sound choices of point estimates. In comparison, the influence of such window function is weaker for $z_{mode}$ (despite with stronger mode collapse), since $z_{mode}$ is determined by the bin with the peak probability rather than the global density distribution. We thus take $z_{mode}$ as the point estimates for our methods and the Baseline method throughout this work.

\FloatBarrier

\section{Modeling the impact of under-populated classes} \label{sec:modeling_underp}

The outcome from Step 2 of our methods suffers from $z_{spec}$-dependent residuals caused by under-populated bins. As discussed in Section~\ref{sec:step3}, the effect of under-populated bins can be characterized by a separation between the actual spectroscopic redshift and the mean of the skewed scattering profile for each instance, which is mutually determined by a initial Gaussian scattering profile and the $z_{spec}$ number density of the subsample under a linearity assumption. We attempt to test this assumption by comparing the actual $<\Delta z> - \,\, z_{spec}$ curves using the subsets of training data from Step 2 against the expected residual curves derived based on skewed scattering profiles. The expected mean residual in each $z_{spec}$ bin is determined as
\begin{equation}
\Delta \overline{z} = \frac{\overline{z} - z_0}{1 + z_0}
\label{eq:res_zmean_sim}
\end{equation}
where $z_0$ is the initial $z_{spec}$ value; $\overline{z}$ is the mean of the skewed scattering profile as in Eq.~\ref{eq:zmean}, regarded as the expected $z_{photo}$ value. We follow the same procedure explained in Section~\ref{sec:step3} to estimate the quadratic error curves $\sigma_2^2(z_{spec}|r_o)$ and the scattering profiles, except that the variances are replaced with $\sigma_2^2(z_{spec}|r_o)$. We also rescale $\sigma_2^2(z_{spec}|r_o)$ by a set of factors $k$ in order to vary the shapes of the skewed scattering profiles and evaluate the effect on the expected residual curves. 

\begin{figure}[ht]
\begin{center}
\centerline{\includegraphics[width=1.2\columnwidth]{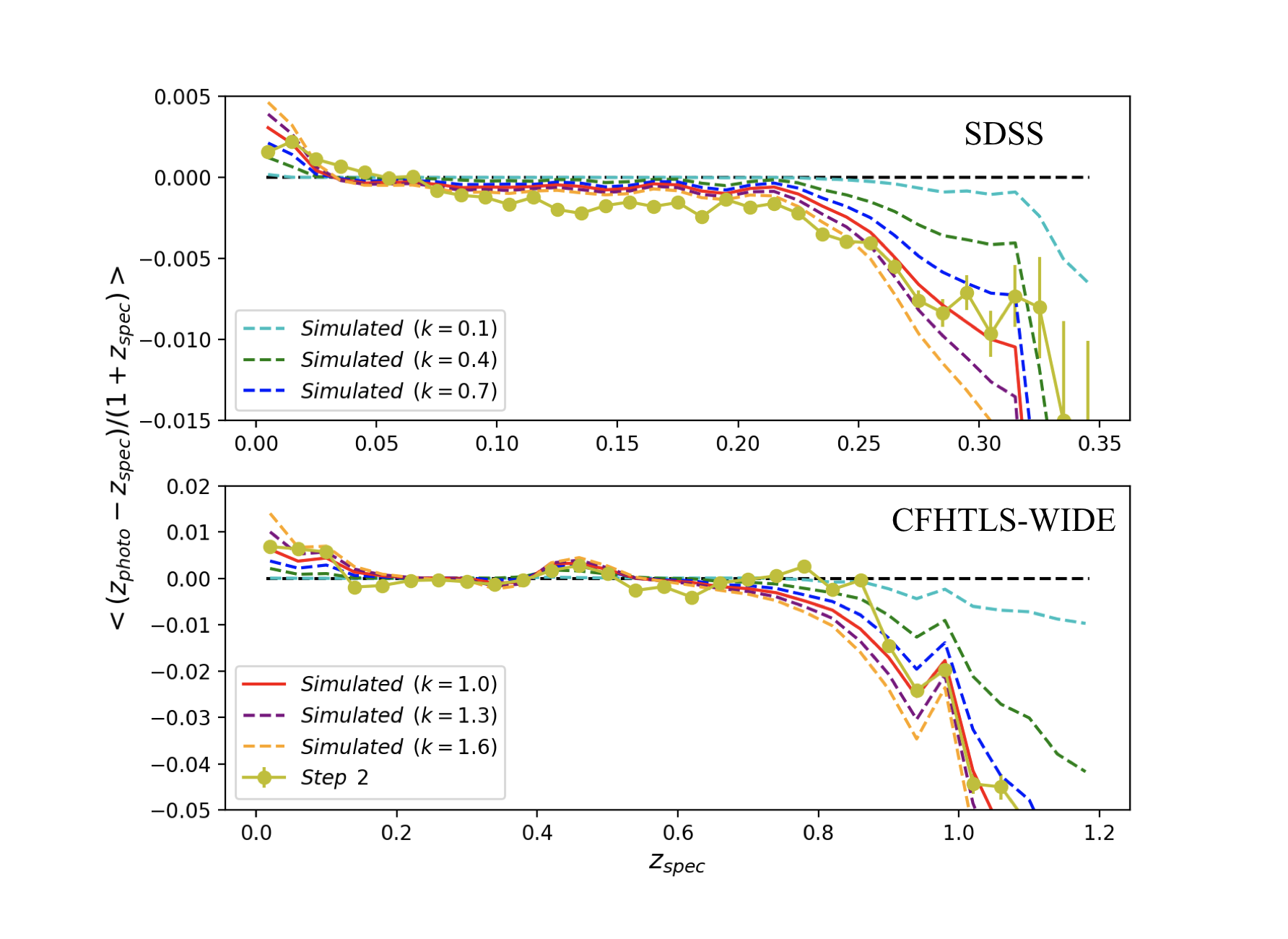}}
\caption{Comparison between the actual and the expected $<\Delta z> - \,\, z_{spec}$ curves for the SDSS data and the CFHTLS-WIDE data, respectively. The actual curves are derived with the subsets of training data from Step 2 (represented by the yellow lines). The expected curves are determined by the means of the scattering profiles as in Eq.~\ref{eq:zmean} yet the variances are replaced with the estimated quadratic error curves and rescaled by a set of different factors $k$ (represented by the dotted lines).}
\label{fig:impact_under_pop}
\end{center}
\end{figure}

Fig.~\ref{fig:impact_under_pop} shows the results with the SDSS data and the CFHTLS-WIDE data. With varying rescaling factors, the amplitudes of the simulated residuals stretch over a wide range, especially for highly under-populated regions that are typically subject to large errors and significant imbalances. This indicates that a proper modeling of skewed scattering profiles is most concerned for bias correction over these regions. The overall rough agreement between the expected curves without rescaling (i.e., $k=1$) and the actual curves strengthens the validity of the assumption of scattering profiles and the procedure for determining the means of the soft labels in Step 3. The assumption may not hold if there exist predominant peaks in number density (Appendix~\ref{sec:impact_th}), but such violation is circumvented in our case as the over-populated bins have been trimmed off in Step 2.

\section{Detailed discussions on bias behaviors} \label{sec:discussions}

In this appendix, we present the details about our investigation on bias behaviors discussed in Section~\ref{sec:discussions_short}. Without losing generality, we conduct experiments only on the SDSS data and the network \texttt{Net\_P} unless otherwise noted. The results are shown with the test data.

\subsection{Impact of the balancing threshold} \label{sec:impact_th}

\begin{figure}[ht]
\begin{center}
\centerline{\includegraphics[width=1.1\columnwidth]{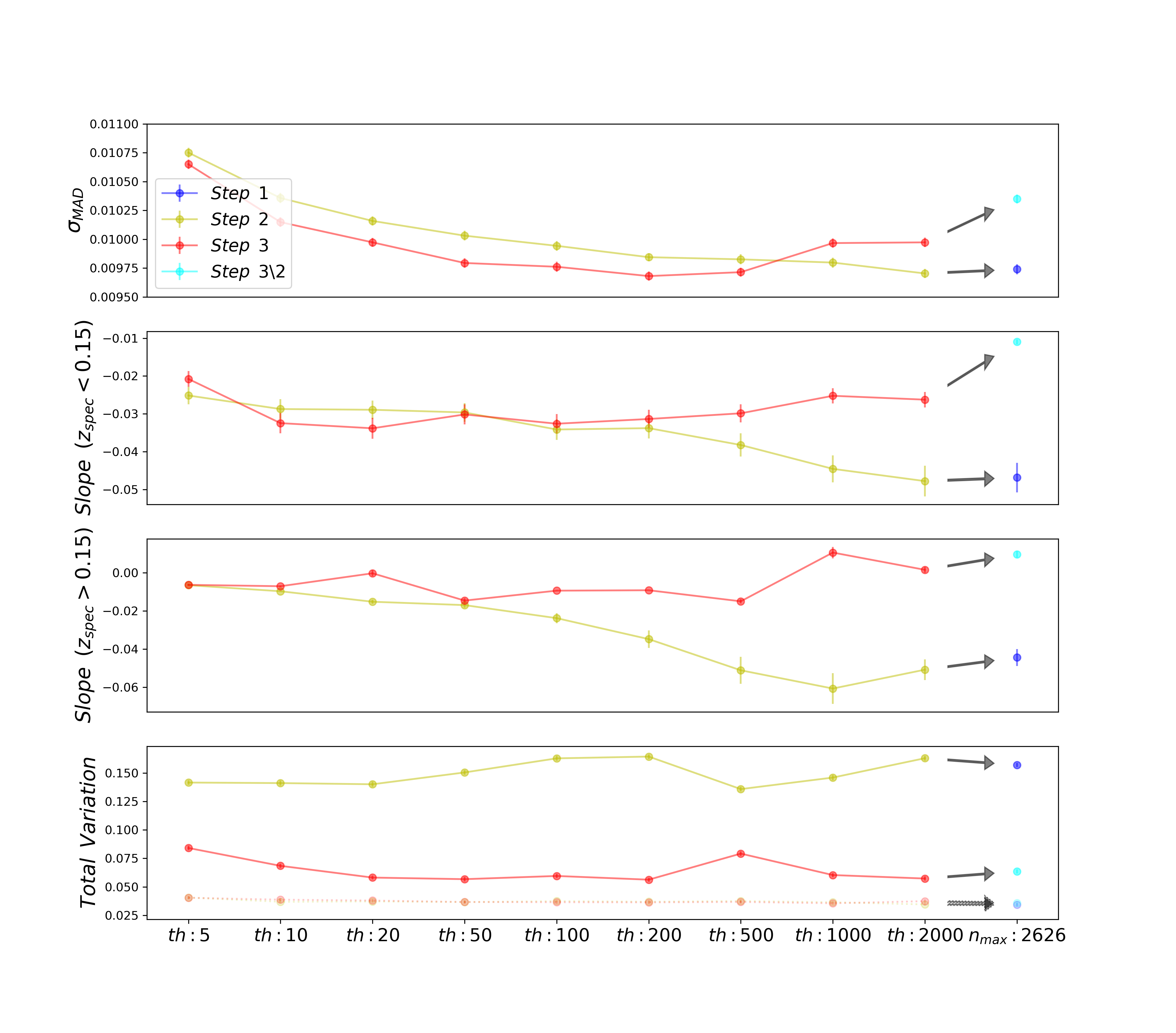}}
\caption{$\sigma_{\mathrm{MAD}}$ (Eq.~\ref{eq:sigma_mad}), the slopes of the $<\Delta z> - \,\, z_{spec}$ piecewise linear fit (Eq.~\ref{eq:resfit}) and the total variation distance between the $z_{photo}$ and $z_{spec}$ sample distributions (Eq.~\ref{eq:d_tv}) as a function of the balancing threshold that defines the near-balanced training subset for Steps 2 and 3 of our methods. As the threshold 
approaches the maximum number density (2,626 in this case), the results from Step 2 are expected to converge to those from Step 1, and those from Step 3 are expected to converge to those from Step $3\backslash2$ in which Step 2 is skipped. The dotted curves show the total variation distances for the simulated $z_{photo}$ sample distributions that are expected to have no mode collapse. Comparison is made among the methods using the SDSS data and the network \texttt{Net\_P}.}
\label{fig:impact_th_SDSS}
\end{center}
\end{figure}

\begin{figure}[ht]
\begin{center}
\centerline{\includegraphics[width=1.0\columnwidth]{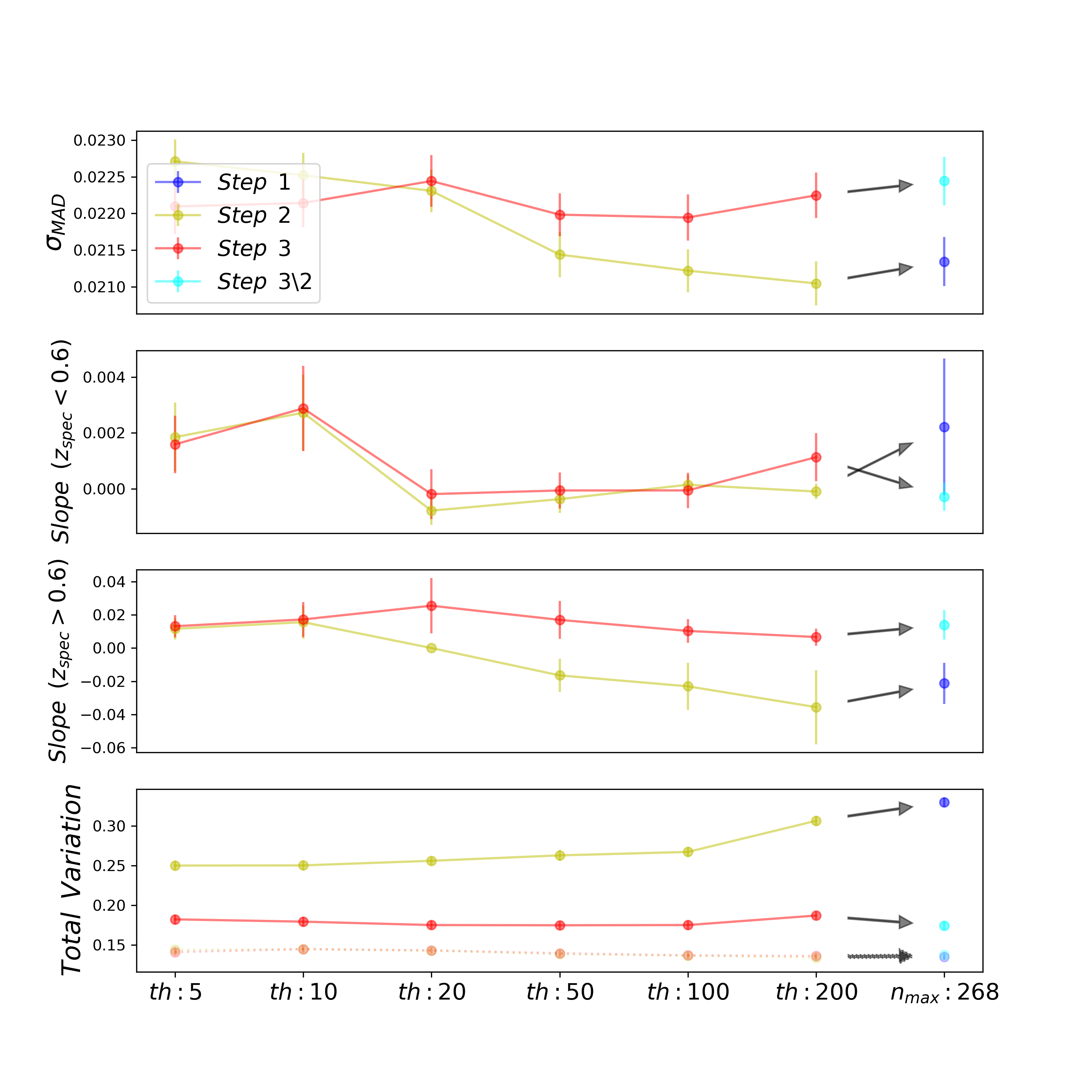}}
\caption{Same as Fig.~\ref{fig:impact_th_SDSS}, except with the CFHTLS-WIDE data.}
\label{fig:impact_th_CFHTW}
\end{center}
\end{figure}

The balancing threshold determines the ``flatness'' of the near-balanced training subset used for suppressing the impact of over-populated bins in Steps 2 and 3, and is one of the key parameters in our methods. A low threshold may lead to insufficient statistics and cause strong overfitting, whereas a high threshold may be ineffective for correcting $z_{spec}$-dependent residuals induced by over-populated classes. In Figs.~\ref{fig:impact_th_SDSS} and \ref{fig:impact_th_CFHTW}, we check the performance of our methods with a set of different thresholds for the SDSS data and the CFHTLS-WIDE data using the network \texttt{Net\_P}. We show the results from Steps 2 and 3, as well as those from Step 1 and Step $3\backslash2$ (with Step 2 skipped), which are expected to be the limits of Steps 2 and 3 respectively as the threshold approaches the maximum number density. The estimation accuracy is indicated by $\sigma_{\mathrm{MAD}}$ (Eq.~\ref{eq:sigma_mad}). The dependence of mean residuals on $z_{spec}$ is indicated by the slopes in the two redshift intervals from the piecewise linear fit (Eq.~\ref{eq:resfit}). The intensity of mode collapse is quantified by the total variation distance (Eq.~\ref{eq:d_tv}). The dotted curves show the total variation distances for the simulated $z_{photo}$ sample distributions which are expected to be free from mode collapse. Ideally, after bias correction, the slopes are expected to be zero, and the total variation distances are expected to be close to the simulated values.

As the threshold evolves, there is a major trade-off between $\sigma_{\mathrm{MAD}}$ and the slope in the high-redshift interval for Step 2. Since the construction of the training subset reduces statistics and causes an increase in estimation errors, one would circumvent this issue by simply lifting up the balancing threshold and regarding all bins except the one with the peak density as ``under-populated'' bins. As discussed in Section~\ref{sec:step3}, the influence from under-populated bins can be modeled by assuming a Gaussian scattering profile linearly modified by the local number density. This assumption might be violated if the peak number density is significantly high, as suggested by the growth of $\sigma_{\mathrm{MAD}}$ and the overly corrected slopes with the SDSS data for Step 3 when the threshold is above 200. Therefore, it is necessary to trim off the peak densities before modeling the scattering profiles. Based on our observations from the figures, we take 200 and 100 as the thresholds in the main experiments for the SDSS data and the CFHTLS-WIDE data, respectively. Each of them places a trade-off between reducing the over-populated bins and avoiding overfitting.

\subsection{Impact of the bin size} \label{sec:impact_binsize}

\begin{figure}[ht]
\begin{center}
\centerline{\includegraphics[width=\columnwidth]{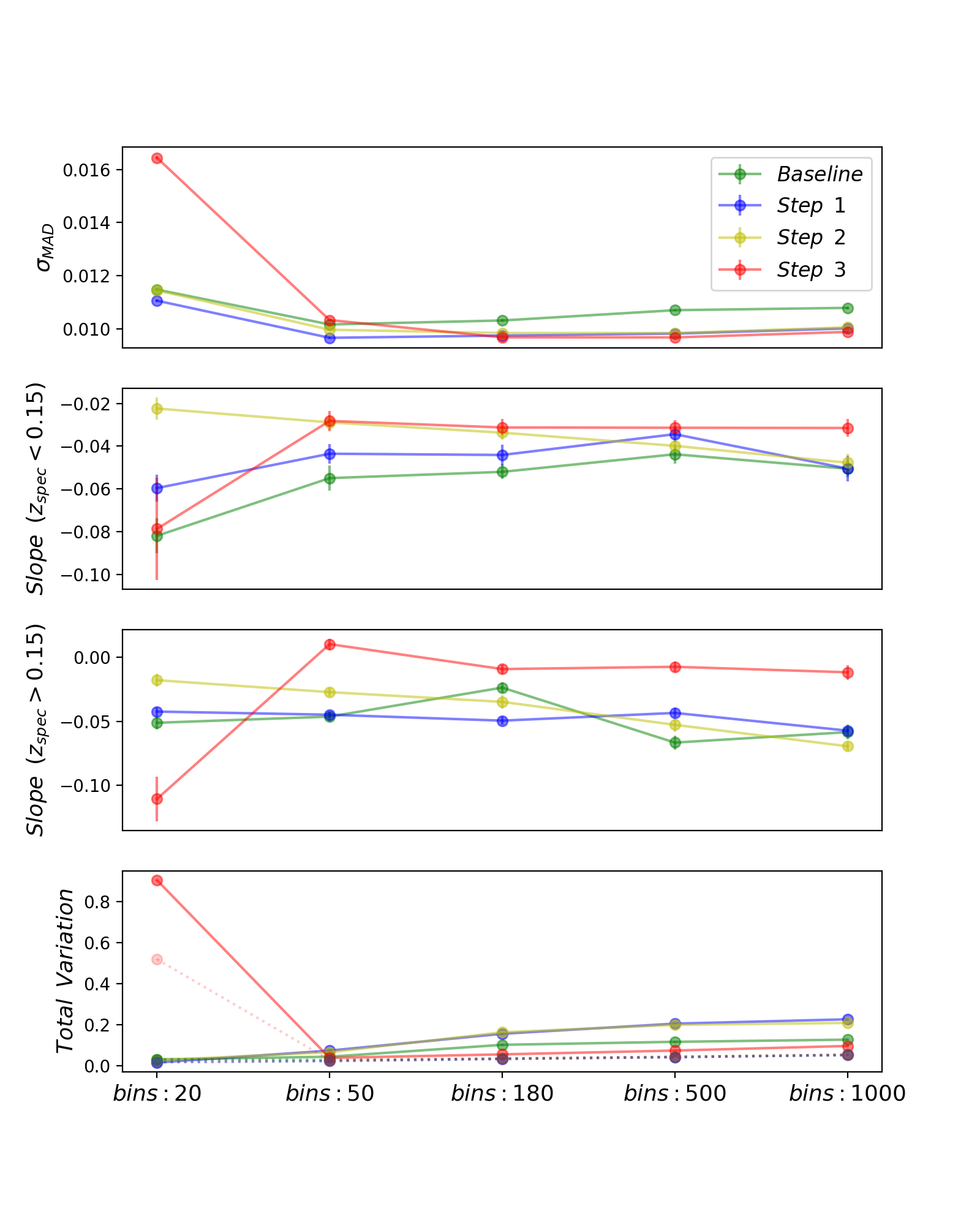}}
\caption{$\sigma_{\mathrm{MAD}}$ (Eq.~\ref{eq:sigma_mad}), the slopes of the $<\Delta z> - \,\, z_{spec}$ piecewise linear fit (Eq.~\ref{eq:resfit}) and the total variation distance between the $z_{photo}$ and $z_{spec}$ sample distributions (Eq.~\ref{eq:d_tv}) as a function of the number of bins in the output redshift density estimate (before being extended in Step 3). The dotted curves show the total variation distances for the simulated $z_{photo}$ sample distributions that are expected to have no mode collapse. The Baseline method and Steps 1 -- 3 of our methods are compared using the SDSS data and the network \texttt{Net\_P}.}
\label{fig:impact_bin}
\end{center}
\end{figure}

A redshift bin defines a small interval within which continuous redshifts are collapsed to a single value. As the predicted redshift point estimates are defined in bins, they can only take discrete values, which would introduce a quantization error. In Fig.~\ref{fig:impact_bin}, we present the result of our experiment on implementing different bin sizes using the SDSS data and the network \texttt{Net\_P}. The bin size is controlled by the number of bins in a redshift output (before being extended in Step 3). For example, a discretization of 20 bins yields a bin size of $0.4/20=0.02$. In Step 3, $n/2$ additional bins are appended to each side of the initial redshift output (where $n$ is the initial number of bins), except that 20 bins are appended to each side in the case of 20 initial bins.

The case with the smallest number of bins (i.e., 20) has the highest $\sigma_{\mathrm{MAD}}$, suggesting a large quantization error. The correction of biases for this case also seems problematic. Meanwhile, $\sigma_{\mathrm{MAD}}$ for the Baseline method appears to increase as the number of bins grows, probably due to a tendency of overfitting. A large number of bins would also make the data sparse, resulting in small-scale spikes in the $z_{photo}$ sample distribution and enlarging the total variation distance. In addition, the slopes of the $<\Delta z> - \,\, z_{spec}$ curves are not strongly affected by the bin size. Same as the choice by \citet{Pasquet2019}, We take 180 bins in the main experiments for the SDSS sample, as this intermediate number is a compromise between avoiding the sparsity of data and limiting the quantization error.

\subsection{Impact of the number of iterations} \label{sec:impact_ite}

The performance of a model is influenced by the number of iterations conducted in training, as it (together with the mini-batch size) determines the average number of times each instance is fed into the model and how well the model fits the training sample. In general, a model trained from scratch would gradually depart from a state of underfitting as the number of training iterations increases, and there might be a trend of overfitting if the volume of information in the training sample does not match the complexity of the model. Using the SDSS data and the network \texttt{Net\_P}, we explore this evolution 
trend by continuously training the network and saving the trained parameters at a set of different iterations for the Baseline method and Step 1. Yet, we do not change the numbers of iterations for Steps 2 and 3.

\begin{figure}[ht]
\begin{center}
\centerline{\includegraphics[width=1.2\columnwidth]{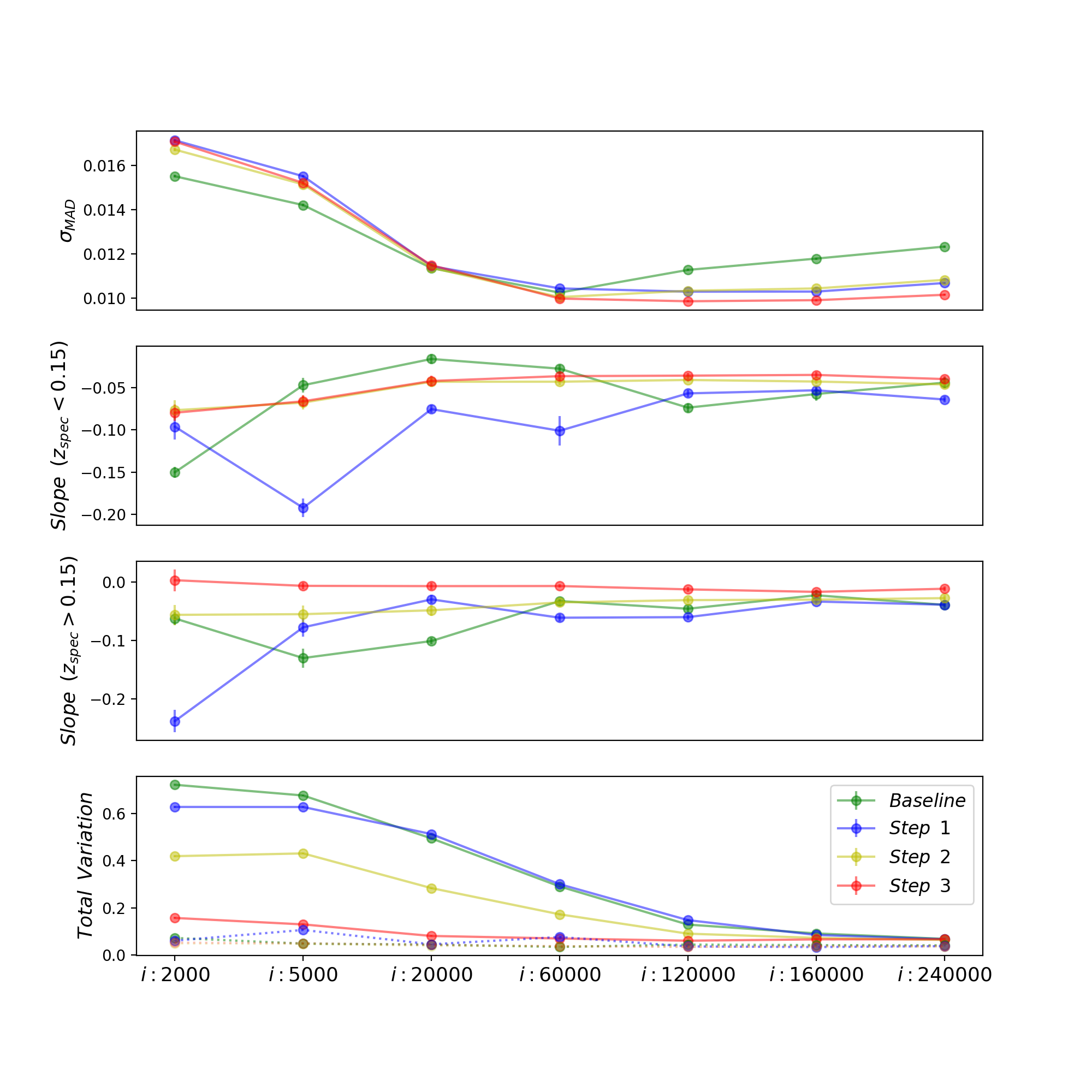}}
\caption{$\sigma_{\mathrm{MAD}}$ (Eq.~\ref{eq:sigma_mad}), the slopes of the $<\Delta z> - \,\, z_{spec}$ piecewise linear fit (Eq.~\ref{eq:resfit}) and the total variation distance between the $z_{photo}$ and $z_{spec}$ sample distributions (Eq.~\ref{eq:d_tv}) as a function of the number of training iterations conducted for the Baseline method and Step 1 of our methods. The numbers of iterations for Steps 2 and 3 remain the same as in the main experiments. The dotted curves show the total variation distances for the simulated $z_{photo}$ sample distributions that are expected to have no mode collapse. Comparison is made among the methods using the SDSS data and the network \texttt{Net\_P}.}
\label{fig:impact_ite}
\end{center}
\end{figure}

As shown in Fig.~\ref{fig:impact_ite}, the estimation accuracy (indicated by $\sigma_{\mathrm{MAD}}$) for the Baseline method improves first but declines after a turning point around 60,000 iterations, implying a shift from underfitting to overfitting. The turning point suggests the number of iterations with which the model reaches a proper fit. The decrease in the estimation accuracy is less significant for Step 1, probably due to the regularizing effect imposed by the multi-channel output unit.

\begin{figure*}
\begin{center}
\centerline{\includegraphics[width=1.0\linewidth]{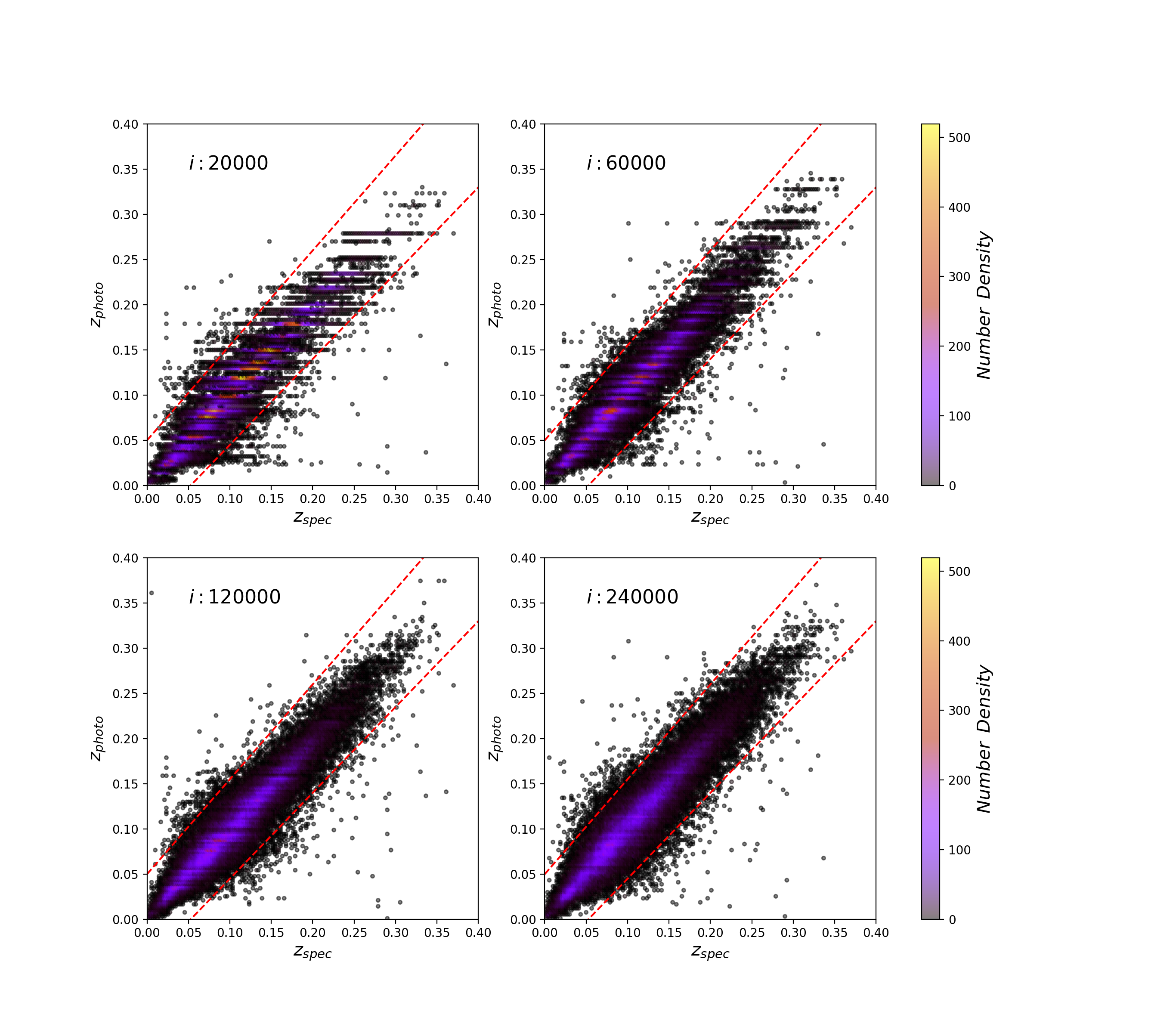}}
\caption{Predicted photometric redshifts as a function of spectroscopic redshifts color-coded with number density for the Baseline method. Comparison is made among cases with different numbers of iterations in training using the SDSS data and and the network \texttt{Net\_P}. The dashed red lines indicate the boundaries $(z_{photo} - z_{spec}) / (1 + z_{spec}) = \pm 0.05 $ for outliers defined for the SDSS data as in \citet{Pasquet2019} and \Treyer{}.}
\label{fig:photoz_vs_specz_ite}
\end{center}
\end{figure*}

Interestingly, the total variation distance between the $z_{photo}$ and $z_{spec}$ sample distributions decreases monotonically as the training proceeds, suggesting that mode collapse is heavily associated with underfitting while overfitting has a positive effect on reducing mode collapse. This trend can also be seen in Fig.~\ref{fig:photoz_vs_specz_ite} for the Baseline method. With a low number of iterations (e.g., 20,000), only a coarse input-output mapping can be established due to underfitting, thus the data points are grouped in big clusters shown as the conspicuous concentrated lines. As the training continues, the grouping of points becomes finer and the plateau on the top gradually shifts upwards. Eventually, with an excessively large number of iterations (e.g., 240,000), the redshift estimates predicted by the overfitted model tend to have large \textit{random} scatters, which smears out the local concentrations caused by mode collapse.

\begin{figure}[ht]
\begin{center}
\centerline{\includegraphics[width=\columnwidth]{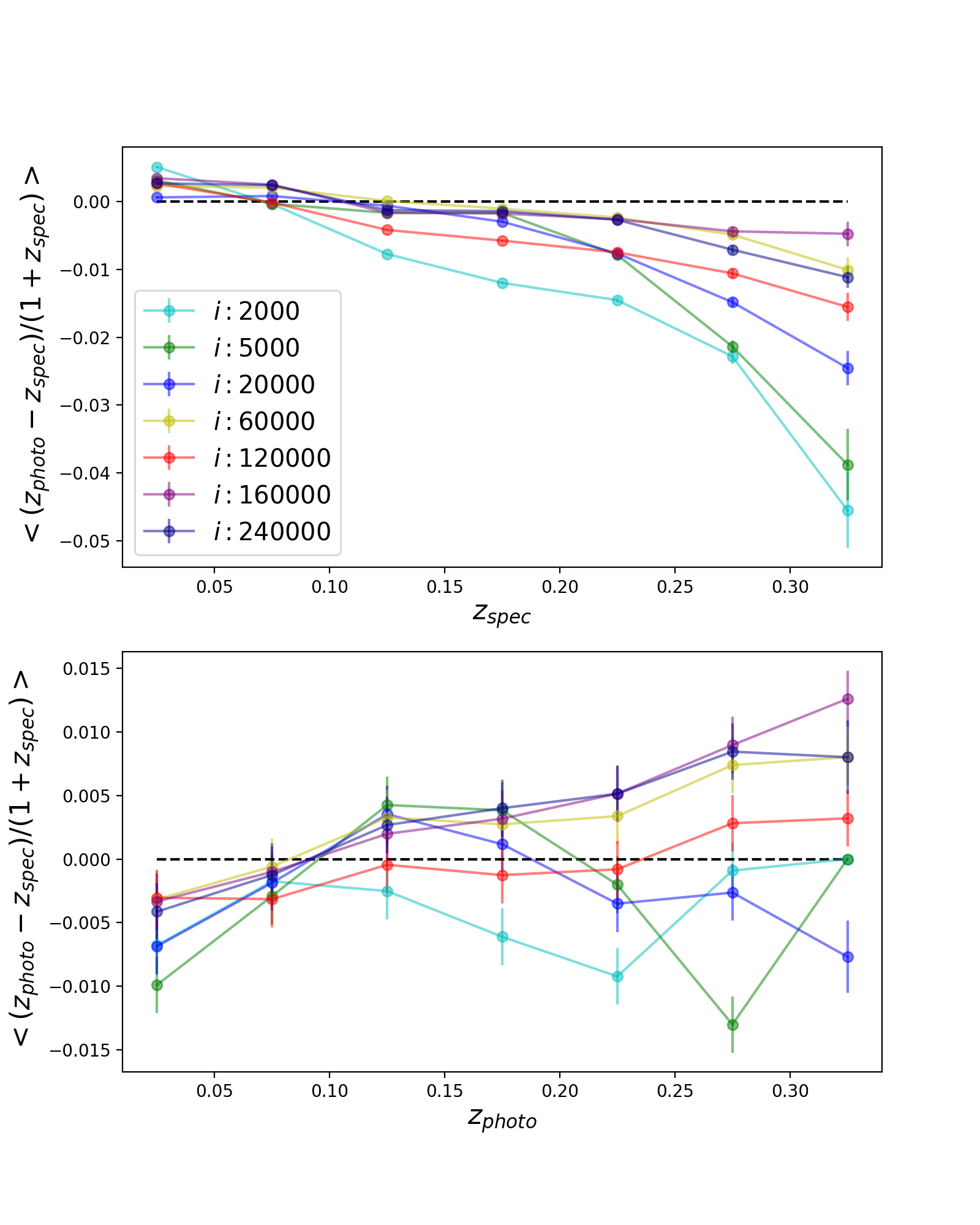}}
\caption{Mean residuals estimated in coarse bins as a function of spectroscopic redshifts or estimated photometric redshifts for the Baseline method. Comparison is made among cases with different numbers of iterations in training using the SDSS data and and the network \texttt{Net\_P}. The horizontal dashed black lines indicate the zero residual.}
\label{fig:residual_vs_z_ite}
\end{center}
\end{figure}

We also examine the influence on mean residuals estimated in spectroscopic redshift bins or photometric redshift bins for the Baseline method. As presented in Fig.~\ref{fig:residual_vs_z_ite}, the $<\Delta z> - \,\, z_{spec}$ curve tends to be steeper with stronger underfitting, which results from the presence of over-populated classes and the hard redshift boundaries set by the model that make the $z_{photo}$ estimates from each $z_{spec}$ bin skewed towards the location of the peak number density. Due to the same reason, the mean residuals for the high-$z_{photo}$ bins are below zero, meaning that each $z_{photo}$ bin is contributed by inward skewness more significantly than outward scattering. However, outward scattering dominates over inward skewness for the low-$z_{photo}$ bins, thus the mean residuals below $z_{photo} \sim 0.1$ are also negative.

When the training passes over the point of proper fit and starts to overfit, the slope of the $<\Delta z> - \,\, z_{spec}$ curve would become more negative again as a result of increasing estimation errors. Meanwhile, the high-$z_{photo}$ tail of the $<\Delta z> - \,\, z_{photo}$ curve shifts from negative to positive, implying that outward scattering is becoming more predominant due to larger random errors. Based on these observations, we suggest that the zero dependence on $z_{photo}$ (though with a drop at the high-$z$ tail) achieved by \citet{Pasquet2019} would not be ensured unless a precise control of training is carried out.

Additionally, in spite of the sub-optimal (mainly underfitted) representations established with different numbers of iterations, the validity of our bias correction methods is verified in all these cases. We use a number of 60,000 iterations for representation learning in the main experiments, though mild underfitting or overfitting would not produce any major impact. We will continue discussing the effect of overfitting in Appendix~\ref{sec:impact_samplesize}.

\subsection{Impact of the sample size} \label{sec:impact_samplesize}

The size of the training sample is one of the most important factors influencing the performance of a model. Predominantly, a complex model would easily overfit on a small training sample. As a complement to the experiment on varying the number of training iterations (discussed in Appendix~\ref{sec:impact_ite}), we investigate the effect of overfitting by controlling the size of the training sample for the Baseline method and the representation learning in Step 1, using the SDSS data and the network \texttt{Net\_P}. Yet still, we use the original training sample to construct a near-balanced subset for Steps 2 and 3.

\begin{figure}[ht]
\begin{center}
\centerline{\includegraphics[width=\columnwidth]{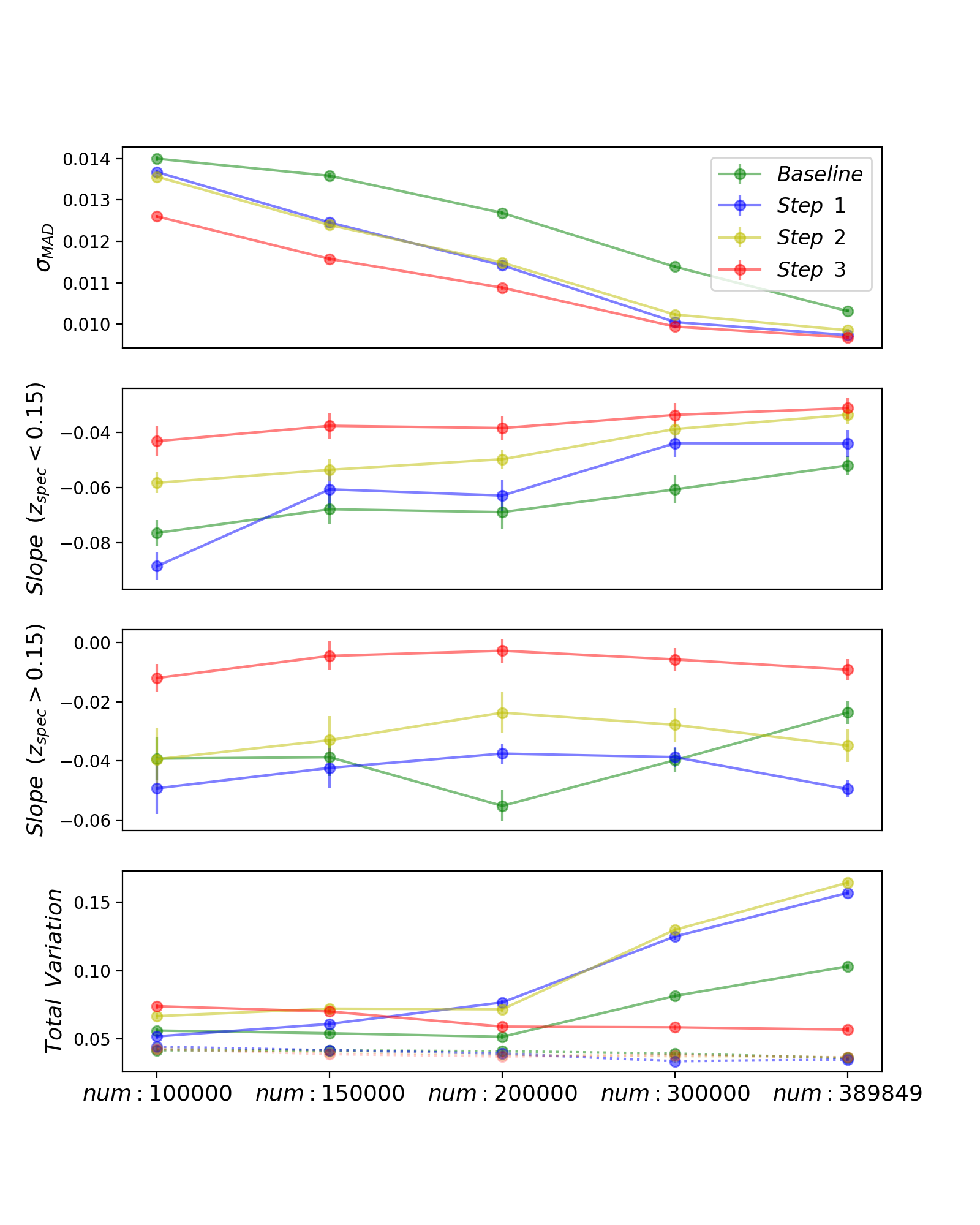}}
\caption{$\sigma_{\mathrm{MAD}}$ (Eq.~\ref{eq:sigma_mad}), the slopes of the $<\Delta z> - \,\, z_{spec}$ piecewise linear fit (Eq.~\ref{eq:resfit}) and the total variation distance between the $z_{photo}$ and $z_{spec}$ sample distributions (Eq.~\ref{eq:d_tv}) as a function of the total number of instances in the training sample for the Baseline method and Step 1 of our methods. The near-balanced training subset in Steps 2 and 3 is constructed using the original training set without reducing the sample size. The dotted curves show the total variation distances for the simulated $z_{photo}$ sample distributions that are expected to have no mode collapse. Comparison is made among the methods using the SDSS data and the network \texttt{Net\_P}.}
\label{fig:impact_datasize}
\end{center}
\end{figure}

As illustrated in Fig.~\ref{fig:impact_datasize}, the estimation accuracy (indicated by $\sigma_{\mathrm{MAD}}$) gradually drops as a result of increasing overfitting when the sample size decreases, in general leading to larger residuals suggested by the steeper $<\Delta z> - \,\, z_{spec}$ curves, but weaker mode collapse suggested by the lower total variation distances between the $z_{photo}$ and $z_{spec}$ sample distributions, which is similar to the consequence of conducting more training iterations. On the other hand, the performance on bias correction appears robust over different degrees of overfitting. Presumably, the training subset contains previously unseen instances that help re-identify salient information from the biased representation and reduce overfitting in the classification phase. We therefore claim that an overfitted representation may still be applicable for bias correction as long as such overfitting can be neutralized with the classification module.

\subsection{Impact of labeling errors} \label{sec:impact_label}

\begin{figure}[ht]
\begin{center}
\centerline{\includegraphics[width=\columnwidth]{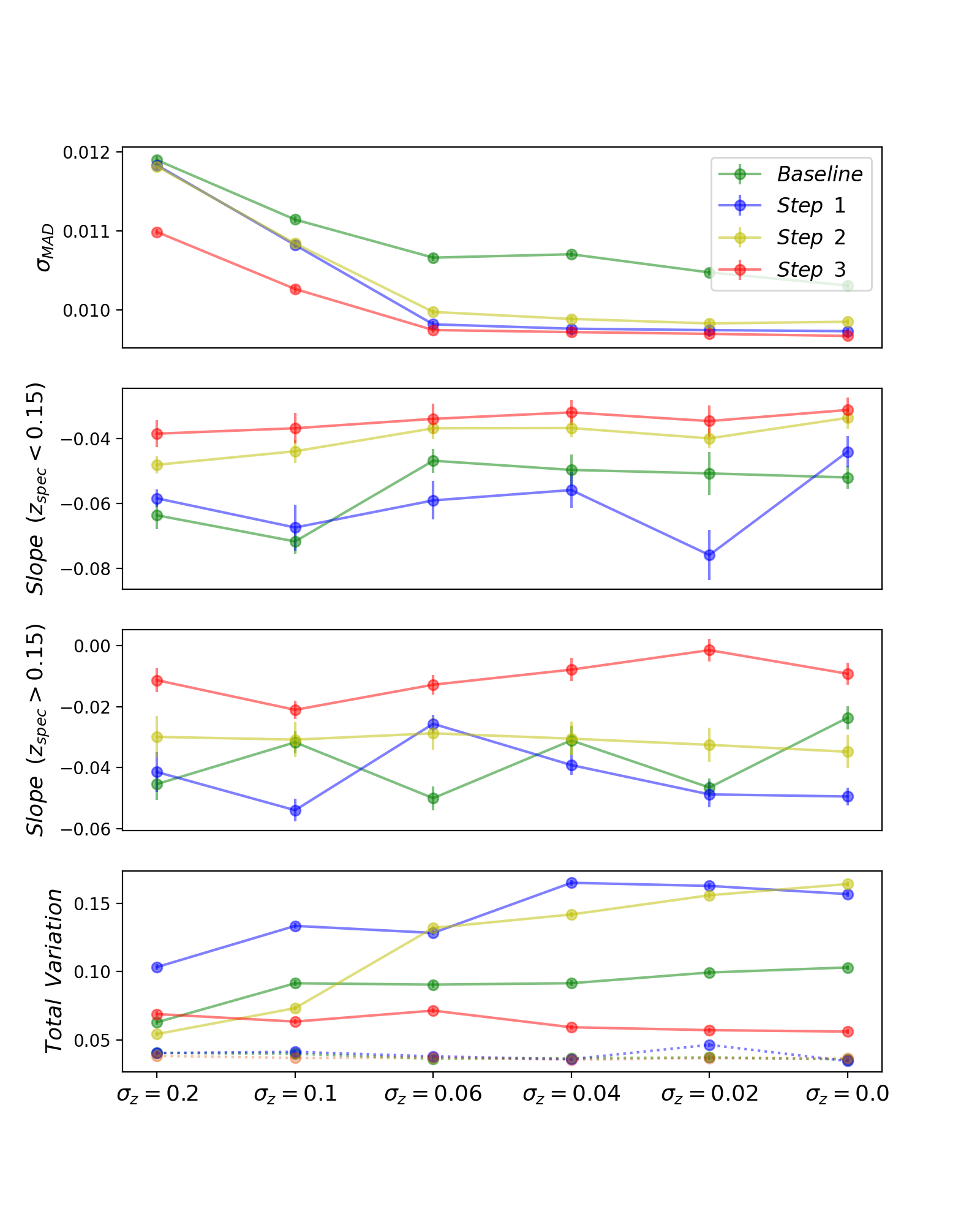}}
\caption{$\sigma_{\mathrm{MAD}}$ (Eq.~\ref{eq:sigma_mad}), the slopes of the $<\Delta z> - \,\, z_{spec}$ piecewise linear fit (Eq.~\ref{eq:resfit}) and the total variation distance between the $z_{photo}$ and $z_{spec}$ sample distributions (Eq.~\ref{eq:d_tv}) as a function of the dispersion of Gaussian errors $\sigma_z$ introduced to the redshift labels for the training data used for the Baseline method and Step 1 of our methods. The Gaussian errors are not applied to the training data in Steps 2 and 3, or the test data. The dotted curves show the total variation distances for the simulated $z_{photo}$ sample distributions that are expected to have no mode collapse. Comparison is made among the methods using the SDSS data and the network \texttt{Net\_P}.}
\label{fig:impact_errlabel}
\end{center}
\end{figure}

In our main experiments, the labels used in training are assumed to contain no errors for the SDSS data and the high-quality CFHTLS data. Using the SDSS data and the network \texttt{Net\_P}, we check whether Gaussian labeling errors would have an effect on the behaviors of the biases and the performance of our methods. We randomly draw Gaussian errors with a series of given dispersions ($\sigma_z$'s) and add to the initial spectroscopic redshifts for the training data (but not the test data). The erroneous spectroscopic redshifts are fixed and those within the redshift range [0, 0.4] are converted to labels used for the Baseline method and the representation learning in Step 1. As shown in Fig.~\ref{fig:impact_errlabel}, introducing labeling errors leads to an increase in estimation errors indicated by $\sigma_{\mathrm{MAD}}$ and may have an impact on the slopes of the $<\Delta z> - \,\, z_{spec}$ curves. Similar to the effect of overfitting, the increasing labeling errors would yield more random incorrect redshift estimates, which helps suppress mode collapse and results in decreasing total variation distances between the $z_{photo}$ and $z_{spec}$ sample distributions.

Since high-quality spectroscopic redshifts are required for bias correction, we then use the initial errorless labels for training in Steps 2 and 3. The outcomes of bias correction remain similar in all the cases even for the one with a substantially large dispersion $\sigma_z=0.2$, given that high-quality labels are provided for bias correction. We thus claim that our methods are valid for a representation established with labels suffering from Gaussian errors, as long as the salient information of the mapping from the input to the correct target output is preserved. While labeling errors should be avoided for bias correction, the requirement on the data quality may be less demanding for representation learning and the data with a variety of systematics may be used simultaneously. Despite this claim, we caution that the labeling errors drawn from a Gaussian distribution may not be representative of the actual distribution of catastrophic failures caused by e.g., misidentification of spectral lines. Further investigation with realistic labeling errors is required in future work.

\subsection{Impact of the model complexity} \label{sec:impact_model}

The complexity of a model determines the volume of information that can be learned from data and thus affects the performance of the model. A more complex model would generally have lower errors and biases in its prediction if not overfitted. We investigate whether our methods are robust with models that have different complexities using the SDSS data. Besides the networks from \citet{Pasquet2019} and \Treyer{} (denoted with ``\texttt{Net\_P}'' and ``\texttt{Net\_T}'', respectively), we experiment on other three simple networks that differ in the number of layers (summarized below). Likewise, the penultimate fully-connected layer of each network is used to obtain a representation.

\begin{enumerate}
\item \textbf{Net\_S1}
    \begin{itemize}
    \item 3 $\times$ 3 convolution with 64 kernels, stride 1, zero padding and the ReLU activation
    \item Global average pooling
    \item Concatenation with $E(B-V)$
    \item Fully-connected with 256 neurons and the ReLU activation
    \item Fully-connected with 256 neurons and the ReLU activation
    \item Fully-connected with 180 neurons or multi-channel output unit
    \end{itemize}

\item \textbf{Net\_S2}
    \begin{itemize}
    \item 3 $\times$ 3 convolution with 64 kernels, stride 1, zero padding and the ReLU activation
    \item 2 $\times$ 2 average pooling with stride 2
    \item 3 $\times$ 3 convolution with 64 kernels, stride 1, zero padding and the ReLU activation
    \item Global average pooling
    \item Concatenation with $E(B-V)$
    \item Fully-connected with 256 neurons and the ReLU activation
    \item Fully-connected with 256 neurons and the ReLU activation
    \item Fully-connected with 180 neurons or multi-channel output unit
    \end{itemize}
   
\item \textbf{Net\_S3}
    \begin{itemize}
    \item 3 $\times$ 3 convolution with 64 kernels, stride 1, zero padding and the ReLU activation
    \item 2 $\times$ 2 average pooling with stride 2
    \item 3 $\times$ 3 convolution with 64 kernels, stride 1, zero padding and the ReLU activation
    \item 2 $\times$ 2 average pooling with stride 2
    \item 3 $\times$ 3 convolution with 64 kernels, stride 1, zero padding and the ReLU activation
    \item Global average pooling
    \item Concatenation with $E(B-V)$
    \item Fully-connected with 256 neurons and the ReLU activation
    \item Fully-connected with 256 neurons and the ReLU activation
    \item Fully-connected with 180 neurons or multi-channel output unit
    \end{itemize}
\end{enumerate}

\begin{figure}[ht]
\begin{center}
\centerline{\includegraphics[width=\columnwidth]{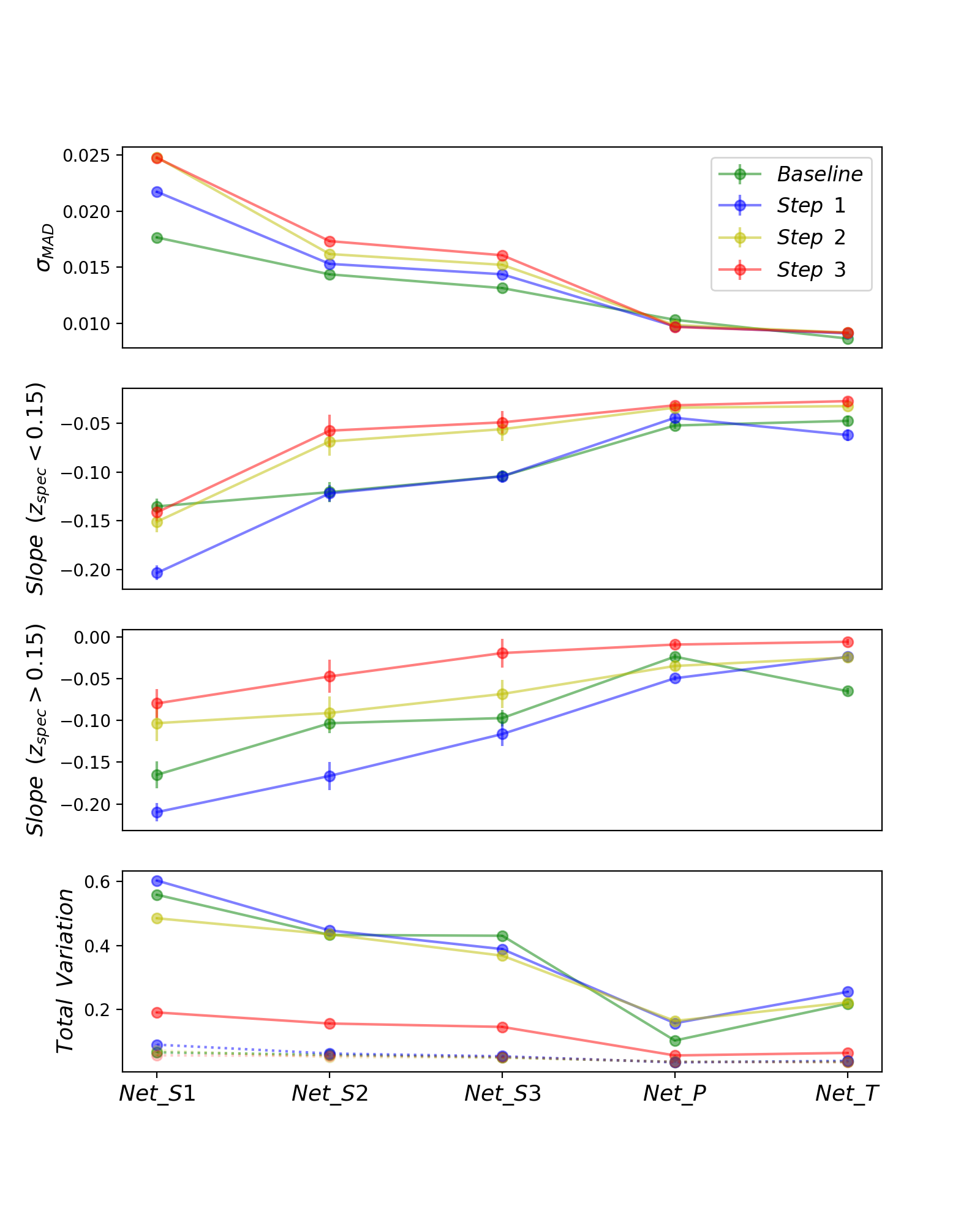}}
\caption{$\sigma_{\mathrm{MAD}}$ (Eq.~\ref{eq:sigma_mad}), the slopes of the $<\Delta z> - \,\, z_{spec}$ piecewise linear fit (Eq.~\ref{eq:resfit}) and the total variation distance between the $z_{photo}$ and $z_{spec}$ sample distributions (Eq.~\ref{eq:d_tv}) as a function of different network architectures with increasing complexity (defined in the text). The dotted curves show the total variation distances for the simulated $z_{photo}$ sample distributions that are expected to have no mode collapse. The Baseline method and Steps 1 -- 3 of our methods are compared using the SDSS data.}
\label{fig:impact_model}
\end{center}
\end{figure}

As shown in Fig.~\ref{fig:impact_model}, $\sigma_{\mathrm{MAD}}$, the slopes of the $<\Delta z> - \,\, z_{spec}$ curves and the total variation distance between the $z_{photo}$ and $z_{spec}$ sample distributions all evolve almost monotonically as the model complexity increases for the Baseline method and Step 1, implying that a model with a higher complexity would be more capable of reducing underfitting and producing a better estimation accuracy and smaller biases. The outcomes of bias correction degrade for the three simple networks, probably due to the fact that the representations in these cases do not contain as enough salient information for determining the output as in more complex models. This suggests that sufficient model complexity is required in order to get a good representation and effectively apply bias correction. In addition, all these cases show similar $<\Delta z> - \,\, z_{spec}$ curves in spite of different slopes, suggesting that it is the uneven data distribution that inherently determines the existence of class-dependent residuals.

\section{Evaluation on estimation quality}  \label{sec:evaluation}

\begin{table*}\footnotesize
\centering
\begin{tabular}{l | c c c | c c c | c c c}
\hline
   &  $<\Delta z_{mode}>$  &  $\sigma_{\mathrm{MAD}} (z_{mode})$  &  $\eta_{>0.05} (z_{mode})$  &  $<\Delta z_{mean}>$  &  $\sigma_{\mathrm{MAD}} (z_{mean})$  &  $\eta_{>0.05} (z_{mean})$  &  $<\Delta z_{median}>$  &  $\sigma_{\mathrm{MAD}} (z_{median})$  &  $\eta_{>0.05} (z_{median})$  \\
\hline
Baseline  &  $-0.00046$  &  0.00975  &  $0.51\%$  &  0.00028  &  0.00905  &  $0.27\%$  &  $-0.00104$  &  0.00899  &  $0.31\%$ \\
\hline
Step 1  &  $-0.00061$  &  0.00941  &  $0.41\%$  &  0.00251  &  0.01008  &  $0.33\%$  &  $-0.00096$  &  0.00893  &  $0.27\%$ \\
\hline
Step 2  &  $-0.00065$  &  0.00951  &  $0.51\%$  &  0.00274  &  0.01002  &  $0.35\%$  &  $-0.00093$  &  0.00904  &  $0.32\%$ \\
\hline
Step 3  &  $-0.00065$  &  0.00927  &  $0.49\%$  &  0.00348  &  0.01023  &  $0.44\%$  &  $-0.00085$  &  0.00915  &  $0.34\%$ \\
\hline
Step 4  &  0.00002  &  0.00919  &  $0.41\%$  &  0.00415  &  0.01052  &  $0.45\%$  &  $-0.00018$  &  0.00916  &  $0.30\%$ \\
\hline
\end{tabular}
\caption{Evaluation on photometric redshift estimation for the SDSS dataset.} \label{tab:eval_S}
\end{table*}

\begin{table*}\footnotesize
\centering
\begin{tabular}{l | c c c | c c c | c c c}
\hline
   &  $<\Delta z_{mode}>$  &  $\sigma_{\mathrm{MAD}} (z_{mode})$  &  $\eta_{>0.15} (z_{mode})$  &  $<\Delta z_{mean}>$  &  $\sigma_{\mathrm{MAD}} (z_{mean})$  &  $\eta_{>0.15} (z_{mean})$  &  $<\Delta z_{median}>$  &  $\sigma_{\mathrm{MAD}} (z_{median})$  &  $\eta_{>0.15} (z_{median})$  \\
\hline
Baseline  &  0.00846  &  0.01853  &  $4.23\%$  &  0.00225  &  0.01742  &  $5.37\%$  &  0.00215  &  0.01627  &  $4.10\%$ \\
\hline
Step 1  &  0.00288  &  0.02155  &  $4.51\%$  &  0.03943  &  0.04118  &  $7.78\%$  &  0.00851  &  0.01884  &  $4.52\%$ \\
\hline
Step 2  &  0.00926  &  0.02163  &  $4.90\%$  &  0.04634  &  0.04185  &  $8.60\%$  &  0.01115  &  0.01836  &  $4.18\%$ \\
\hline
Step 3  &  0.00626  &  0.02611  &  $7.76\%$  &  0.08473  &  0.06903  &  $17.80\%$  &  0.01276  &  0.02485  &  $7.41\%$ \\
\hline
Step 4  &  0.00347  &  0.02481  &  $7.25\%$  &  0.08193  &  0.07169  &  $20.91\%$  &  0.00996  &  0.02399  &  $7.19\%$ \\
\hline
\end{tabular}
\caption{Evaluation on photometric redshift estimation for the CFHTLS-DEEP dataset.} \label{tab:eval_D}
\end{table*}

\begin{table*}\footnotesize
\centering
\begin{tabular}{l | c c c | c c c | c c c}
\hline
   &  $<\Delta z_{mode}>$  &  $\sigma_{\mathrm{MAD}} (z_{mode})$  &  $\eta_{>0.15} (z_{mode})$  &  $<\Delta z_{mean}>$  &  $\sigma_{\mathrm{MAD}} (z_{mean})$  &  $\eta_{>0.15} (z_{mean})$  &  $<\Delta z_{median}>$  &  $\sigma_{\mathrm{MAD}} (z_{median})$  &  $\eta_{>0.15} (z_{median})$  \\
\hline
Baseline  &  0.00125  &  0.02047  &  $1.26\%$  &  0.00063  &  0.01970  &  $1.00\%$  &  $-0.00108$  &  0.01899  &  $0.99\%$ \\
\hline
Step 1  &  $-0.00098$  &  0.02094  &  $1.32\%$  &  0.03425  &  0.03809  &  $3.93\%$  &  $-0.00037$  &  0.01969  &  $1.02\%$ \\
\hline
Step 2  &  $-0.00013$  &  0.02057  &  $1.30\%$  &  0.03391  &  0.03755  &  $3.97\%$  &  $-0.00023$  &  0.01980  &  $1.02\%$ \\
\hline
Step 3  &  $-0.00172$  &  0.02128  &  $1.61\%$  &  0.06822  &  0.05964  &  $14.45\%$  &  0.00066  &  0.02160  &  $1.39\%$ \\
\hline
Step 4  &  0.00107  &  0.02052  &  $1.37\%$  &  0.07101  &  0.05985  &  $15.19\%$  &  0.00346  &  0.02167  &  $1.27\%$ \\
\hline
\end{tabular}
\caption{Evaluation on photometric redshift estimation for the CFHTLS-WIDE dataset.} \label{tab:eval_W}
\end{table*}

Despite a prioritized focus on bias correction in this work, we present a broader evaluation on the quality of estimated photometric redshifts. For each dataset (i.e., SDSS, CFHTLS-DEEP and CFHTLS-WIDE), we leverage three evaluation metrics utilized by \citet{Pasquet2019} and \Treyer{} --- the global mean residual $<\Delta z>$, $\sigma_{\mathrm{MAD}}$ (Eq.~\ref{eq:sigma_mad}), and the outlier fraction $\eta_{>\alpha}$ with $|\Delta z|>\alpha$. We adopt $\alpha=0.05$ for the SDSS dataset and $\alpha=0.15$ for the CFHTLS datasets. These metrics are applied to the three point estimates --- $z_{mode}$, $z_{mean}$ and $z_{median}$, obtained with the network \texttt{Net\_P}.

The results are listed in Tabs.~\ref{tab:eval_S}, \ref{tab:eval_D} and \ref{tab:eval_W}. Due to limited data, the $z_{photo}$ estimation for the CFHTLS-DEEP dataset has the worst accuracy, which makes our bias correction methods unreliable. As demonstrated in our work, the accuracy may be compromised by bias correction. In particular, there is a dramatic boost of errors for $z_{mean}$ with Step 3 especially for the CFHTLS datasets. We note that this only indicates the consequence of $z_{mode}$-based bias correction, yet similar trends would be found for $z_{mean}$-based or $z_{median}$-based correction. Although $\sigma_{\mathrm{MAD}}$ and $\eta_{>\alpha}$ for $z_{mode}$ may be worse than those for $z_{mean}$ or $z_{median}$, we adopt $z_{mode}$ to achieve a better control of biases (Appendix~\ref{sec:zcomparison}).

Finally, we caution that it might not be statistically meaningful to evaluate global properties for heterogeneous samples with these metrics. Given sufficient statistics, it would be more preferable to evaluate the behaviors of subsamples in which data instances are homogeneously sampled.

\end{appendix}
\end{document}